\documentclass[prd,
tightenlines,superscriptaddress,showpacs,
nofootinbib,eqsecnum,amsfonts,amsmath]{revtex4}
\usepackage{bm,graphics,graphicx,epsfig,color,ulem}
\allowdisplaybreaks
\def\be{\begin{equation}}
\def\ee{\end{equation}}
\def\ba{\begin{eqnarray}}
\def\ea{\end{eqnarray}}

\def\bse{\begin{subequations}}
\def\ese{\end{subequations}}
\def\bit{\begin{itemize}}
\def\eit{\end{itemize}}

\begin{document}
\title{Parameter estimation of neutron star-black hole binaries using an advanced 
gravitational-wave detector network: Effects of the full post-Newtonian waveform}
\author{Hideyuki Tagoshi}\email{tagoshi@vega.ess.sci.osaka-u.ac.jp}
\affiliation{Department of Earth and Space Science, Graduate School of Science, 
Osaka University, Osaka, 560-0043, Japan}
\author{Chandra Kant Mishra}\email{chandra@icts.res.in}
\altaffiliation[Presently at]{ the International Centre for Theoretical Sciences, 
Tata Institute of Fundamental Research, Bangalore 560012, India}
\affiliation{Indian Institute of Science Education and Research Thiruvananthapuram, 
Computer Science Building, College of Engineering Campus, Trivandrum, Kerala, 
695016, India} 
\author{Archana Pai}\email{archana@iisertvm.ac.in}
\affiliation{Indian Institute of Science Education and Research Thiruvananthapuram, 
Computer Science Building, College of Engineering Campus, Trivandrum, Kerala, 
695016, India}
\author{K. G. Arun}\email{kgarun@cmi.ac.in}
\affiliation{Chennai Mathematical Institute, Siruseri, Tamilnadu, 603103, India}

\begin{abstract} We investigate the effects of using the full waveform
(FWF) over the conventional restricted waveform (RWF) of the inspiral
signal from a coalescing compact binary system in extracting the
parameters of the source, using a global network of second generation
interferometric detectors. We study a hypothetical population of
(1.4-10)$M_\odot$ neutron star-black hole (NS-BH) binaries (uniformly distributed and oriented in the
sky) by employing the full post-Newtonian waveforms, which not only include
contributions from various harmonics other than the dominant one (quadrupolar
mode) but also the post-Newtonian amplitude corrections associated with each
harmonic, of the inspiral signal expected from this system. It is expected that
the GW detector network consisting of the two LIGO detectors and a Virgo
detector will be joined by KAGRA (a Japanese detector) and by proposed
LIGO-India. We study the problem of parameter estimation with all 16 possible
detector configurations. Comparing medians of error distributions obtained
using FWFs with those obtained using RWFs (which only include contributions
from the dominant harmonic with Newtonian amplitude) we find that the
measurement accuracies for luminosity distance and the cosine of the
inclination angle improve almost by a factor of 1.5-2 depending upon the
network under consideration. We find that this improvement can be attributed to
the presence of additional inclination angle dependent terms, which appear in
the amplitude corrections to various harmonics, which break the strong
degeneracy between the luminosity distance and inclination angle.  Although the
use of FWF does not improve the source localization accuracy much, the global
network consisting of five detectors will improve the source localization
accuracy by a factor of 4 as compared to the estimates using a three-detector
LIGO-Virgo network for the same waveform model.  \end{abstract}
\date{\today}  \maketitle
\section{Introduction} \label{sec:intro} Coalescing compact binary (CCB)
systems, composed of NSs and/or stellar mass BHs, are among the prime targets
for the second generation of GW detectors such as advanced LIGO \cite{aligo}
and advanced Virgo \cite{avirgo}. On the other hand, the proposed space-based
detector eLISA \cite{2012arXiv1201.3621A} shall be primarily looking at GW
signals from super massive BHs. In addition, although there are no
observational evidences for the existence of CCBs with intermediate mass BHs
(with masses of few tens to few hundred solar masses), if at all such systems
exist they should be observed by advanced ground-based detectors (see
\cite{lrr-2011-5} for a review on detection of GW sources from ground and
space).\footnote{We do not have observational evidence even for CCBs with
stellar mass BHs but the models related to the formation of stellar mass black
holes in close binaries are well supported by stellar evolution models (see for
example \cite{Fryer:2001fv}).} The GW observation of stellar/intermediate mass
CCB systems in advanced GW detectors will not only provide the first direct
evidence for the existence of GWs but also will reveal a great deal of
information about the source properties which cannot be accessed through
conventional electromagnetic observations. Hence, apart from the problem of
detection one is interested in estimating the parameters which characterize the
source. In the case of ground-based detectors, in general one would have a
situation when the GW signal is completely buried in the noise. Hence, in order
to be able to detect or to extract parameters of the source one employs data
analysis techniques such as matched filtering \cite{Helstrom68, Th300,
Schutz89}, which in turn requires accurate modeling of the dynamics of sources
emitting the signal. This has led to the development of many analytical and
numerical techniques which are used to model various stages of CCB evolution,
namely, the early inspiral phase, the late inspiral, the merger phase and the
final ringdown phase. For instance, the early inspiral phase can be very well
modeled using approximation schemes in General Relativity (GR) such as the
post-Newtonian (PN) approximation~\cite{Bliving}. The late inspiral and merger
phase can be computed by using Numerical Relativity~\cite{Pretorius07Review}
whereas the final ringdown phase can be accurately modeled using black hole
perturbation theory~\cite{TSLivRev03}. 

Although, in general it is believed that at the time of their formation all CCB
systems possess eccentric orbits, it is reasonable to assume that in late
stages of their evolution (this is precisely the stage when signals would be
visible in earth-bound detectors), their orbits would become circular due to
radiation reaction \cite{Pe64}.  During this phase the signal from a
nonspinning CCB can be approximated by a template whose frequency and amplitude
steadily increases until the last stable orbit is reached.  The
phase~\cite{BDIWW95,BFIJ02,BDEI04} and
amplitude~\cite{BIWW96,ABIQ04,KBI07,BFIS08} of GW signals from CCBs in this
stage has been computed to very high accuracies using the post-Newtonian
approximations in GR. Further, the fact that the amplitude of the signal in
this phase varies much more slowly as compared to the phase of the signal and
also because most of the signal power is contained in the dominant harmonic
(quadrupolar mode), it seems reasonable to approximate the signal to a template
which neglects contributions from harmonics other than the dominant one and
various post-Newtonian amplitude corrections associated with each harmonic. A
waveform obtained in this fashion is called the restricted waveform (RWF) and
contains only the dominant harmonic at twice the orbital frequency, with phase
which includes all the PN corrections to the leading phase term but only the
Newtonian amplitude. Note that modes other than the dominant one are suppressed
as they contribute to the waveform at a higher post-Newtonian order. In the
light of this argument we refer these additional modes together with the higher
order post-Newtonian corrections to the amplitude of the dominant harmonics as
subdominant  modes and would follow this terminology in rest of the paper. It
has been argued in a number of investigations that RWFs are good enough as far
as the detection of low mass binaries (M$<$10$M_\odot$) are concerned (see e.g.
\cite{ChrisAnand06}). Even CCBs as massive as 25$M_\odot$ can be detected using
template bank constructed using restricted waveform approximation of the
inspiral signals, however, the efficiency of extracting parameters reduces as
the mass of the binary increases (see the discussion in
Ref.~\cite{Farr:2009pg}). It was discussed in the case of single ground-based
detectors \cite{Chris06, ChrisAnand06} and in the case of space-based detector
LISA \cite{AISS07} that the mass reach of GW detectors can be significantly
increased by including contributions from subdominant  harmonics. Such a
waveform which includes contributions from various subdominant  harmonics and
the post-Newtonian amplitude corrections associated with each harmonic is refer
to as the full waveform (FWF).\footnote{In some places in the literature the
term FWF is used for inspiral-merger-ringdown waveforms. Here we simply call
such waveforms as IMR waveforms or complete waveforms and reserve the term FWF
for inspiral waveforms including the contributions from the subdominant
harmonics and amplitude corrections associated with different harmonics.}
Although, as we move towards the higher mass end, even subdominant  harmonics
fail to penetrate the frequency band where the detector is most sensitive. In
that case it becomes important to include the contributions from the merger
phase of the binary evolution. 

A recent work by Capano {\it et al}.~\cite{Capano:2013raa} suggests that,
inspiral-merger-ringdown (IMR) waveforms based on just the contributions from
dominant harmonic will be sufficient for detecting signals from binaries with
total mass up to 360$M_\odot$. However, it was also mentioned that, for systems
with total mass $>100M_\odot$ and with mass ratios $>4$, indeed the sensitivity
of the search improves if the waveform includes contributions from subdominant
modes . Another recent study based purely on numerical waveforms
\cite{Pekowsky:2012sr} suggests that with the inclusion of subdominant  modes
of the waveform the detection volume can be significantly increased (by about
30\%) as compared to what could be achieved by using waveforms based on the RWF
approximation of the inspiral signal. Some of the previous studies showed that
that the inclusion of subdominant  modes in the model of the GW signal not
only improves the mass-reach and the detection rates of future GW detectors but
also provides a more powerful template to match with the signal in order to
extract the parameters of the source accurately in context of single ground
based detectors \cite{SinVecc00a, ChrisAnand06b, Littenberg:2012uj} and in the
case of space-based LISA \cite{SinVecc00b, MH02, HM03, AISSV07,
TriasSintes07,PorterCornish08} for nonspinning binaries (see also
Ref.~\cite{KKS95,AISS05} which use RWF to investigate the quality of parameter
estimation). Effects of the use of FWF over RWF on parameter estimation for
{\it precessing} binaries was discussed in a recent paper by O'Shaughnessy et
al. \cite{O'Shaughnessy:2014dka}, where they show how the inclusion of
subdominant  modes improves the parameter estimation for precessing NS-BH
systems observed in next generation of ground-based GW detectors. This is
possible as the FWF, by the virtue of contributions from subdominant  modes,
has a great deal of structure, which enables one to extract parameters of the
source more efficiently as compared to the case when RWF is used (see
Ref.~\cite{ChrisAnand06b} for a discussion). Further, since the inclusion of
subdominant  modes  in the waveform brings explicit dependences on the
inclination angle of the binary, the degeneracy between the inclination angle
and the distance of the source, which persists in the case of RWF, finally
breaks. This leads to better measurement of the inclination angle of the
source. Since inclination angle and distance are strongly correlated, an
improvement in the measurement of the inclination angle further improves the
distance measurement. In addition, as we shall see below, with FWF the
polarization angle measurement also improves. This together with the
inclination angle measurement enables one to constrain the orientation of the
binary significantly. 

Since in the future we shall have a network of five ground based detectors, one
can analyze the data from different detectors coherently \cite{Pai:2000zt}.
Such an analysis shall not only enable one to have larger detection volume but
also help one to estimate the parameters of the sources much more accurately as
compared to the accuracies that can be achieved using the single detector data.
Most importantly, networks with three or more detectors will be able to
localize the source very accurately, which is of great importance to
astrophysics and fundamental physics (see \cite{Schutz:2011tw} for a detailed
discussion). The problem of parameter estimation in context of the future
network of ground based detectors has been studied extensively in the past
\cite{JK94, JKKT96, Ajith:2009fz, 2010PhRvD..81h2001W, Nissanke:2011ax,
Klimenko:2011hz, Fairhurst:2012tf, Schutz:2011tw, Sathya.LIGOIndia}. All of
these studies used RWF approximation of the GW signal to show how a network of
three or more detectors shall improve the localization (or in general the
measurements of parameters of the source) of the CCB system observed in the
earth-bound detectors. However, Rover {\it et al}.~\cite{Rover:2006bb} considered a
network consisting the initial LIGO detectors and the Virgo and investigated
the accuracies with which parameters of a BNS system can be measured. They used
inspiral waveforms with 2PN amplitude and phase up to 2.5PN order and used
their Markov chain Monte Carlo (MCMC) routine for coherent parameter
estimation. Recently, the effect of higher signal harmonics on parameter
estimation of a BH-NS system was investigated in \cite{Cho:2012ed,
O'Shaughnessy:2013vma} in context of a fiducial (idealized) network of two
interferometric detectors using an effective Fisher matrix approach
introduced in \cite{Cho:2012ed}. 

In this work we aim to study the effects of using the FWF over RWF on the
parameter estimation for a typical nonspinning CCB system, in context future GW
interferometric detectors using the Fisher information matrix approach
\cite{Finn92,FinnCh93}.  For this purpose we consider a population of NS-BH
systems (with component masses as (1.4, 10 M$_\odot$)), all placed at a
luminosity distance of 200 Mpc and distributed uniformly over the sky surface.
We run simulations for about 12800 realizations obtained by randomly choosing
the angular parameters giving the location and orientation of the binary. We
make use of an inspiral waveform which includes amplitude corrections to
various harmonics consistent up to 2.5PN order and phasing up to 3.5PN order
\cite{ABIQ04}.\footnote{Inspiral waveforms with amplitude corrections to
various harmonics consistent up to 3PN order are already available
\cite{BFIS08} but in the present study we chose to work with a waveform which
is 2.5PN accurate in amplitude.} Since it is convenient to use the waveforms in
frequency domain in the Fisher information matrix approach, we use the
frequency domain waveform obtained with the stationary phase approximation
\cite{Th300} of the Fourier transformation of the time domain waveform of
\cite{ABIQ04}. This was already computed in \cite{ChrisAnand06} and here we
just use the waveform obtained there.

The organization of the paper is as follows. In Sec.~\ref{sec:pe}, we first
discuss the future network of advanced detectors along with the noise curves
for individual detectors used in the present study. Next, we introduce our
waveform model and discuss various coordinate frames which have been chosen to
obtain the response of the each detector of the network. We discuss briefly our
parameter estimation strategy which broadly includes the details of Fisher
matrix formalism. Finally we close this section by providing the details of the
system under investigation and other analysis details.  In
Sec.~\ref{sec:results} we list main features of the improvement in parameter
estimation due to the use of FWF and compare the results for various
multidetector networks. We have added a subsection to address the implications
of including the LIGO-India in the global network of detectors. Finally, in
Sec.~\ref{sec:discn} we summarize our results and give some future directions.  
\section{Parameter Estimation} 
\label{sec:pe} 
\subsection{The advanced network}
\label{subsec:advnet} 

It is expected that the future worldwide network of
interferometric GW detectors would consist of five kilometer-scale
detectors (with 3-4 km-long arms) at five distinct locations across the globe.
Initially, the United States hosted three of the LIGO detectors at two different sites.
Two of the LIGO detectors (one 4 km long and other 2 km long) were installed at
the Hanford site and shared the same vacuum system. The third detector was
installed at Livingston and had 4-km-long arms. Currently, the LIGO detectors
are undergoing major upgrades to second generation detectors (advanced LIGO)
\cite{aligo} and are expected to become operational by the end of the year
2015. Virgo is a French-Italian detector with 3-km-long arms and has been
installed at Cascina, Italy.  Similar to the LIGO detectors it is also going
through major upgrades towards the construction of advanced Virgo \cite{avirgo}
and is expected to start taking data by early 2016. The Japanese  detector,
KAGRA (with 3km long arms), has been funded and is being constructed. This is
expected to be operational by the end of year 2015 with initial configuration.
The KAGRA with full configuration using cryogenic mirrors is expected to be
operational by the year 2018 \cite{kagra, Aso:2013eba}. In addition, there is a
proposal for a 4-kilometer-long-arm detector in India by the year 2022
(LIGO-India) \cite{indigo}.\footnote{Note that, in the advanced era, two 4-km-long
arm-length detectors, namely, one in Livingston and the other in Hanford will
be operational known as aLIGO detectors. The third detector which was
originally planned to be in Hanford with 4-km arm length would move to India, if
the LIGO-India project is approved.} Hence, in less than a decade time we 
might have a fully operational network of five second generation detectors
which will include, LIGO-Livingston (L), LIGO-Hanford (H), advanced Virgo (V),
KAGRA (K), and LIGO-India (I). Having five detectors at five sites means that
in total we shall have 16 different network configurations (of three or more
detectors) which will include ten 3-detector networks (LHV, LHK, LHI, LVK, LVI,
LKI, HVK, HVI, HKI, VKI), five 4-detector networks (LHVK, LHVI, LHKI, LVKI,
HVKI) and one 5-detector network (LHVKI). Hence, as compared to the LIGO-Virgo
network, which shall have just one 3-site network (assuming the duty cycle of
the two detectors at Hanford site are not independent), the future network
shall have 16 different configurations with three or more detectors. Assuming
that each detector in the network shall have a duty cycle of 80\%, the LHV
network would have a duty cycle of $(0.8)^3\sim 51\%$, the net duty cycle of
all possible 16 network combinations (with five detectors at five locations)
reaches to $(0.8)^5+5(0.2)(0.8)^4+10(0.2)^2(0.8)^3\sim 94\%$ (see
\cite{Schutz:2011tw} for a detailed discussion). This will ensure that most of
the time at least three or more detectors will be taking data. This is of prime
importance when one is interested in localizing the source which requires a
minimum of three site network.

\begin{figure}[ht]
\includegraphics[width=0.70\textwidth,angle=0]{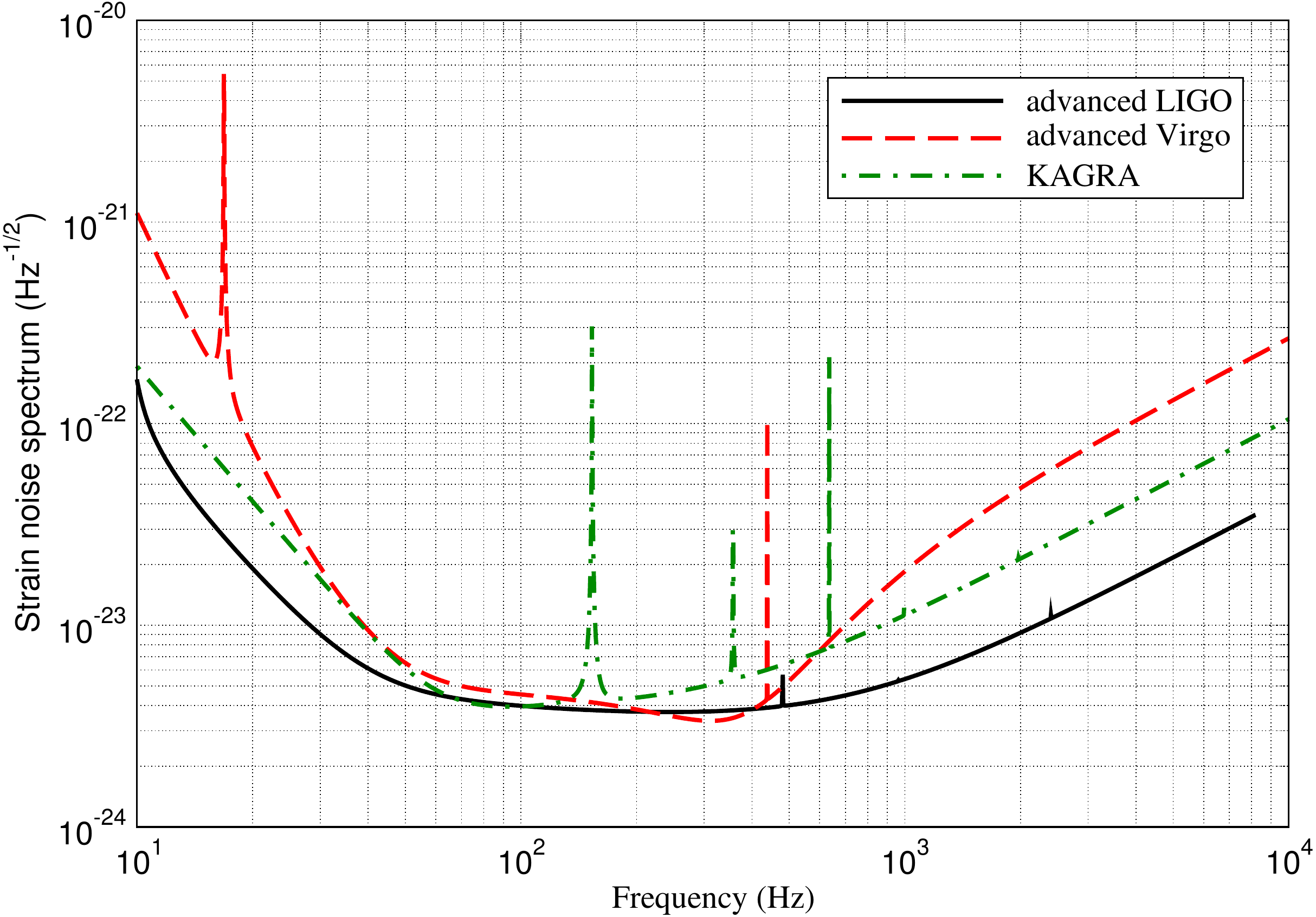}
\caption{One-sided noise power spectral density for advanced LIGO, advanced
Virgo, and KAGRA.} \label{fig:noisecurve} \end{figure}

Figure~\ref{fig:noisecurve} displays the expected one-sided noise power
spectral density for advanced LIGO, advanced Virgo and KAGRA. For all three
LIGO detectors (L, H, I) we use the sensitivity curve labeled "Zero
Det, High P" which can be found in \cite{ligopsd}. For KAGRA we use the curve
labeled "VRSE(B)" which can be found in \cite{kagrapsd}, whereas
the advanced Virgo noise can be found at the advanced Virgo project home page
\cite{avirgo}.

\subsection{The Waveform model} \label{subsec:wfmodel} 

The amplitude corrected post-Newtonian (PN) waveforms in the two polarizations 
(plus and cross), up to the 2.5PN order, were first computed in 
Ref.~\cite{ABIQ04} and take the following form, 

\be h_{+,\times}=\frac{2M\nu}{D_L} x \, \left\{H^{(0)}_{+,\times} +
x^{1/2}H^{(1/2)}_{+,\times} + x H^{(1)}_{+,\times} +
x^{3/2}H^{(3/2)}_{+,\times} + x^2 H^{(2)}_{+,\times} + x^{5/2}
H^{(5/2)}_{+,\times} \right\}\,.  \label{eq:hpluscross} \ee

Here, $M=(m_1+m_2)$ is the total mass of the binary where as $\nu=(m_1
m_2/M^2)$ is a dimensionless mass parameter which is termed as the symmetric
mass ratio and $D_L$ denotes the distance to the binary (or luminosity
distance). $x$ is the dimensionless PN expansion parameter and is related to
the binary's instantaneous orbital frequency, $F(t)$, as $x=(2 \pi M
F(t))^{2/3}$ (in units where $G=c=1$).  Finally, the coefficients
$H^{(n/2)}_{+, \times}$ where $n=0, \cdots, 5$, are linear combinations of
various harmonics with prefactors that are functions of the inclination angle
($\iota$) of the binary's angular momentum vector with respect to the line of
sight and the symmetric mass ratio $\nu$ (see \cite{ABIQ04} for explicit
expressions).

The strain in the detector arms due to the signal also depends on the location
and orientation of the binary through detector beam pattern functions ( $F_+$
and $F_\times$) and can be given as 

\be h(t)=F_+ h_+(t) + F_\times h_\times(t) \,, \label{eq:strain} \ee

where $F_+$ and $F_\times$ in terms of the angular parameters ($\theta, \phi$)
giving location of the binary and the polarization angle ($\psi$) giving the
binary's orientation in the plane of sky take the following form

\ba F_+(\theta,\phi,\psi) &=&
\frac{1}{2}\left(1+\cos^2(\theta)\right)\cos(2\phi)\cos(2\psi) 
- \cos(\theta)\sin(2\phi)\sin(2\psi)\,,\\ F_\times(\theta,\phi,\psi) &=&
  \frac{1}{2}\left(1+\cos^2(\theta)\right)\cos(2\phi)\sin(2\psi) +
\cos(\theta)\sin(2\phi)\cos(2\psi) \,.  \label{eq:beampatternfunctions} \ea \\

After combining Eqs.~\eqref{eq:hpluscross}-\eqref{eq:beampatternfunctions}
along with expressions for $H^{(n/2)}_{+, \times}$ listed in Ref.~\cite{ABIQ04}
one can write the expression for the strain in the detector arms as a linear
combination of different harmonics of the orbital phase ($\Psi$) in the
following way

\be h(t) = \sum_{k=1}^7 \sum_{n=0}^5 A_{(k,n/2)} x^{n/2}(t) \cos(k\Psi(t) +
\varphi_{(k,n/2)})\,, \label{lincomb} \ee

where $k$ runs over various harmonics and $n/2$ denotes the PN order. Note that
at the 2.5PN order, apart from the dominant harmonic ($k$=$2$), six additional
harmonics ($k$=$\{1,3,4,5,6,7\}$) contribute to the waveform. The coefficients
$A_{(k,n/2)}$ and the phase offsets $\varphi_{(k,n/2)}$ are functions of the
parameters ($D_L$, $M$, $\nu$, $\theta$, $\phi$, $\iota$, $\psi$) for the
signal observed in the detector and can be assumed to be constants for a given
ground-based detector for the duration of the observed signal
\cite{ChrisAnand06, ChrisAnand06b}. 

Since for the present analysis we shall be using Fisher information
matrix approach, it is convenient to use the waveforms in frequency domain.
The waveform (2.5PN accurate in amplitude and 3.5PN accurate in phase) in the
frequency domain is computed by using the stationary phase approximation, and
is given in Ref.~\cite{ChrisAnand06, ChrisAnand06b}. We simply recall it here,re

\begin{eqnarray} \tilde{h}(f) &=&  \frac{2M\nu}{D_L}
\,\sum_{k=1}^{7}\,\sum_{n=0}^5\, \frac{A_{(k,n/2)}
\,e^{-\imath\,\varphi_{(k,n/2)} }\,x^{\frac{n}{2}+1}\left (t\left (f_k\right
)\right )} {2\sqrt{k\dot{F}\left (t\left (f_k\right )\right )}} \times
e^{\imath(2 \pi f t_c - \pi/4 + k \Psi\left (f_k \right ))} \,, \label {eq:FT}
\end{eqnarray}

where \footnote{Note that here $f$ is the Fourier transform variable and should
not be confused with the instantaneous orbital frequency $F$ of the signal.}
$f_k=f/k$ and the Fourier phase $\Psi(f)$ \cite{BFIJ02} is given by

\begin{equation} \Psi(f) = - \Phi_c + {3\over 256\,\nu} \sum_{j=0}^7\psi_j\,(2
\pi M f)^{(j-5)/3} \,, \label{eq:FourierPhase} \end{equation}

where the coefficients $\psi_j$ read

\begin{eqnarray} \psi_0 &=&1,\nonumber \\   \psi_1 &=&0,\nonumber \\  \psi_2
&=&\frac{3715}{756}+\frac{55}{9}\nu,\nonumber \\   \psi_3 &=&-16 \pi,\nonumber
\\  \psi_4 &=&\frac{15293365}{508032}+\frac{27145}{504} \nu+\frac{3085}{72}
\nu^2, \nonumber \\  \psi_5 &=&\pi
\left(\frac{38645}{756}-\frac{65}{9}\nu\right) \left[1+\ln\,\left({f\over
F_{LSO}}\right)\right], \nonumber \\ \psi_6
&=&\frac{11583231236531}{4694215680}-\frac{640}{3}\pi ^2-\frac{6848}{21}C
+\left(-\frac{15737765635}{3048192}+ \frac{2255}{12}\pi ^2\right)\nu
\nonumber\\&+&\frac{76055}{1728}\nu^2-\frac{127825}{1296}\nu^3
-\frac{6848}{63}\ln\left(128\,\pi M\, f\right),\nonumber\\ \psi_7 &=&\pi\left(
\frac{77096675}{254016}+\frac{378515}{1512} \nu -\frac{74045}{756}\nu
^2\right).  \label{eq:phasingcoefficients2} \end{eqnarray}

Here $t_c$ and $\Phi_c$ appearing in above expressions denote the time and
phase at the coalescence epoch. $t_c$ can be freely specified in any
calculation, and we choose $t_c$=$0$. On the other hand, there is a dependence
on $\Phi_c$ in signal-to-noise ratio and in the Fisher matrix defined in Eq. 
(\ref{Fishermatrix}) below. This dependence comes from the cross products of
different $k$ modes in $\tilde{h}(f)$. However, such cross product terms are
highly oscillating in frequency domain, and their contribution to the integral
of Eq. (\ref{eq:inner product}) become very small, and the dependence of the final
results on $\Phi_c$ is not very large. We thus choose $\Phi_c=0$ in this paper.
To add more to this, we find in our simulations that if we randomly choose our
$\Phi_c$ in the interval of $[0, 2\pi]$, maximum relative change in the error
estimation is not more than about $2$-$7$\% for any given detector combination
or parameter under study. Also note that the quantity $F_{\rm LSO}$ denotes the
orbital frequency of the binary at the last stable orbit (LSO) and can be
approximated as $F_{\rm LSO}={1/(6^{3/2}\,2\,\pi\,M})$, the orbital frequency
at LSO of a test particle moving in Schwarzschild geometry of an object with
mass as the total mass ($M$) of the binary in $G=c=1$ units. It turns out that
most of the terms (except the ones which are proportional to the factor $\ln
f$) appearing in the expression for $\psi_5$ given by
Eq.~\eqref{eq:phasingcoefficients2} can be absorbed into a new definition of
$\Phi_c$ while performing computations as they have no frequency dependence.
Finally, the PN expressions for $\dot{F}$ is given in \cite{ChrisAnand06,
ChrisAnand06b} and we simply recall it here, 

\ba \dot{F} (f) &=& \frac{96}{5\pi M^2}(2 \pi M f)^{11/3} \left[1 -
\left(\frac{743}{336} +\frac{11}{4}\nu\right)(2 \pi M f)^{2/3} + 4\,\pi (2 \pi
M f)  \right. \nonumber\\ && \left. \quad +
\left(\frac{34103}{18144}+\frac{13661}{2016}\nu + \frac{59}{18}\nu^2 \right)(2
\pi M f)^{4/3} - \left(\frac{4159\pi}{672} + \frac{189\pi}{8}\nu\right)(2 \pi M
f)^{5/3}\right]\,.  \label{freqsweep} \ea

Before we proceed it is important to note that the term $\dot{F}$ can be
treated in many different ways which would lead to small numerical differences
in the results. For instance, one can reexpand the factor $1/\sqrt{\dot{F}}$
in the amplitude and then truncate the resulting amplitude at the working PN
order \cite{AISS07, AISSV07} or one may completely skip performing this
reexpansion. In this work we follow the latter treatment and use the
expression for $\dot{F}$ at the same PN order as that of the signal amplitude.
For instance, when using FWF with 2.5PN amplitude corrections we use the
$\dot{F}$ expression which is 2.5PN accurate but we do not perform any
reexpansion.  
We find that the difference to the parameter estimation errors discussed below
caused by the difference of this treatment are around at most 7\%. 
Thus, the absolute value of the parameter estimation accuracy 
might have errors at this level due to this choice.  
However, the different treatment results in both FWF and RWF in the same way.  
Thus, the conclusions about the comparison between FWF and RWF do not change.

\subsection{Coordinate frames and the detector response} 

In the previous subsection, we listed expressions for the strain in detector arms
(response of the detector to the incoming GW signal) due to the presence of the
signal, both in time and frequency domains. It was mentioned there that the
response of the detector to the incoming signal also depends on the binary's
position and orientation through beam pattern functions given by
Eq.~\eqref{eq:beampatternfunctions}. When dealing with a network of detectors
which consists of detectors at different locations around the globe, response
of each detector to the signal will be different.  Reference \cite{Pai:2007ms}
shows how a set of rotation transformations between appropriately chosen
coordinate systems can tell us the response of each detector. In this section
we shall recall the related result of Ref.~\cite{Pai:2007ms} for the
completeness of the text and refer to the paper for definitions and details.
The main idea is as follows.

Let us choose three coordinate frames associated with the wave, detector and
the Earth denoted by $\mathbf{x_w}\equiv(x_w,y_w,z_w)$,
$\mathbf{x_d}\equiv(x_d,y_d,z_d)$ and $\mathbf{x_E}\equiv(x_E,y_E,z_E)$,
respectively (see IIIA of Ref.~\cite{Pai:2007ms} for definitions). Let ${\cal
O}$ be the rotation operator which transforms one frame to other given three
Euler angles. Then if the set ($\phi_e, \theta_e, \psi_e$) characterizes the
transformation between the Earth frame and the wave frame and the set ($\alpha,
\beta, \gamma$) characterizes the transformation for the detector-Earth frame,
we can have (see Fig.~1 of Ref.~\cite{Pai:2007ms} for a graphical display of
these transformations)

\ba \mathbf{x}_w &=& {\cal O}(\phi_e, \theta_e, \psi_e) \mathbf{x}_E,
\label{WtoE}\\ \mathbf{x}_d &=& {\cal O}(\alpha, \beta, \gamma ) \mathbf{x}_E.
\label{DtoE} \ea

In the present convention the source Euler angles ($\phi_e, \theta_e, \psi_e$)
in terms of the angular parameters describing the location ($\theta, \phi$) and
the polarization angle ($\psi$) in the Earth frame are given as

\begin{align} \label{eq:pts} \phi_e&= \phi-\pi/2,& \theta_e&= \pi-\theta,&
\psi_e&=\psi, \end{align}

On the other hand, the detector Euler angles $(\alpha,\beta,\gamma)$ are given
in terms of the location and orientation of the detector as:

\begin{align} \label{eq:alpha} \alpha &= L+\pi/2,\\ \label{eq:beta} \beta &=
\pi/2-l,\\ \label{eq:gamma} \gamma &= \frac{a_1+ a_2}{2}+ \frac{3 \pi}{2} &
\text{if $|a_1-a_2| >   \pi$}, \\ &= \frac{a_1+ a_2}{2}+ \frac{  \pi}{2} &
\text{if $|a_1-a_2| \le \pi$}, \label{eq:abg} \end{align}

where $l$ and $L$ are the latitude and longitude of the detector site. The
angles $a_1$ and $a_2$ describe the orientation of the first and second arm,
respectively.  In Table~\ref{tab:det_euler} of this paper we provide the
information about the location and orientation of various detectors considered
in this analysis.

\begin{table*}
\caption{Location and orientation information of the future Earth-based interferometric GW detectors \cite{LIGO-T0980044-v1, Pai:2000zt}. 
  The location of each detector is given in terms of the latitude and longitude. The orientation of 
  an arm is given by the angle through which one must rotate it clockwise (while viewing from top) 
  to point the local North. The corresponding detector Euler angles $(\alpha,\beta,\gamma)$ are 
  listed. Note that, for the location of the LIGO-India detector we use 
  the values listed in Table~I of Ref.~\cite{Schutz:2011tw}. This location was chosen as a fiducial site for 
  the detector.
}
\begin{center}
\label{tab:det_euler}
\begin{tabular}{lccccccc}
\hline\\
Detector & Vertex & Vertex & Arm 1 & Arm 2 & $\alpha$ & $\beta$ & $\gamma$ \\ 
& latitude (N) & longitude (E) & {$a_1$} &{$a_2$}   & & & \\ \hline 
LIGO Livingston (L) & $30^\circ 33' 46.4196'' $ & $-90^\circ 46' 27.2654''$ & $107.72^\circ $ & $197.72^\circ$ & $-0.77^\circ$  &$59.44^\circ$& $242.72^\circ$ \\
LIGO Hanford (H) & $46^\circ 27'18.528''$ & $-119^\circ 25'27.5657''$  & $36^\circ$ & $126^\circ$ & $-29.41^\circ$ & $43.55^\circ$ & $171.0^\circ$ \\
VIRGO (V) & $43^\circ 37' 53.0921''$  & $10^\circ30'16.1878''$   & $340.57^\circ$ & $70.57^\circ$ & $100.5^\circ$  & $46.37^\circ$ & $116.0^\circ$\\
KAGRA (K) & $36^\circ 14' 60''$ & $137^\circ 10' 48''$  & $295^\circ$ & $25^\circ$ &$227.18^\circ$ & $53.75^\circ$ & $70^\circ$ \\
LIGO-India (I) & $19^\circ 05' 47''$ & $74^\circ 2' 59''$    & $45^\circ$& $135^\circ$ &$164.05^\circ$ & $70.90^\circ $ & $180^\circ$ \\
\hline\\
\end{tabular}
\end{center}  
\end{table*}

The coordinate transformation between the wave frame and the detector frame can
be obtained by combining Eq.~\eqref{WtoE} and Eq.~\eqref{DtoE}

\be \mathbf{x}_w = {\cal O}(\phi'_e, \theta'_e, \psi'_e) \mathbf{x}_d \,,
\label{eq:WtoD} \ee

where ${\cal O}(\phi'_e, \theta'_e, \psi'_e) \equiv {\cal O}(\phi_e, \theta_e,
\psi_e) {\cal O}^{-1}(\alpha,\beta,\gamma)$.

It should be evident from the above that, transformations associating the
detector frame with the wave frame can be split into two rotations: one from the
detector frame to the Earth frame and one from the Earth frame to the wave frame.
These two successive transformations can be translated into the addition
theorem of Gel'fand functions \cite{Gelfand1}, which reads as 

\be \label{eq:add} T_{mn}(\phi'_e,\theta'_e,\psi'_e)=\sum_{l=-2}^{2}
T_{ml}(\phi_e,\theta_e,\psi_e) T^*_{nl} (\alpha,\beta,\gamma) \,, \ee

where $T_{ij}$ denotes the Gel'fand functions.  The detector response due to the
incoming GW (or the strain induced by the signal in the detector arms) is given
by Eq.~\eqref{eq:strain} which in a more compact notation can be written as

\be \label{eq:hcomp} h(t)\equiv \Re[f_c^*\,h_c]\,, \ee

where $f_c=F_{+}+i\,F_{\times}$ and $h_c=h_{+}+i\,h_{\times}$ are defined as
complex antenna pattern function and complex GW signal, respectively  (see the
discussion in Sec. IIB and in Appendix A of Ref.~\cite{Pai:2007ms}), where the (*)
indicates the complex conjugate of $f_c$, and $\Re$ represents the real part.\\

In addition to this, since detectors in the network will be located at
different places around the globe, the incoming GW signal shall arrive at
various detector sites at different instances. In order to correctly account
for the time delays between the arrival times at different detectors one has to
choose a reference frame with respect to which all the time measurements are
performed. Following Ref.~\cite{Pai:2007ms} we choose this reference frame to
be the frame attached to the center of the earth. In such case the response of
$I$-th detector (after folding in the effect of delays) 

\be \label{eq:hcompwithdelay} h^I(t)\equiv \Re[f_c^{I
*}h_c(t-\tau_I(\theta,\phi))]\,, \ee

where $\tau_I(\theta, \phi)=({\bf r_I}-{\bf r_E})\cdot {\bf w}(\theta,
\phi)/c$, denotes the time-delay in the arrival times of the incoming signal at
the detector and at the center of the Earth.  Quantities ${\bf r_I}$ and ${\bf
r_E}$ denote the vectors directed to locations at the detector and the Earth's
center, from the origin of the reference frame chosen (here it is Earth's
center itself). ${\bf w}(\theta, \phi)$ is the unit vector along the
propagation of the wave with $\theta$, $\phi$ again giving the source location
in a frame attached to the center of the Earth and $c$ denotes the speed of
light.

It was discussed in detail in the appendix of Ref.~\cite{Pai:2007ms} that one
can write the complex pattern function ($f_c$ above) in terms of Gel'fand
functions as (see Eq.~(B13) there) 

\be \label{eq:ant} f_c^I = \sum_{s=-2}^2 i T_{2s}
(\phi_e,\theta_e,\psi_e)[T_{2s}(\alpha,\beta,\gamma)-T_{-2s}(\alpha,\beta,\gamma)]^*.
\ee

Given the source Euler angles ($\phi_e,\theta_e,\psi_e$) and the $I$-th
detector Euler angles ($\alpha^I,\beta^I,\gamma^I$) given by Eq.~\eqref{eq:pts}
and Eq.~\eqref{eq:abg}, along with definition of Gel'fand functions, one can
calculate $f_c^I$ for a given $I$-th detector and hence the response of
individual detectors to the signal both in time and frequency domain. With
these inputs we go on to describe our parameter estimation strategy in the next
section.

\subsection{Error Estimation} 
\label{subsec:errest} 

The inspiral signal from the nonspinning compact binary systems can be characterized in terms of total
nine parameters (see Sec.~\ref{subsec:wfmodel} above). This means we have a
nine dimensional parameter space which reads 

\be \label{eq:param} \mathbf{p}=\{\ln D_L\,, {\cal M}_c, \delta, t_c, \Phi_c,
\cos(\theta), \phi, \psi, \cos(\iota)\}\ \, , \ee

where, ${\cal M}_c=M\,\nu^{3/5}$ is termed as the {\it Chirp Mass} and
$\delta=|m1-m2|/m$ is called the difference mass ratio parameter.  We  employ
the Fisher matrix approach~\cite{Finn92,FinnCh93} to see how well we can
constrain these parameters. Below we briefly discuss our strategy for
estimating various parameters of the source which is based on the Fisher matrix
approach.  We first define the matched filter signal-to-noise ratio (SNR) of a
network of $N$ detectors, $\rho$, as

\begin{eqnarray} \label{eq:defSNR} \rho&=&\left[\sum_{I=1}^N
(h^I|h^I)_I\right]^{1/2}.  \end{eqnarray}

Here, $(\,|\,)_I$ denotes the noise weighted inner product for $I$-th detector.
In general, for any two functions $g$ and $h$, their inner product is defined
as:

\begin{equation} (g\,|\,h)_I\equiv 4 \Re \int_{f_{min}}^{f_{max}}df\,
\frac{\tilde g^{*}(f)\,\tilde h(f)}{S_h^{(I)}(f)}.\label{eq:inner product}
\end{equation}

Here $S_h^{(I)}(f)$ represent the one-sided noise power spectral density of
$I$th detector. The limits of integration $[f_{min}, f_{max}]$ are determined
by both the detector and by the nature of the signal. Since we are using
inspiral waveform, which is usually not reliable beyond the last stable orbit
we can choose to terminate the integrals when the last stable orbit is reached.
For instance, we assume that the contribution from $k$th harmonic to the
waveform is zero above the frequency $k F_{\rm LSO}$, where $F_{\rm LSO}$ is
the orbital frequency at the last stable orbit~\cite{ChrisAnand06}. Since the
amplitude-corrected waveform we are using in this work has seven harmonics, we
set the upper cutoff to be 7$F_{\rm LSO}$ when we use the FWF in the analysis.
For lower cutoff, as power spectral densities $S_h(f)$ tend to rise very
quickly below a certain frequency $f_s$ where they can be considered infinite
for all practical purposes, we may set it to be $f_s$. 

Let $\tilde{\theta}_{a}$ denote the `true values' of the parameters and let
$\tilde{\theta}_{a}+\Delta\theta_{a}$ be the best-fit parameters in the
presence of some realization of the noise.  Then for large SNR, error in the
estimation of parameters $\Delta\theta_{a}$ obey a Gaussian probability
distribution~\cite{Helstrom68,Wainstein,Finn92,FinnCh93} of the form

\begin{equation} p({\bf \Delta\theta})=p^{(0)} \exp\left
[-\frac{1}{2}\Gamma_{bc}\Delta\theta_{b}\Delta\theta_{c}\right ],
\label{eq:prob-dist} \end{equation}

where ${{\bf \Delta \theta} = \{ \Delta \theta_a \}}$ and repeated indices are
summed up. The $p^{(0)}$ is a normalization constant.  The quantity
$\Gamma_{ab}$ appearing in Eq.~\eqref{eq:prob-dist} is the Fisher information
matrix and is given by,

\begin{equation} \Gamma_{ab}=(h_{a}\,|\,h_{b}) \label{Fishermatrix}
\end{equation}

where $h_{a}\equiv \partial h/\partial \theta^a$.  Using the definition of the
inner product, one can reexpress the Fisher matrix associated with the
$I$-{th} detector $\Gamma_{ab}^{I}$ more explicitly as

\begin{equation} \Gamma_{ab}^I=(h_{a}^I\,|\,h_{b}^I)_I\equiv 4
\int_{f_{s}}^{k\,F_{LSO}}df\,\frac{\Re([\tilde{h}_{a}^I(f)]^*
[\tilde{h}_{b}^I(f)])}{S_h(f)}\,, \label{eq:gamma-eqn} \end{equation}

The Fisher matrix for a network of $N$ detectors is simply the sum of individual
Fisher matrices associated with different detectors and is given by 

\begin{equation} \Gamma_{ab}=\sum_{I=1}^{N}\Gamma_{ab}^I \end{equation}

The covariance matrix, defined as the inverse of the Fisher matrix, is given by 

\begin{equation} \Sigma_{ab} \equiv \langle \Delta \theta_a \Delta \theta_b
\rangle = ( {\Gamma_{ab}})^{-1}, \label{sigma_a} \end{equation}

where $\langle \cdot \rangle$ denotes an average over the probability
distribution function in Eq.~(\ref{eq:prob-dist}).  The root-mean-square error
$\sigma_a$ in the estimation of the parameters $\theta_{a}$ is

\begin{equation} \sigma_a = \bigl\langle (\Delta \theta_a)^2 \bigr\rangle^{1/2}
= \sqrt{\Sigma_{aa}}\,, \label{eq:sigma_a} \end{equation}

\subsection{Numerical simulations}
\label{subsec:systeminputs}

As discussed in Sec.~\ref{sec:intro}, in this paper we investigate the
parameter estimation problem for a compact binary system consisting a NS
($1.4M_\odot$) and a BH ($10M_\odot$).  Despite the fact that BNS systems are
expected to be seen more often in ground-based detectors as compared to the
NS-BH systems, here we chose to study asymmetric systems. This is because the
contribution from odd harmonics (k=1,3,5,7) is directly proportional to the
asymmetry of the system described by the parameter $\delta=|m_1-m_2|/M^2$. This
would mean that for symmetric or nearly symmetric systems (such as BNS systems)
such terms either would not contribute or shall have small effects. Since one
of the prime goals of the present study is to investigate the improvements in
parameter estimation accuracies due to inclusion of subdominant  modes of the
signal we must choose a system which is sufficiently asymmetric. Hence, we
expect that with increasing asymmetry of the binary, subdominant  modes of the
signal become more and more important (as also odd ones would then start
contributing significantly) which eventually leads to better estimation of
parameters. Moreover, effects of subdominant  harmonics are expected to be
more important for heavier systems as the dominant mode fails to enter the
sensitive part of detector bandwidth \cite{ChrisAnand06b, AISS07}. However, it
should be noted that parameter estimation shall in general be poor for such
systems as they will be observed with smaller SNRs (since the dominant harmonic
either does not contribute or its contribution is negligible).  Another reason
is related to the question of correctness of the PN waveform itself for systems
heavier than 12$M_\odot$ and with larger mass ratios (as different approximants
start showing deviations from each other)~\cite{BIOPS2009}. We could have
considered even more asymmetric NS-BH systems, for which the effect of
subdominant  modes  would be even more. But one should bear in mind that, for
heavier NS-BH binaries, the neglect of merger and ringdown waveforms are going
to be even more important than PN subdominant  modes and hence we do not
consider them here.  Keeping the above constraints in mind we choose to study a
population of NS-BH system with neutron star mass as $1.4M_\odot$ and BH mass
as $10M_\odot$. 

We assume a population of NS-BH systems ((1.4-10)$M_\odot$), all placed at a
luminosity distance of 200 Mpc. 
The choice of the distance is rather arbitrary. 
The results of the parameter estimation error for 9 parameters are inversely
proportional to the distance for both FWF and RWF, and for 
any detector combinations. 
Let the parameter estimation errors (\ref{eq:sigma_a}) with distance $r$ 
to be $\sigma_a^r$.  
The parameter estimation errors with other distance $r=r'$
is given as $\sigma_a^{\rm 200Mpc}r'/(200$Mpc).

In total we consider 12800 realizations of the
source uniformly distributed over the sky and obtained by randomizing the
angular parameters specifying the location ($\cos(\theta), \phi$) and
orientation ($\cos(\iota), \psi$) of the binary. The nine-dimensional parameter
space given by Eq.~\eqref{eq:param} shall lead to the $9\times9$ Fisher matrix
which is further used to compute errors in various parameters for each one of
these realizations. However, errors in $\cos (\theta)$ and $\phi$ can be
combined to give error in the solid angle ($\Omega$) centred around the source.
Following \cite{BarackCutler04}, we define

\be \label{eq:omegaAdef} \Delta\Omega_S=2\pi\sqrt{\sigma^2_{\cos(\theta)}
\sigma^2_\phi - \Sigma^2_{\cos(\theta)\,\phi}} \ee

where $\Sigma_{\cos(\theta)\,\phi}$ is the covariance between $\cos (\theta)$
and $\phi$. 

As discussed in \cite{BarackCutler04}, the probability that the source lies
outside an error ellipse enclosing solid angle $\Delta\Omega$ is
$e^{-\Delta\Omega/\Delta\Omega_S}$.  We then adopt $\Delta\Omega_{95}\equiv
3\Delta\Omega_S$ as our definition of the source localization error, which
represents approximately 95\% confidence region of the localization error
ellipses.

\subsection{Accuracy of the numerical computation}
The covariant matrix is obtained by inverting the Fisher matrix. In this paper,
this is done numerically with the LU decomposition in the GSL
library~\cite{GSL}. Some of the results are also computed and are confirmed
with MATHEMATICA \cite{MATHEMATICA}. Numerical inversion of matrices often
suffer from the problem of accuracy due to the ill-conditioned Fisher matrices.
We check the accuracy of the matrix inversion by multiplying the inverse with
the original matrix, and check the deviation of it from the identity matrix.
Similar to Berti {\it et al}. \cite{BBW05a}, we define $\epsilon_{\rm
inv}=\max_{i\neq j}|\left(\Gamma\, \Sigma \right)_{ij}|$, and use it as a
measure of the accuracy of the matrix inversion.

We find that in the case of FWF, $\epsilon_{\rm inv}$ is distributed in a
Gaussian-like form with mean value of around $10^{-12}$, and the maximum is
about $10^{-10}$.  Since the numerical computation is done with double
precision, the round off error is around $10^{-15}$, we can say that this
accuracy is good enough.  On the other hand, in the case of RWF, the
distribution of $\epsilon_{\rm inv}$ has a tail at larger value up to $\sim
10^{-3}$. In addition, we also find the correlation between ($\sigma_{\ln
D_L}$, $\sigma_{\Phi_c}$, $\sigma_{\psi}$, $\sigma_{\cos\iota}$) and
$\epsilon_{\rm inv}$.  

The error of the matrix inversion for RWF is caused by the ill condition of the Fisher matrix. 
It mainly occurs when $\iota$ is near $0$ or $\pi$. In such cases, 
the derivatives of $\tilde{h}(f)$ with respect to $\ln D_L$ and $\cos(\iota)$ 
become nearly proportional to each other, which makes the Fisher matrix 
ill conditioned. 
On the other hand, because of the complex dependence of the amplitude of FWF 
on $\cos(\iota)$, such a problem does not occur in the case of FWF. 

Since we can not trust the results of the cases with
large $\epsilon_{\rm inv}$, we decided not to use the results with
$\epsilon_{\rm inv}>10^{-8}$.  With this prescription, around 5\% of the cases
for RWF are removed and are not used in the final results.  We checked that if
we change the criteria to $\epsilon_{\rm inv}>10^{-10}$, the median of
$\sigma_{\ln D_L}$, $\sigma_{\psi}$, and $\sigma_{\cos\iota}$ are changed at
most about 30\%.  The changes of the median error of $\sigma_{\Phi_c}$ is at
most 13\%.  The changes of the median error of $\Delta\Omega$ is at most 16\%.
The changes of the median errors of other parameters are at most 10\%.  We
conservatively adopt these value as estimate of the accuracy of the median of
the error of the parameter estimation. 

In addition to the accuracy of the matrix inversion, in the Fisher matrix analysis,
there is a problem in low SNR cases.  We find that, for a small fraction of the
source population, the network SNR is smaller than the value 8.  Since, the
Fisher matrix approach can not be trusted for weak signals (those with smaller
SNRs), we remove such cases from our final results.  As a result, about $5$ \%
of cases for both FWF and RWF are removed for the 3 detector cases.  Note
however that this does not change the median of the error of all parameters
significantly. The change is only about $8$\% for all parameters.  When the
number of detector is 4 or 5, the network SNR is larger than 3 detector cases.
Thus, the effect of this SNR threshold is smaller than these value.

\section{Results}\label{sec:results}

The results of our exhaustive parameter estimation exercise and interpretations
of the trends observed are discussed in this section. The  improvement in the
parameter estimation due to the use of FWF in the multidetector framework
comes from a combination of two independent contributions: the improvement due
to additional features of FWF and the effect of additional detectors which
observe the signal. Hence the first part (\ref{subsec:RWF&FWF}) of the section
discusses the effect of FWF on parameter estimation as compared to the RWF and
in the second part (\ref{subsec:compNetwork}) we compare our results for
various detector combinations with three or more detectors.  We choose to
quantify the measurement accuracy of various parameters by the {\it median}
values of the error distributions since the median is unaffected by the tail of
the distribution. Further, the width of the distribution is given by the {\it
inter-quartile range}. The inter-quartile range (denoted by Q3-Q1) is defined
as the difference between the third (Q3: upper quartile) and the first quartile
(Q1: lower quartile) and represents the width of the distribution around the
median.\footnote{Q1, median and Q3 represent error values which would contain
25\%, 50\% and 75\% of source population.} Thus the two numbers collectively
give the range in which error in the measurement of a parameter varies about
the {\it median error} for 50\% of the population.

\subsection{Effect of the use of FWF over RWF on parameter accuracy}
\label{subsec:RWF&FWF}
\subsubsection{LHV}
\label{subsubsec:LHV}

\begin{figure*}[!htbp]
\includegraphics[width=0.40\textwidth,angle=0]{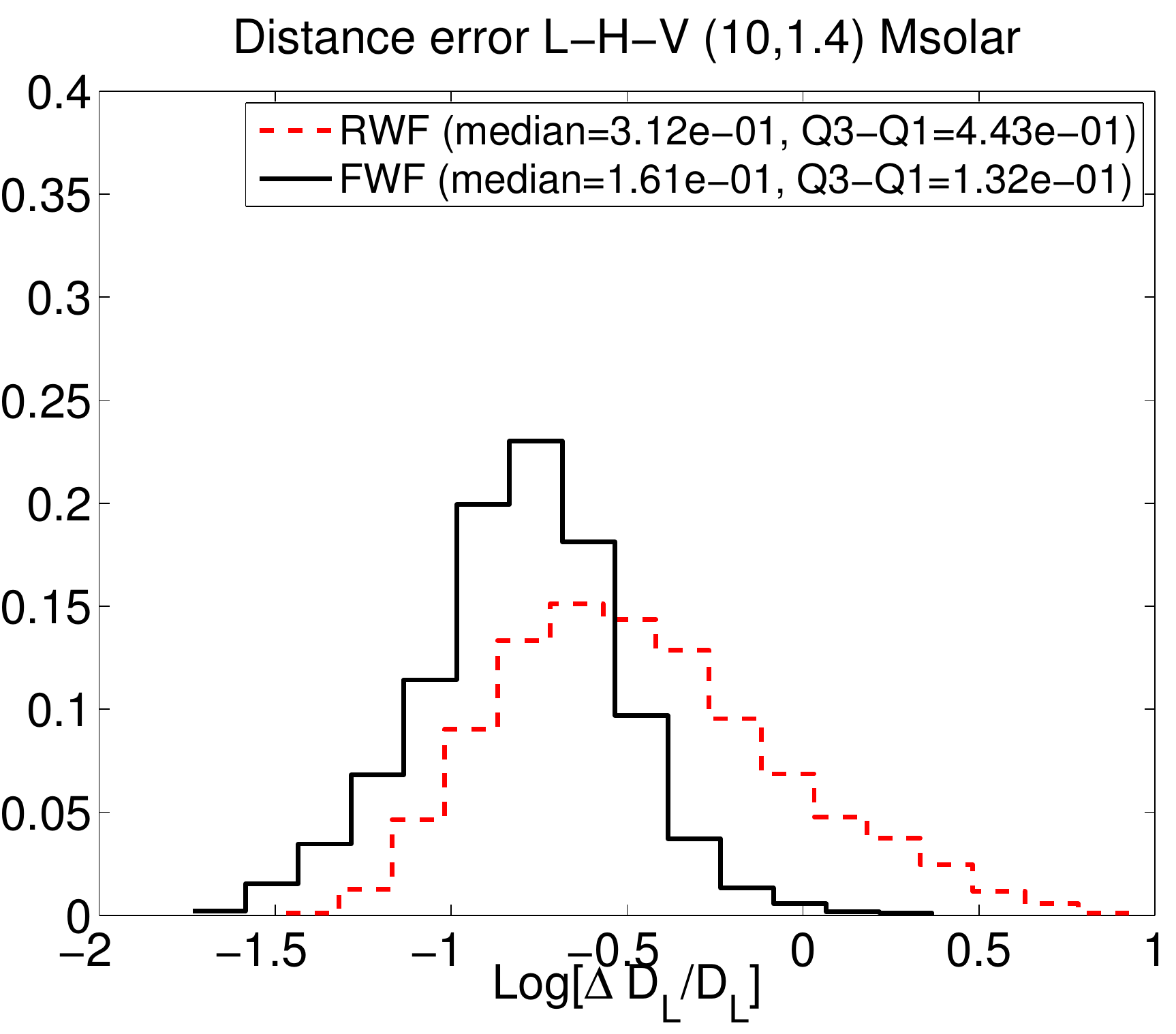}
\hskip 0.2cm
\includegraphics[width=0.40\textwidth,angle=0]{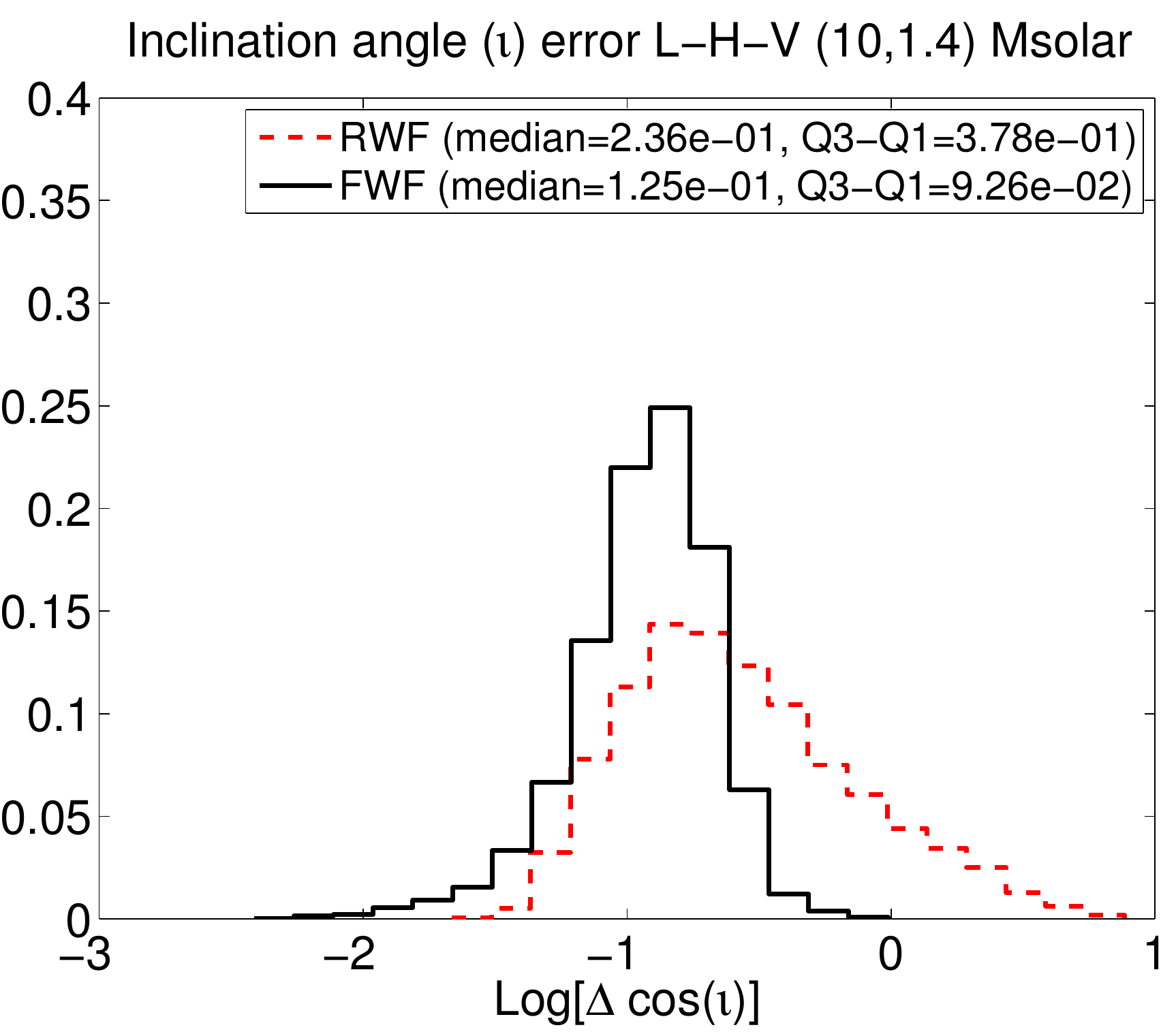} \vskip
0.2cm \includegraphics[width=0.40\textwidth,angle=0]{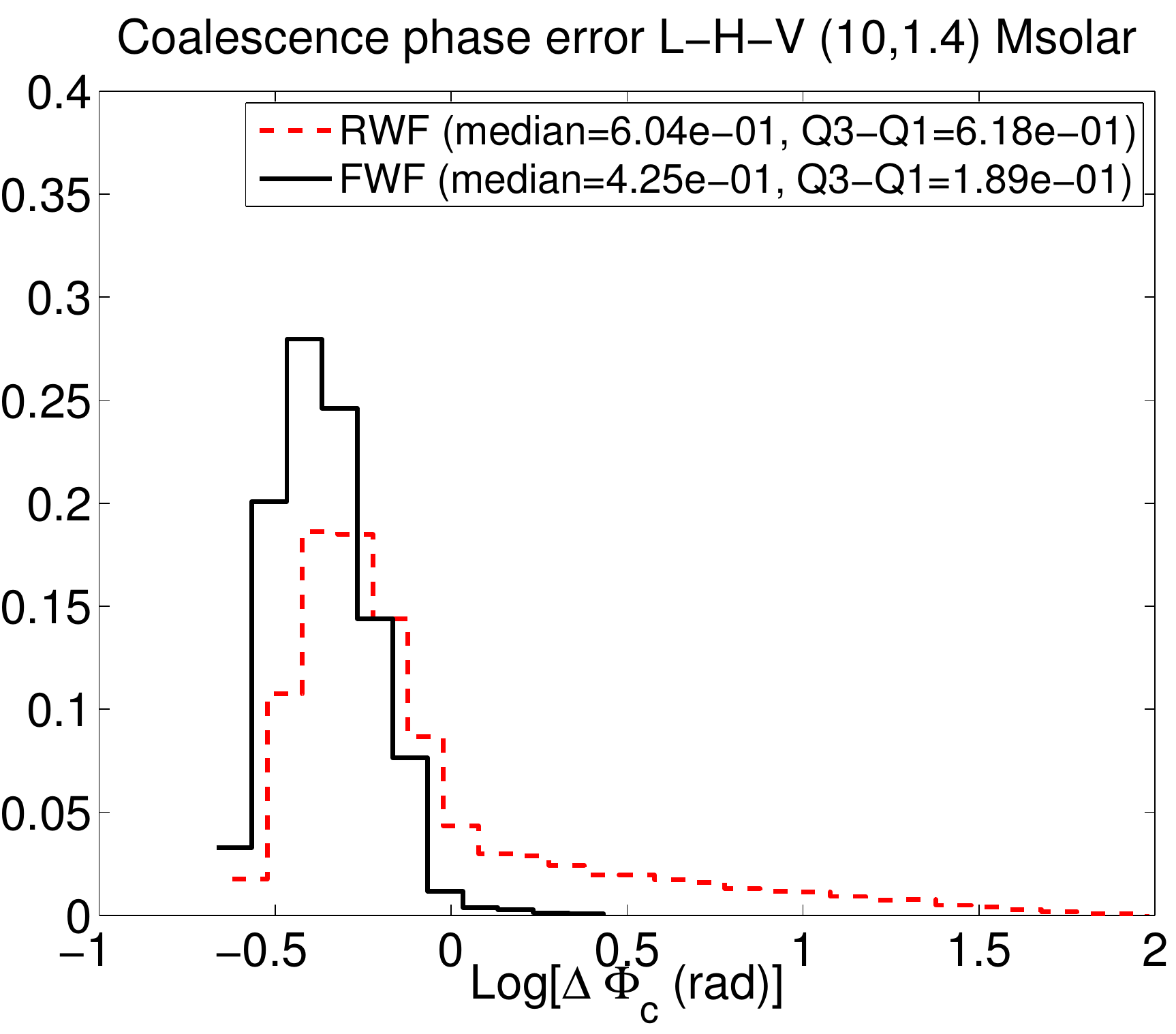}
\hskip 0.2cm
\includegraphics[width=0.40\textwidth,angle=0]{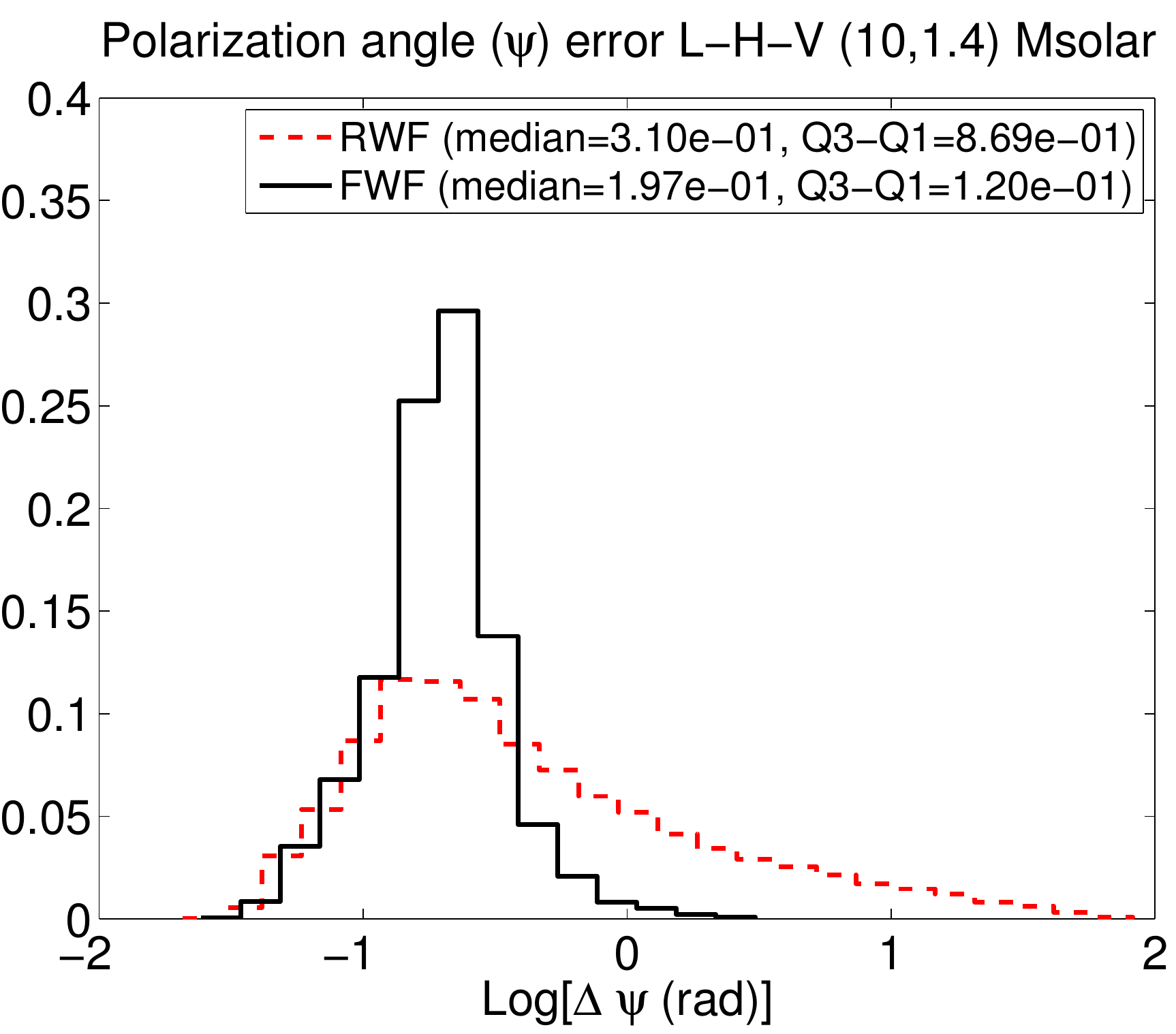}
\caption{Figure displays error distributions corresponding to the luminosity
distance ($D_L$), cosine of the inclination angle ($\cos (\iota)$), phase at
the time of coalescence ($\Phi_c$) and the polarization angle ($\psi$), in
context of the LIGO-Virgo network (LHV). The study has been performed for a
population of ($1.4-10$)$M_\odot$ NS-BH binaries, all placed at 200 Mpc and
distributed and oriented uniformly over the sky surface. The error
distributions obtained by choosing two different waveform models for signals
from the source, the restricted waveform (RWF) and the full waveform (FWF), are
compared. Since the error distributions are asymmetric (usually with a long
tail) we chose "median" as a reliable measure of accuracies with which
parameters are measured. In addition we also wish to give the interquartile
range, represented by Q3-Q1 [ difference between the upper quartile (Q3) and the
lower quartile (Q1) ], corresponding to each error distribution. Median and the
interquartile range (Q3-Q1) corresponding to each parameter has been displayed
in various panels. Note that out of nine parameters listed in
Eq.~\eqref{eq:param} we are displaying graphical results only for four of them
since effect of the use of FWF over RWF is only significant in these four
cases. However, in Table~\ref{tab:medianerror} we list medians corresponding to
error distribution of all the parameters.} \label{fig:compRvsFLHV}
\end{figure*}

In this section we aim to study the effects of using the FWF over the RWF on
measurement accuracies of various parameters in context of the LIGO-Virgo (LHV)
network.  Note that here we choose to display the error distributions for only
four of the nine parameters ($D_L$, $\cos(\iota)$, $\Phi_c$, and $\psi$) (see
Fig.~\ref{fig:compRvsFLHV}). This is mainly to avoid proliferation of graphical
details, as in the case of other parameters the error distributions
corresponding to the two cases (RWF and FWF) largely are same both in shape and
in positioning.  However, we display medians of error distributions
corresponding to all nine parameters in Table~\ref{tab:medianerror}. Different
panels in Fig.~\ref{fig:noisecurve} also display two numbers corresponding to
the median and the interquartile range.

It should be obvious from the shifts observed in different panels of
Fig.~\ref{fig:compRvsFLHV} that the FWF indeed significantly improves the
measurements of the parameters ($D_L$, $\cos(\iota)$, $\Phi_c$, and $\psi$).
This is not surprising as in general the FWF, by the virtue of contributions
from subdominant  modes, has a great deal of structure, which enables one to
extract parameters of the source more efficiently as compared to the case when
RWF is used (see Ref.~\cite{ChrisAnand06b} for a discussion). Comparing the
median of distributions related to the errors in $D_L$ and $\cos(\iota)$ we
find that the accuracies with which the two parameters will be measured will
improve by almost a factor of about 2 and those related to $\Phi_c$ and $\psi$
improve by a factor of about 1.5. It is noteworthy that we find such
improvements despite slightly smaller signal-to-noise ratio (SNR) for the FWF
cases. 

\begin{figure*}[tbp]
\includegraphics[width=0.40\textwidth,angle=0]{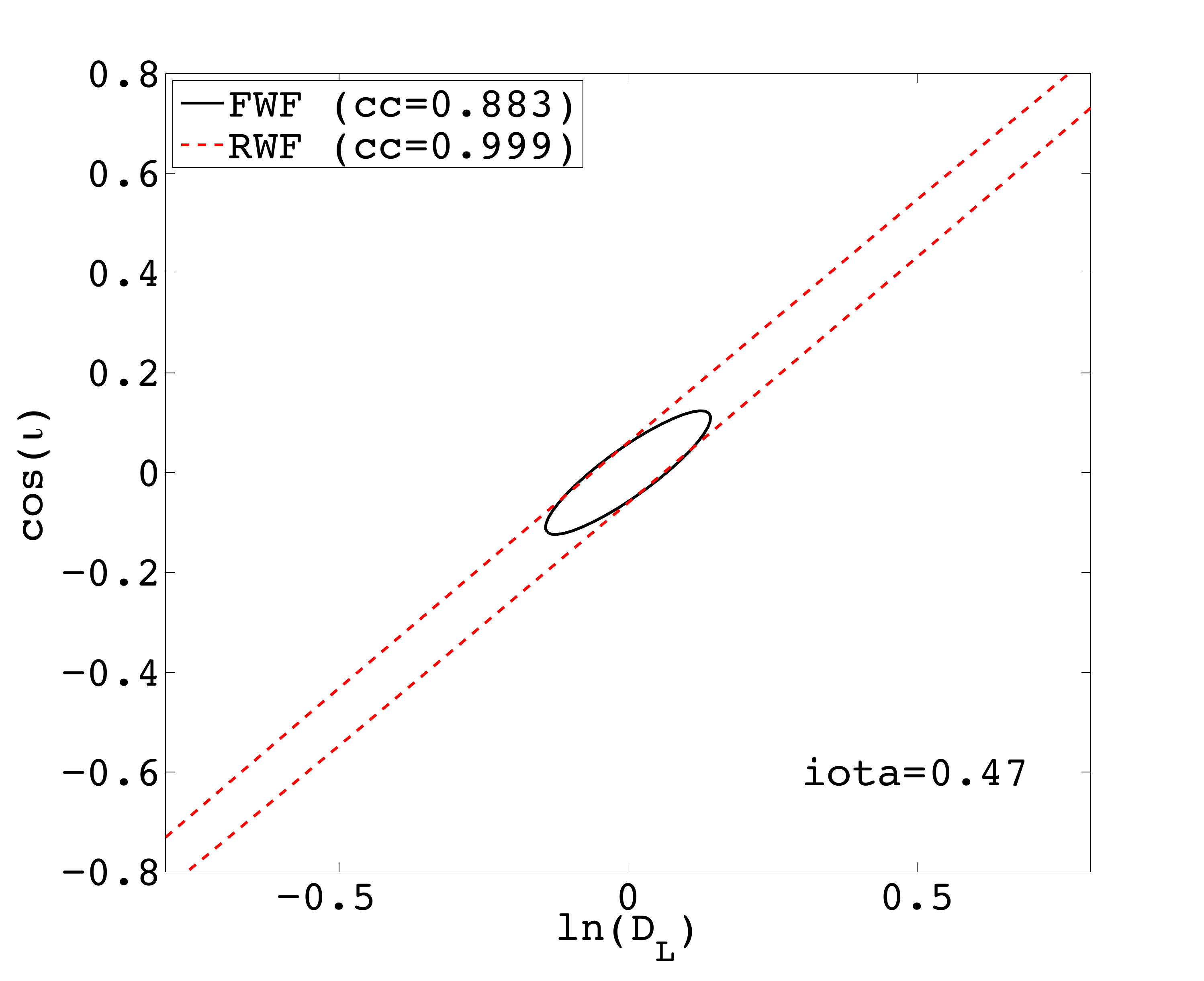}
\hskip 0.2cm
\includegraphics[width=0.40\textwidth,angle=0]{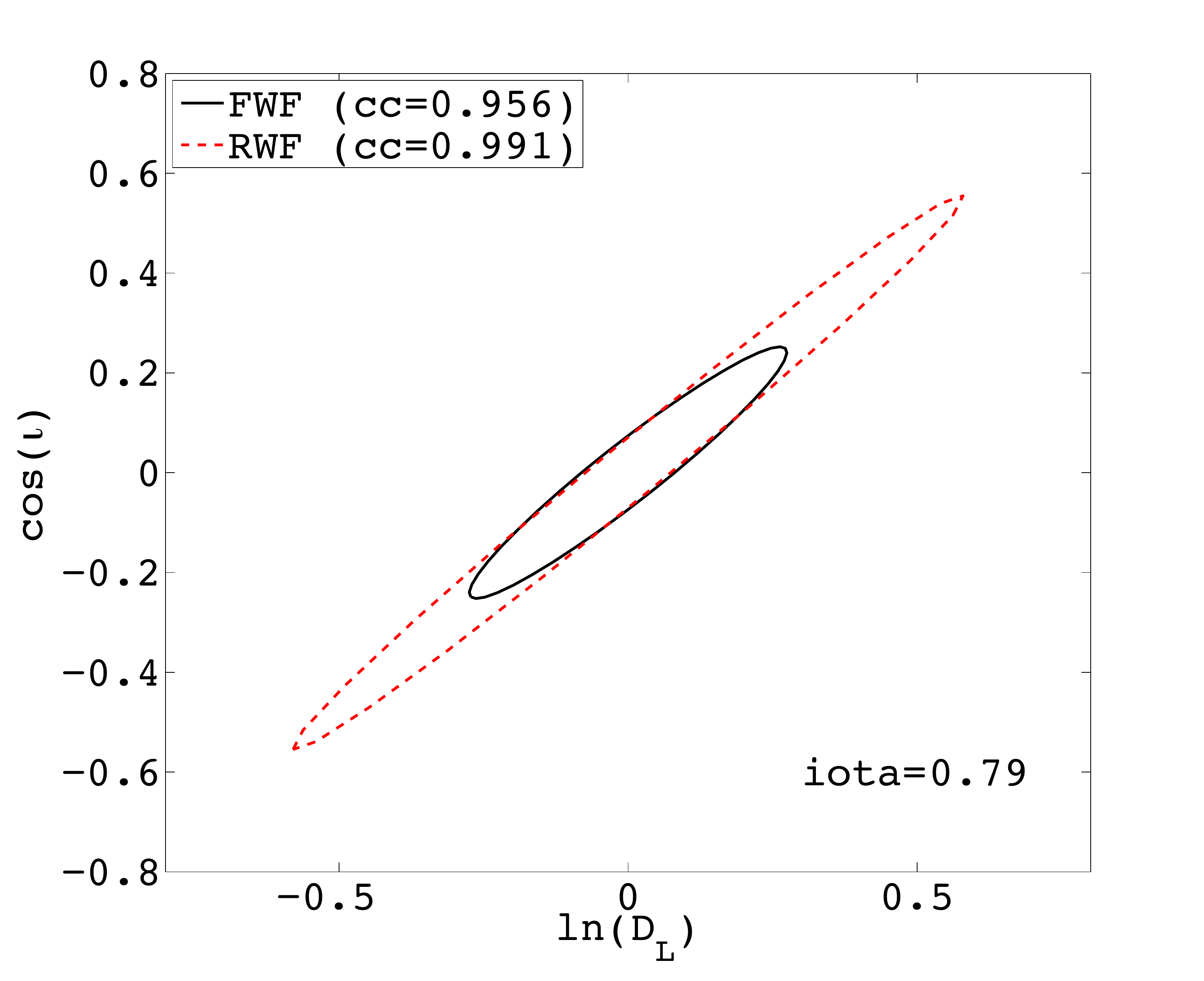}
\vskip 0.2cm 
\includegraphics[width=0.40\textwidth,angle=0]{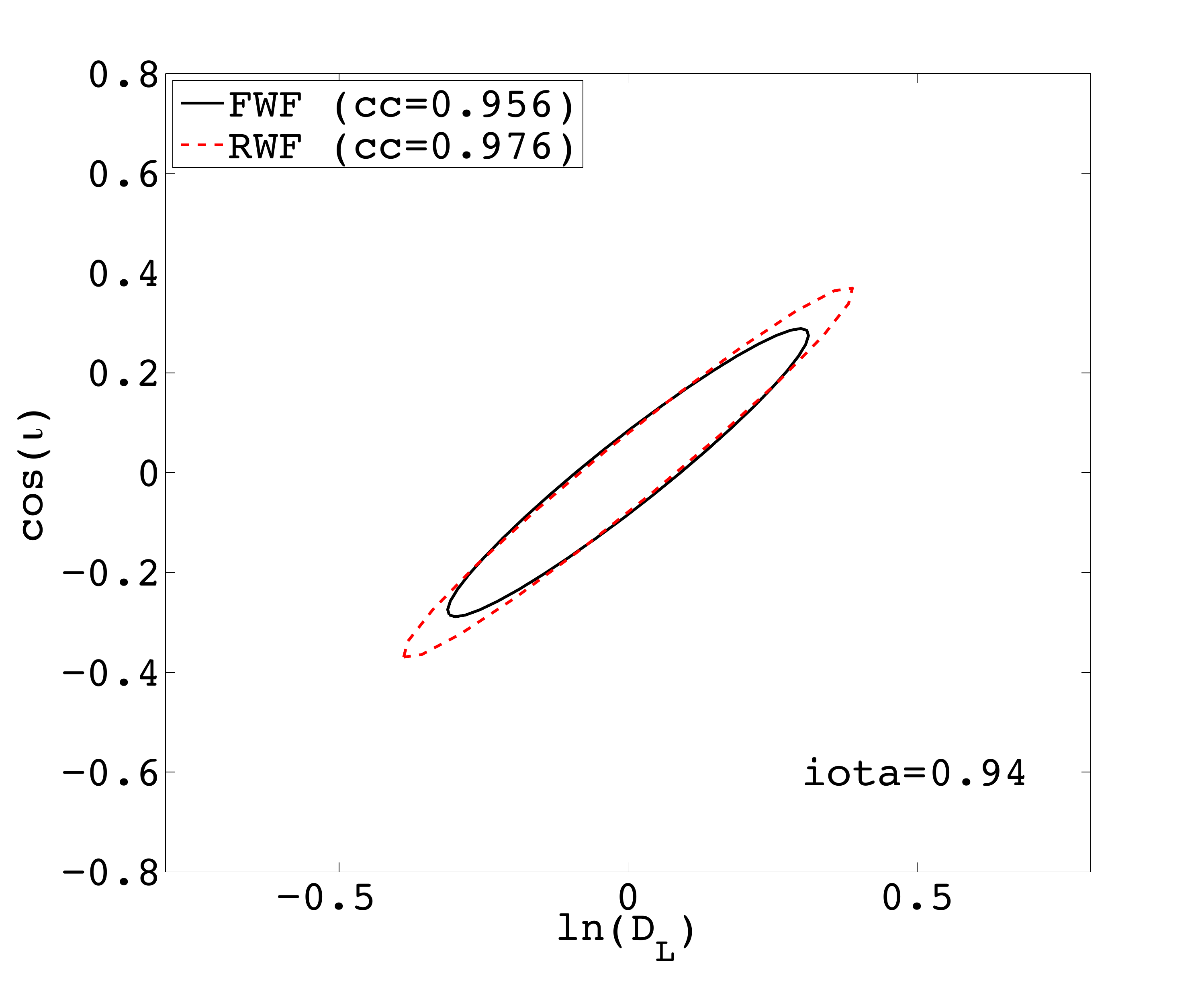}
\hskip 0.2cm
\includegraphics[width=0.40\textwidth,angle=0]{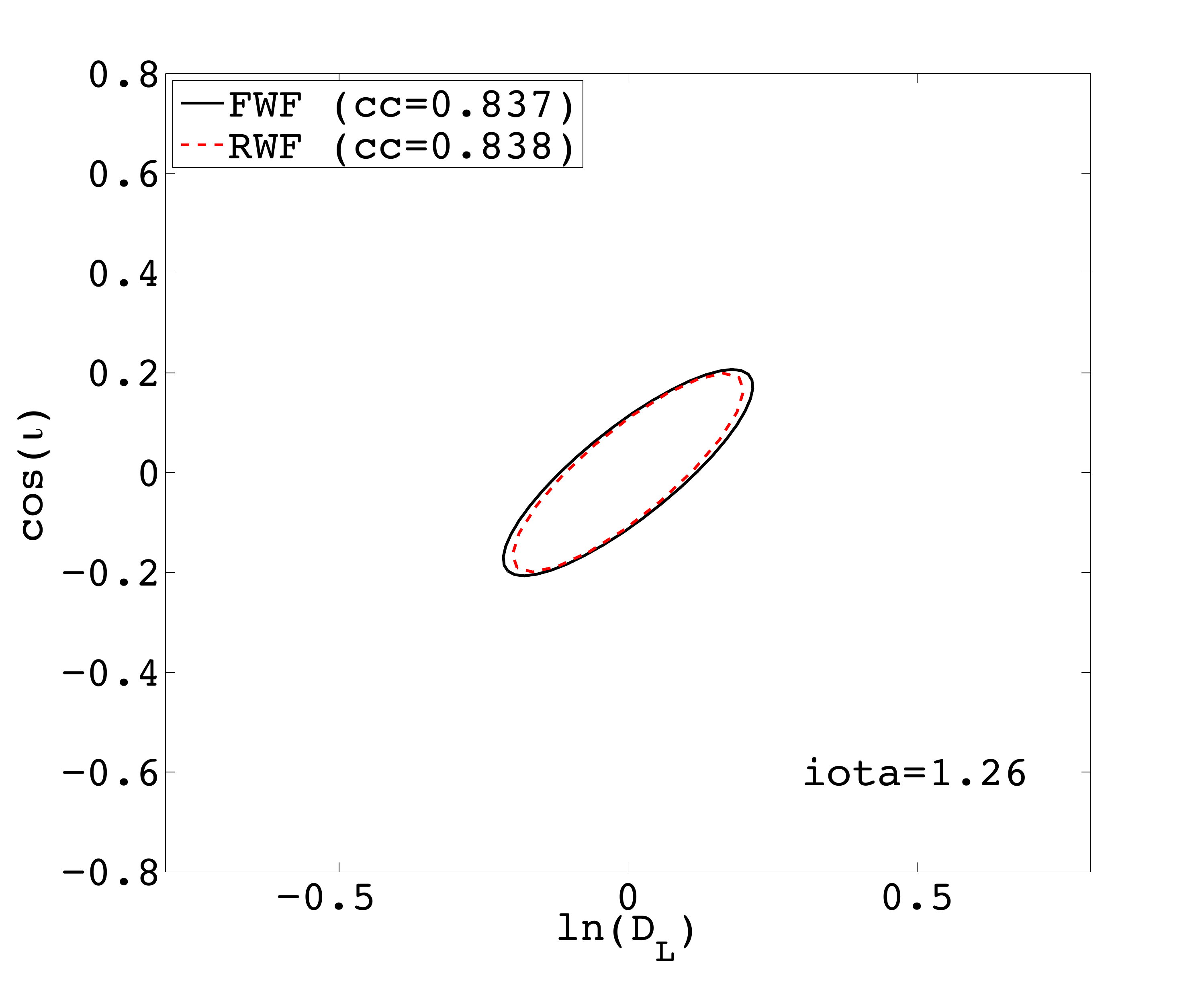}
\caption{Figure displays two-dimensional contour of the error of $D_L$ and $\cos(\iota)$ 
for four different values of $\iota$ in the LHV case.
The other angle parameters are $\theta=\pi/3$, $\phi=\pi/3$, and $\psi=\pi/4$. 
$cc$ denotes the value of the correlation coefficient between $D_L$ and $\cos(\iota)$ computed 
from the covariant matrix $\Sigma_{ab}$. The contour corresponds to the 68\% error region. }
\label{fig:contourLHV}
\end{figure*}

To quantify this we rescale errors to values that correspond to a SNR of 20.
Median errors for the fixed SNR case has been given in
Table~\ref{tab:medianerror-fixedsnr}. After comparing RWF and FWF errors for
$D_L$, and $\cos(\iota)$ we find that improvement factors are still about 2.
The main reason for such improvements in measurement of $D_L$ and
$\cos(\iota)$, when FWF is used, is the fact that, in the RWF case, there
persists a degeneracy between the two parameters which breaks when one uses the
FWF. To elaborate more, the FWF in contrast to the RWF contains additional
information about the inclination angle of the binary through amplitude
corrections, which enables one to measure the inclination angle parameter with
much better accuracy. Further, since inclination angle and the distance to the
binary are strongly correlated with each other, accuracy of distance
measurement also improves. It was argued in Ref.~\cite{AISS05} that the trends
in the measurement of parameters which are strongly correlated can be
understood in terms of the related correlation coefficients. It was argued
there that a decrease (increase) in correlation coefficients indicates better
(worse) measurement of related parameters. 
As an example, we compare the median of the correlation coefficient (absolute value) 
in context of LHV network which are shown in the top in Tables \ref{tab:corrcoef} 
and \ref{tab:corrcoefRWF}.
The correlation coefficient between $D_L$ and $\cos(\iota)$
is 0.95 for the RWF case. 
We find that this value decreases slightly to a value of 0.91 for the FWF case. 
We have checked that,
by considering the accuracy of inversion of Fisher matrices and 
the number of simulation of order $10^4$, 
the numerical and the statistical errors are much less than this difference, 
and this difference is significant. 
Although the difference is small, this is effective to reduce the error of $D_L$ and $\cos(\iota)$
for the FWF case. 
We show examples of 
the 2 dimensional contour of the error in the $D_L-\cos(\iota)$ plane 
for the LHV case in Fig. ~\ref{fig:contourLHV}.
We can see that the reduction of the correlation coefficient helps the improvement of the 
distance and the inclination angle measurement.
We also find that when $\iota\lesssim 0.8$, 
the difference between FWF cases  and RWF cases become very large. 
One the other hand, when $\iota\gtrsim 0.9$, the difference between FWF and RWF is small. 
This is because when $\iota$ is closer to $\pi/2$, $\cos(\iota)$ becomes smaller and 
the difference between FWF and RWF become smaller. 

At this stage we would like to
point out that GWs from binary systems with at least one component as NS, will
be observed with some electromagnetic counterpart 
(as a recent reference, see \cite{Singer2014}). 
In such a situation,
electromagnetic (EM) observations can be used to fix the location as well as
the distance to the binary (using redshift measurements), which completely
breaks the $D_L$-$\iota$ degeneracy and hence further significantly improves
the $\iota$ measurements. An analysis under the assumption of coincidence GW-EM
observations has been performed in the case of binary NS (BNS) and BH-NS
systems (which are strong candidates for progenitors of short-hard gamma ray
bursts (SGRBs)) in Ref.~\cite{ATPM14}. It has been shown there that once the
information about the source location and its distance is folded in the
analysis, one can put tight constraints on the inclination angle measurements,
which further can help us understand various aspects of SGRB science.

Improvement in the measurement of the coalescence phase ($\Phi_c$) can be
understood as an effect of the fact that the FWF has more information about
this parameter as compared to that present in the RWF as different harmonics
enter the sensitivity band of the detector at different times. Next, we find
that the $\Phi_c$-$\psi$ component of the correlation coefficient matrix,
reduces to a value of 0.43 for FWF from its RWF value of 0.58. This explains
why we see an improvement in the measurement of $\psi$ when the FWF is used
over the RWF.

As far as other parameters are concerned we do not see much improvement due to
the use of the FWF over the RWF (see Table~\ref{tab:medianerror}). For
instance, the mass parameters can be very well measured using the phase
information which is already present in the RWF and hence additional
information about the mass parameters present in the amplitude leads to minor
improvements in the measurement accuracies of mass parameters. On the other
hand, measurement of $t_c$, $\theta$ and $\phi$ basically depend on the
time-delays between different detector sites which for a given network are same
irrespective of the waveform model involved. However, since the polarization
angle is better measured when FWF is used, improvements in the measurement of
location angular parameters ($\theta$, $\phi$) are expected, since they enter
the waveform in more or less similar ways through the antenna pattern functions
(see Eq.~\eqref{eq:beampatternfunctions}), and hence they are expected to be
strongly correlated (see also the related discussion in \cite{ChrisAnand06b}).
Upon comparing correlation coefficients related to $\theta$-$\phi$-$\psi$ pairs
we find that for the FWF case correlations are significantly small as compared
to the RWF case. However, one should also keep in mind that the correlations
between these parameters are not so strong for the network case. This is
expected as in the case of a network various degeneracy among angular
parameters break which makes various quantities relatively independent of each
other. This would mean that although when going from RWF to FWF correlations
are significantly reduced, the measurements of one parameter would affect
weakly the measurement of the other. This is why we only see small a
improvement in $\theta$ and $\phi$ which further leads to small improvements in
angular resolution. In addition, we notice that $t_c$ has moderately strong
correlations with ${\cal M}_c$, $\delta$, $\Phi_c$, $\theta$ and $\phi$. We
find that when going from RWF to FWF, for some pairs correlations decrease
(which would lead to better in parameter estimation (PE)) and for the rest it
increases (worsening the PE). It is the combined effect of various correlations
that we see an effective minor improvement in $t_c$.
   
\subsubsection{LHVK}
\label{subsubsec:LHVK}

\begin{figure*}[htbp!]
\includegraphics[width=0.40\textwidth,angle=0]{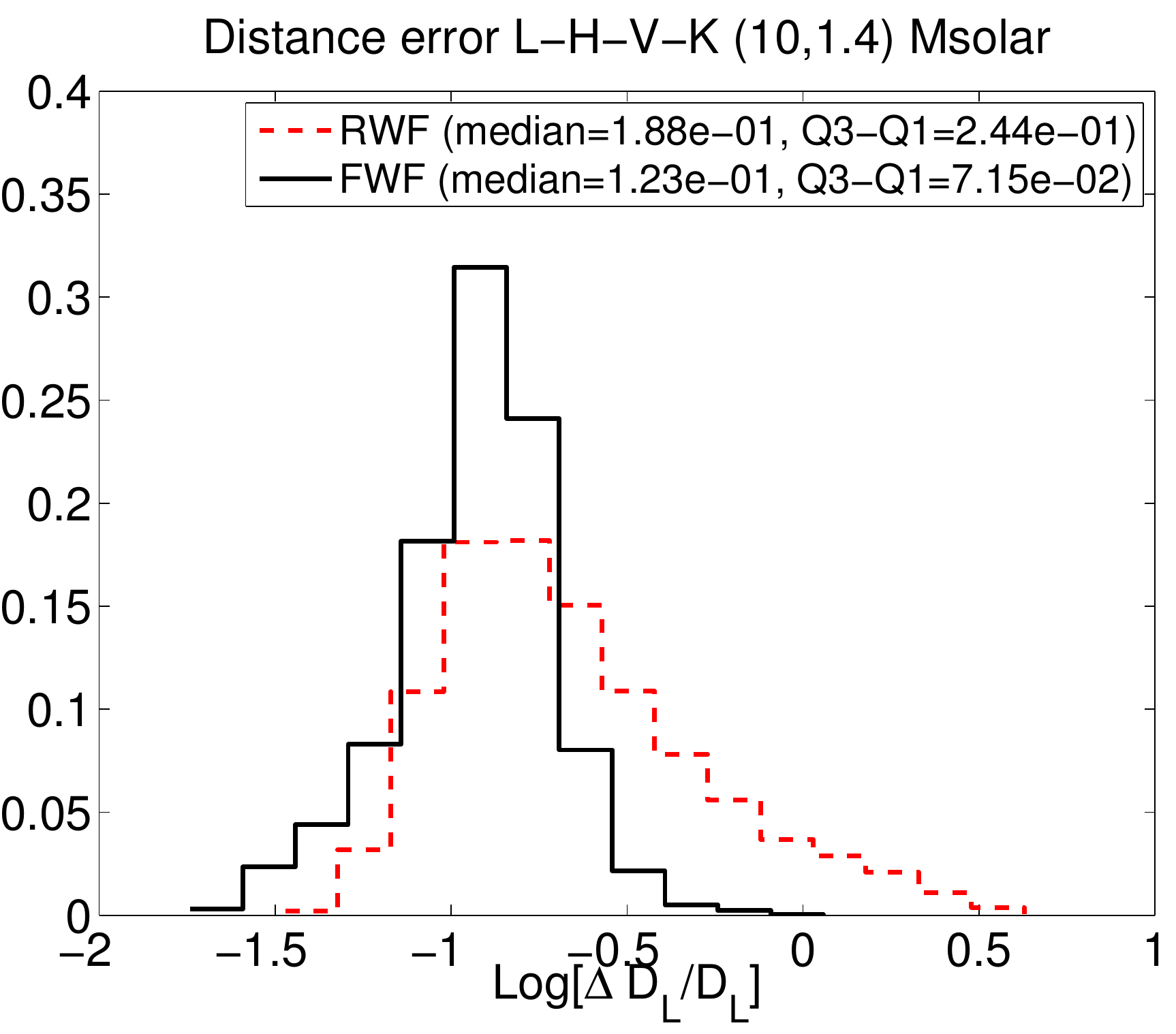}
\hskip 0.2cm
\includegraphics[width=0.40\textwidth,angle=0]{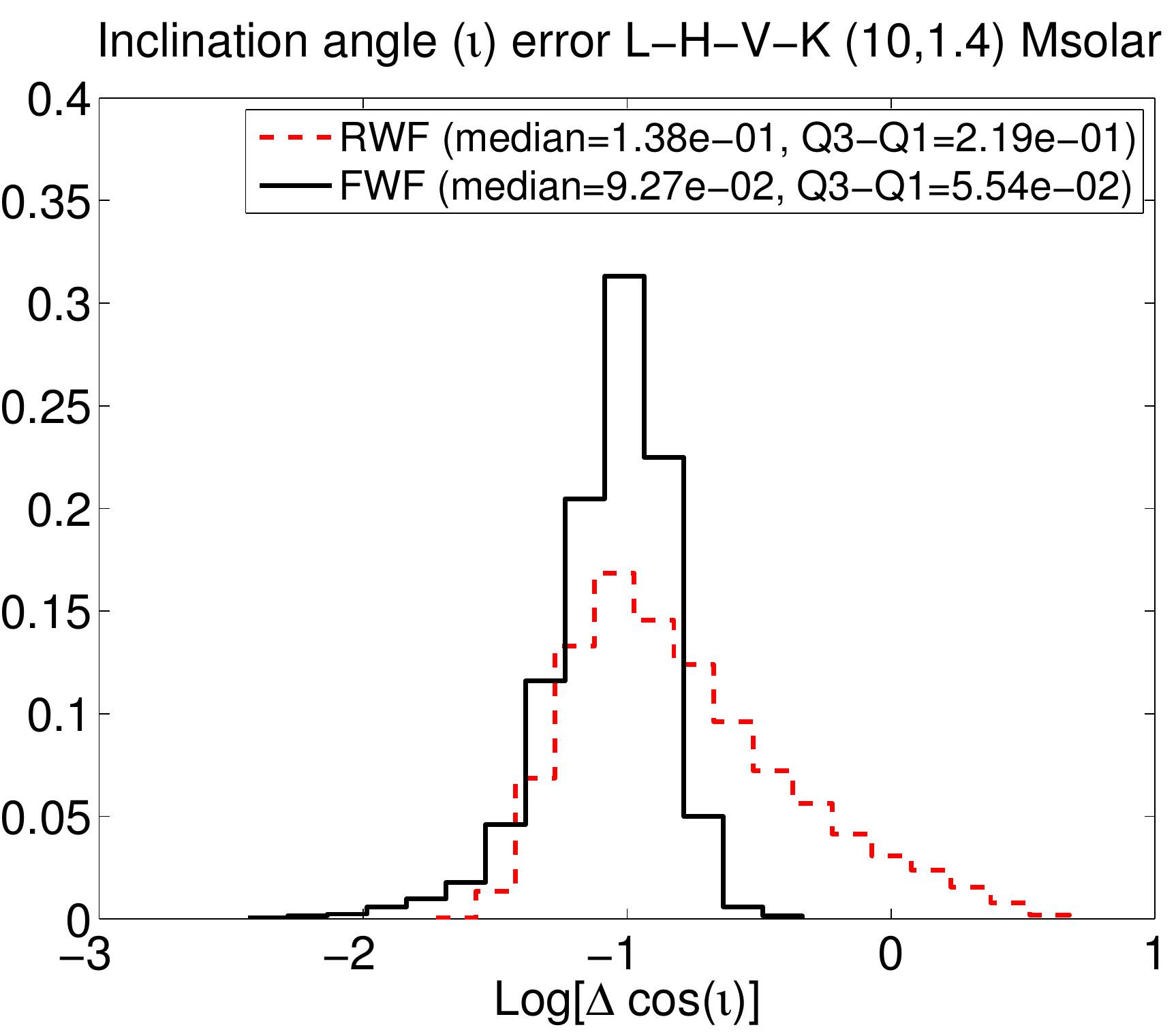}
\vskip 0.2cm
\includegraphics[width=0.40\textwidth,angle=0]{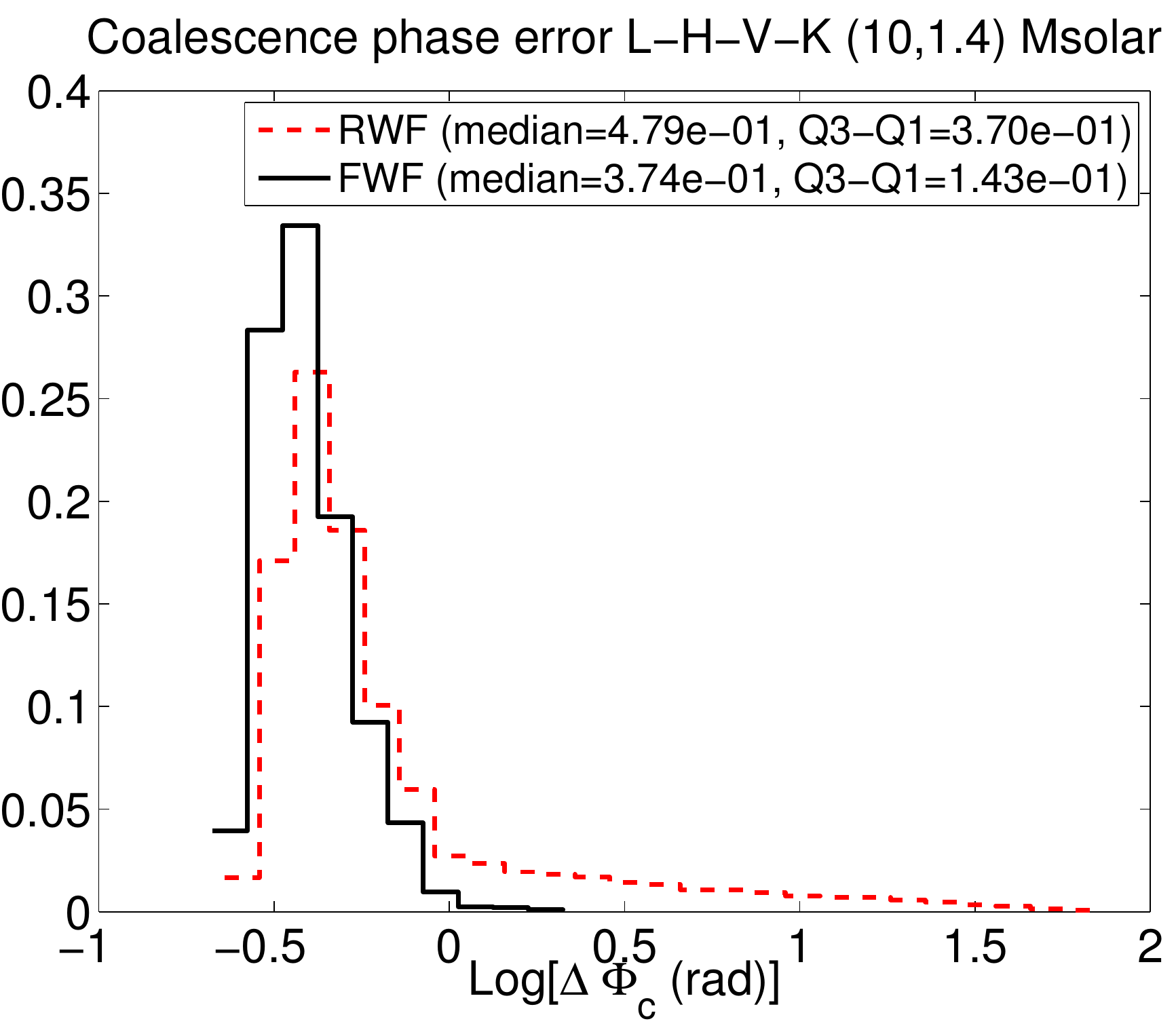}
\hskip 0.2cm
\includegraphics[width=0.40\textwidth,angle=0]{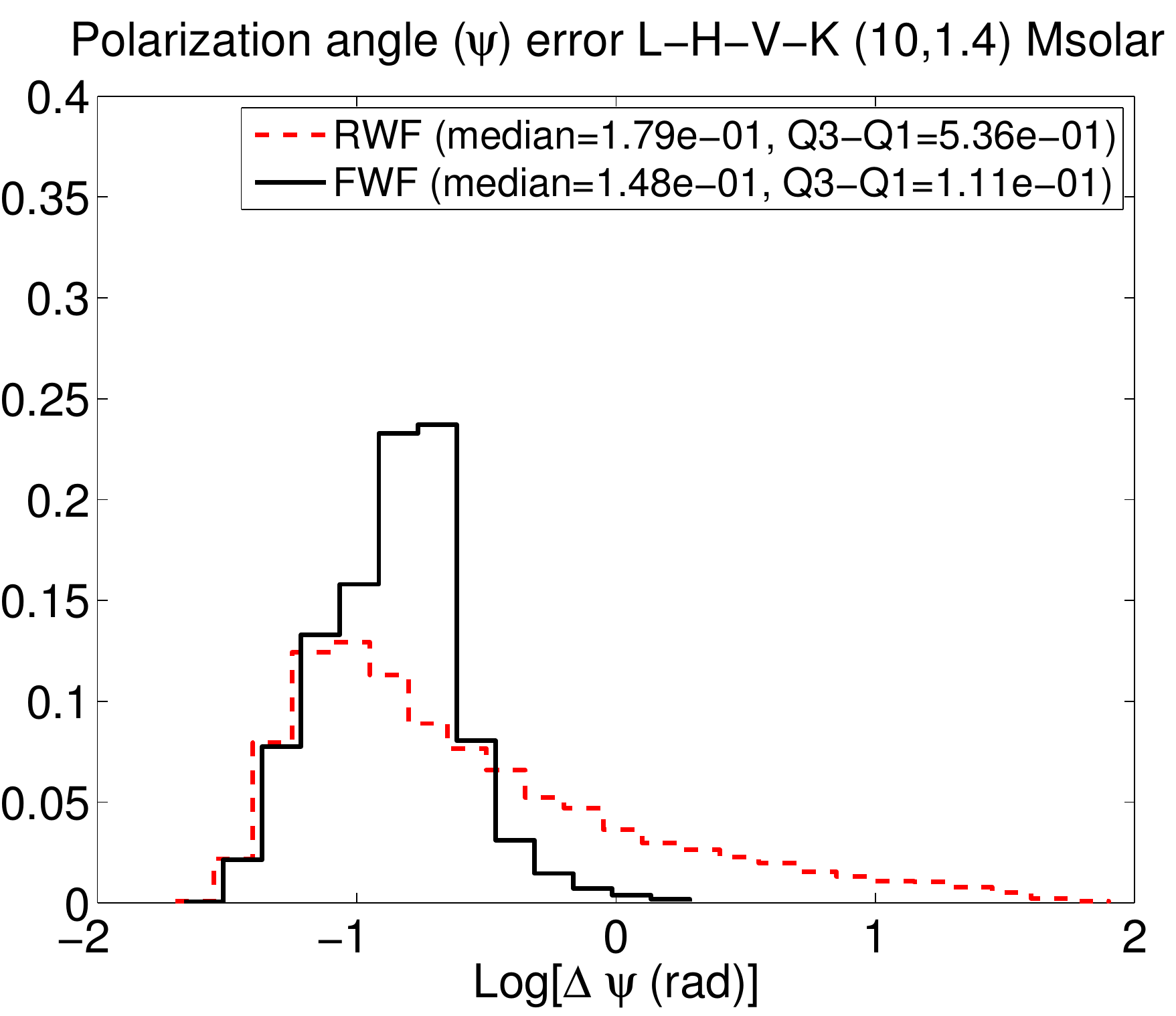}
\caption{Figure displays error distributions corresponding to the ($D_L, \cos
(\iota), \Phi_c, \psi$), in context of the LIGO-Virgo-KAGRA network (LHVK). The
study has been performed for a population of (1.4-10)$M_\odot$ NS-BH binaries,
all placed at 200 Mpc and distributed and oriented uniformly over the whole
sky. The error distributions for signals with RWF and FWF, are compared. Median
and the inter-quartile range (Q3-Q1) corresponding to each parameter has been
displayed in various panels. Note that out of nine parameters listed in
Eq.~\eqref{eq:param} we are displaying graphical results only for four of them
since effect of the use of FWF over RWF is only significant in these four
cases. However, in Table~\ref{tab:medianerror} we list medians corresponding to
error distribution of all the parameters.} \label{fig:compRvFLHVK}
\end{figure*}

In the previous subsection we discussed the accuracies with which various
parameters are measured in the context of the LIGO-Virgo network (LHV). We also
tried to understand possible reasons for the improvements in estimating various
parameters when FWF is used as compared to the RWF in LHV network. The LHV
network is expected to be operational by early 2016. However, as discussed in
Sec.~\ref{subsec:advnet}, the Japanese detector KAGRA is expected to be fully
operational by the end of year 2018, and hence by that time we might have a
4-detector network, LIGO-Virgo-KAGRA (LHVK).  The addition of the fourth
detector would not only increase the duty cycle of the detector networks but
also would improve the localization of the source (see below and the discussion
in Sec.~\ref{subsec:compNetwork}). Error distributions corresponding to
parameters $D_L$, $\cos(\iota)$, $\Phi_c$ and $\psi$, in the context of the
LIGO-Virgo-KAGRA (LHVK) network, has been displayed in
Fig.~\ref{fig:compRvFLHVK}. 

Median errors displayed in each panel of Fig.~\ref{fig:compRvFLHVK} suggest
that the use of FWF over RWF shall improve the measurements of $D_L$, and
$\cos(\iota)$ by a factor of about 1.5 and those of $\Phi_c$ and $\psi$ by
factors of 1.3 and 1.2, respectively. As far as the measurement of other
parameters are concerned, the improvement is still very small and we do not
wish to show graphical results corresponding to these parameters for the reason
mention in the previous subsection. However, we list median errors in
Table~\ref{tab:medianerror}. Note that here also we can find that the effects
of the SNR is only minor in error estimation as was seen in the LHV case (see
Table~\ref{tab:medianerror}- \ref{tab:medianerror-fixedsnr}). The reason behind
the improvements in various parameters is again similar to those discussed in
the previous section. However, note that as compared to LHV case the
measurement accuracies with LHVK case are much better. As we shall discuss in
detail in the Sec.~\ref{subsec:compNetwork}, this is due to the fact that the
coherent SNR for LHVK is larger than the LHV case. In particular, angular
resolution improves significantly with the inclusion of the fourth detector in
the network as LHVK would have larger effective area as compared to the one LHV
case, which in turn guarantees better localization.  We postpone the discussion
related to the angular resolution to Sec.~\ref{subsec:compNetwork}.
\subsubsection{LHVKI} \label{subsubsec:LHVKI}

\begin{figure*}[t]
\includegraphics[width=0.40\textwidth,angle=0]{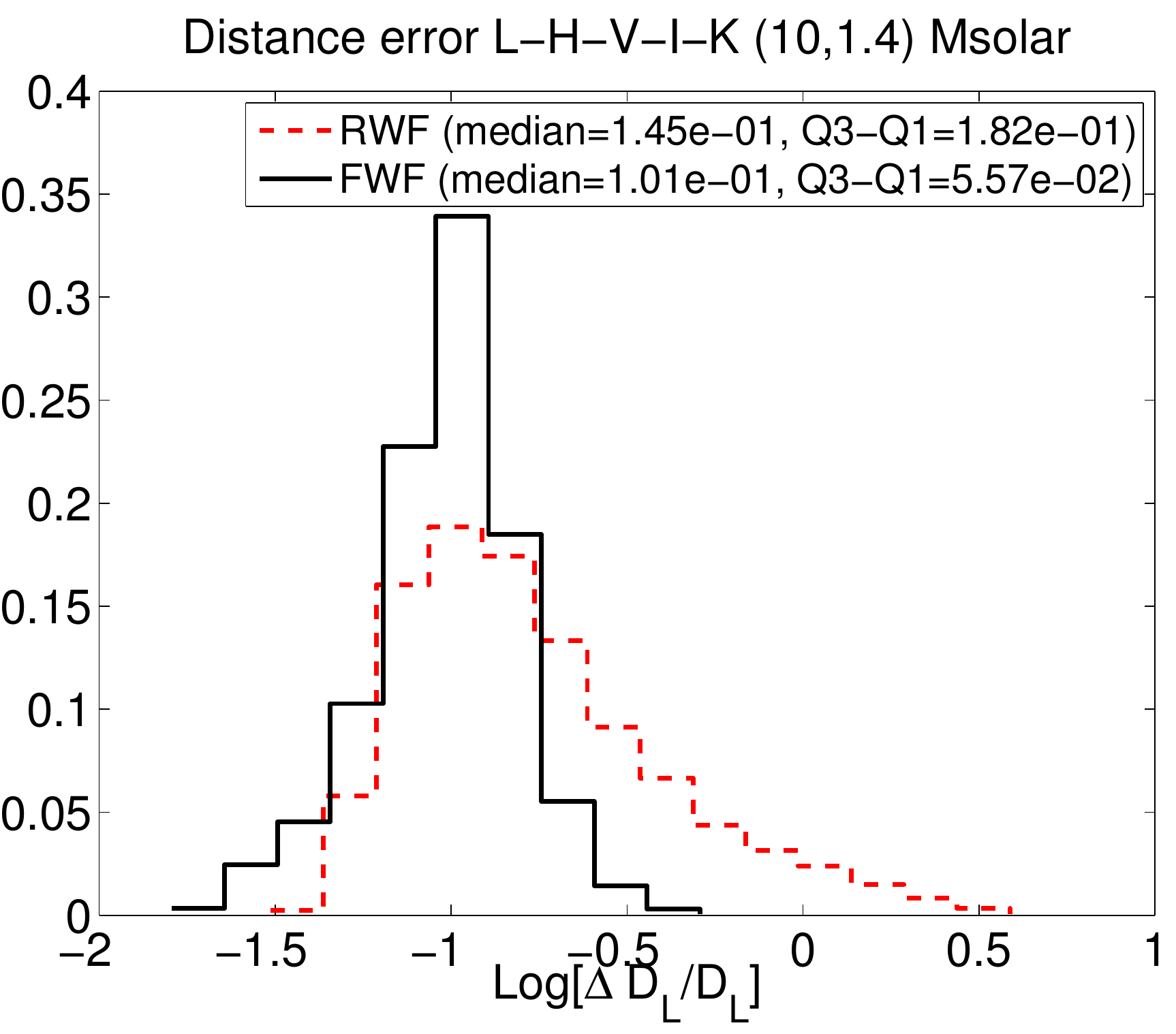}
\hskip 0.2cm
\includegraphics[width=0.40\textwidth,angle=0]{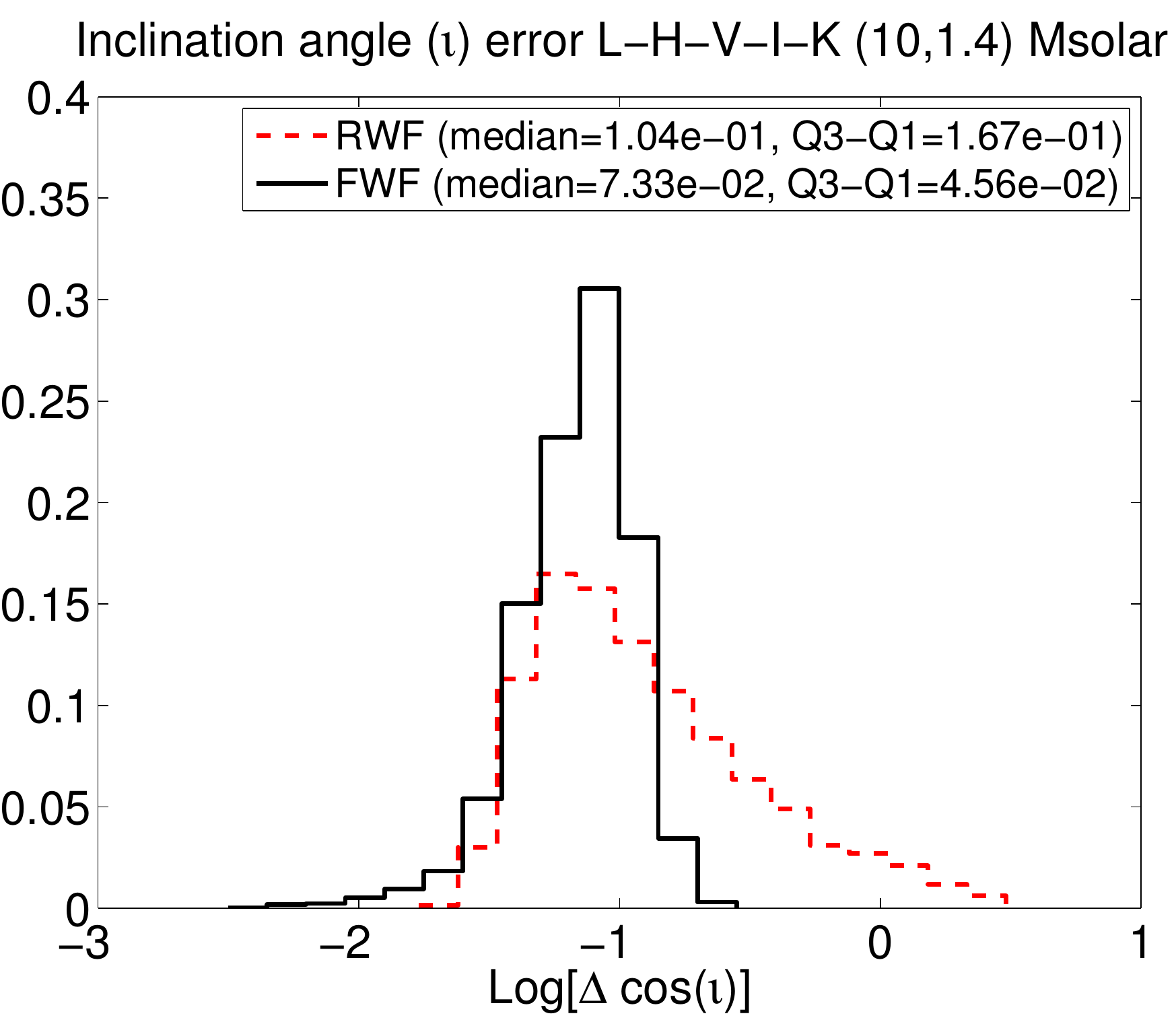}
\vskip 0.2cm
\includegraphics[width=0.40\textwidth,angle=0]{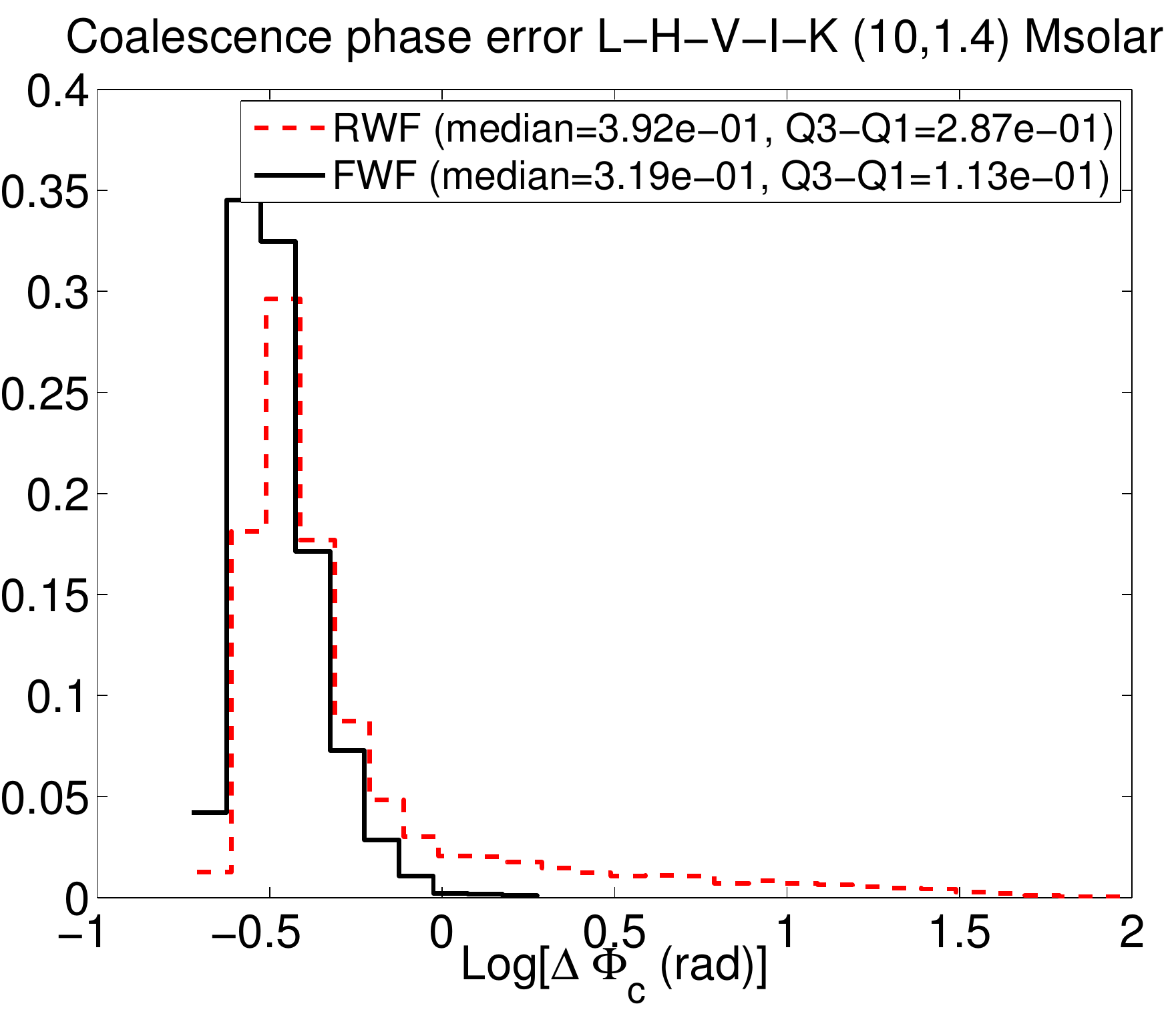}
\hskip 0.2cm
\includegraphics[width=0.40\textwidth,angle=0]{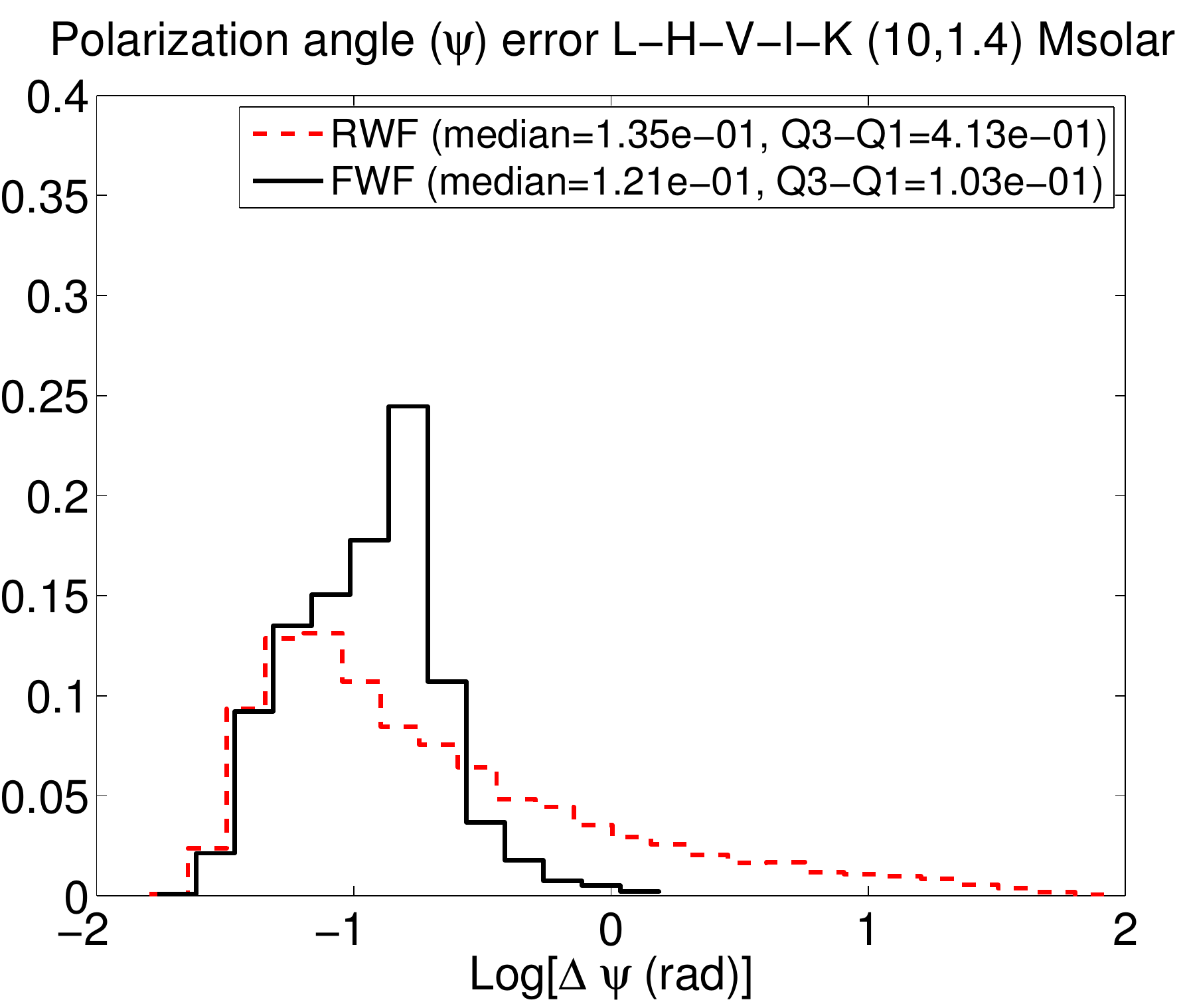}
\caption{Figure displays error distributions corresponding to the ($D_L, \cos
(\iota), \Phi_c, \psi$), in context of the LIGO-Virgo-LIGO-India-KAGRA network
(LHVKI). The study has been performed for a population of (1.4-10)$M_\odot$
NS-BH binaries, all placed at 200 Mpc and distributed uniformly over the sky
surface. The error distributions for signals with RWF and FWF, are compared.
Median and the inter-quartile range (Q3-Q1) corresponding to each parameter has
been displayed in various panels. Note that out of nine parameters listed in
Eq.~\eqref{eq:param} we are displaying graphical results only for four of them
since effect of the use of FWF over RWF is only significant in these four
cases. However, in Table~\ref{tab:medianerror} we list medians corresponding to
error distribution of all the parameters.} \label{fig:compRvFLHVKI}
\end{figure*}

Just as adding the Japanese detector KAGRA to the LIGO-Virgo (LHV) network 
improves measurements of various parameters as well increases the duty 
cycle of the detector networks, addition of LIGO-India will guarantee 
better measurement of various parameters as compared to the three and four 
detector networks. Similar to Figs.~\ref{fig:compRvsFLHV}-\ref{fig:compRvFLHVK}
the error distributions for $D_L$, $\cos(\iota)$, $\Phi_c$ and $\psi$ is displayed 
in Fig.~\ref{fig:compRvFLHVKI} in context of the 5-detector network LHVKI.
For this case median errors in $D_L$, and $\cos(\iota)$ improve by a factor of about
1.4 and those for $\Phi_c$ and $\psi$ by factors of about 1.2 and 1.1, respectively. 
The measurements of all other parameters improve by even smaller factors when FWF is 
used as compared to the RWF. Note that although the median errors suggest that 
using RWF one can measure parameters with almost similar accuracies as with FWF, 
for a number of cases RWF still gives very large errors. This suggests that 
using FWF would make sure that systematic effects do not bias the measurements.
In the next section we shall compare the benefits of having a network with 
large number of detectors in context of parameter estimation by taking examples of 
the three different combinations (LHV, LHVK, and LHVKI). 

When looking at localization error for various detector combinations in
Table~\ref{tab:medianerror} we notice that for all 3-detector cases the
localization is better when FWF is used, whereas for all 4-detector and the
5-detector cases, the use of RWF leads to better localization. However, when we
look at the localization errors for fixed SNR cases
(Table~\ref{tab:medianerror-fixedsnr}), we do not see these two opposite
trends; for all detector combination the use of FWF gives better localization.
Let us try understanding first the two opposite trends we see in
Table~\ref{tab:medianerror}. We notice, for all 3-detector cases, FWF works
better (in localizing the source), despite the fact that the FWF SNR is smaller
than RWF SNR. This can be understood by recalling the arguments presented in
Sec.~\ref{subsubsec:LHV}, in context of better measurement of location angle
parameters ($\theta$, $\phi$) with FWF, which further leads to better
localization. Trends in Table~\ref{tab:medianerror} suggest that for all the
3-detector cases, whatever degradation happens because of smaller SNR in FWF
cases is in fact compensated by the better measurement of location angle
parameter. Also it is noteworthy that the difference between the RWF and FWF
SNR is very small, hence more or less SNR does not play a significant role in
the case of 3-detector networks. However, when we add fourth and fifth detector
to the network, coherent SNR for RWF cases become significantly larger than the
coherent SNR for FWF cases. However, as was argued in Sec.~\ref{subsubsec:LHV},
as more detectors are included in the network, various degeneracy between the
angular parameters are resolved and hence measurements of different angular
parameters becomes almost independent of each other even in the case of RWF,
and hence milds down the effect of FWF which played an important role
in three detector cases. These two arguments combined explain why we see two
opposite trends in the Table~\ref{tab:medianerror}. However, when we look at
the fixed-SNR table (Table~\ref{tab:medianerror-fixedsnr}), the SNR does not
play a role and in that case the FWF of course would perform better, and this
is why the use of FWF gives better localization for all detector combinations
as can be seen in Table~\ref{tab:medianerror-fixedsnr}. Note that the addition
of fourth and fifth detector to the network will anyway improve the
localization irrespective of the waveform used. 

\subsection{Comparison of effects of various multidetector networks on parameter accuracies}
\label{subsec:compNetwork}

\begin{figure*}[!htbp]
\includegraphics[width=0.40\textwidth,angle=0]{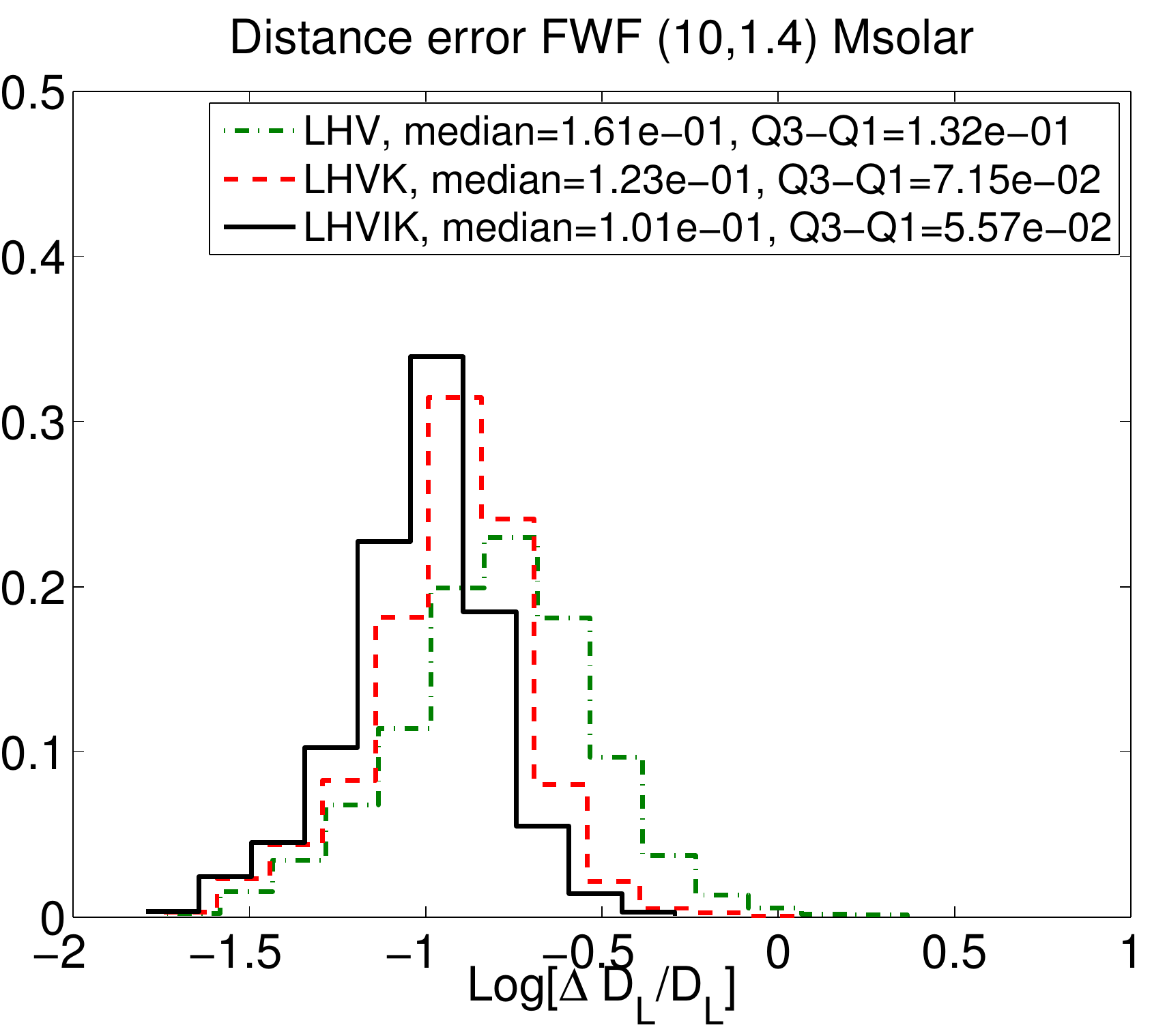}
\hskip 0.2cm
\includegraphics[width=0.40\textwidth,angle=0]{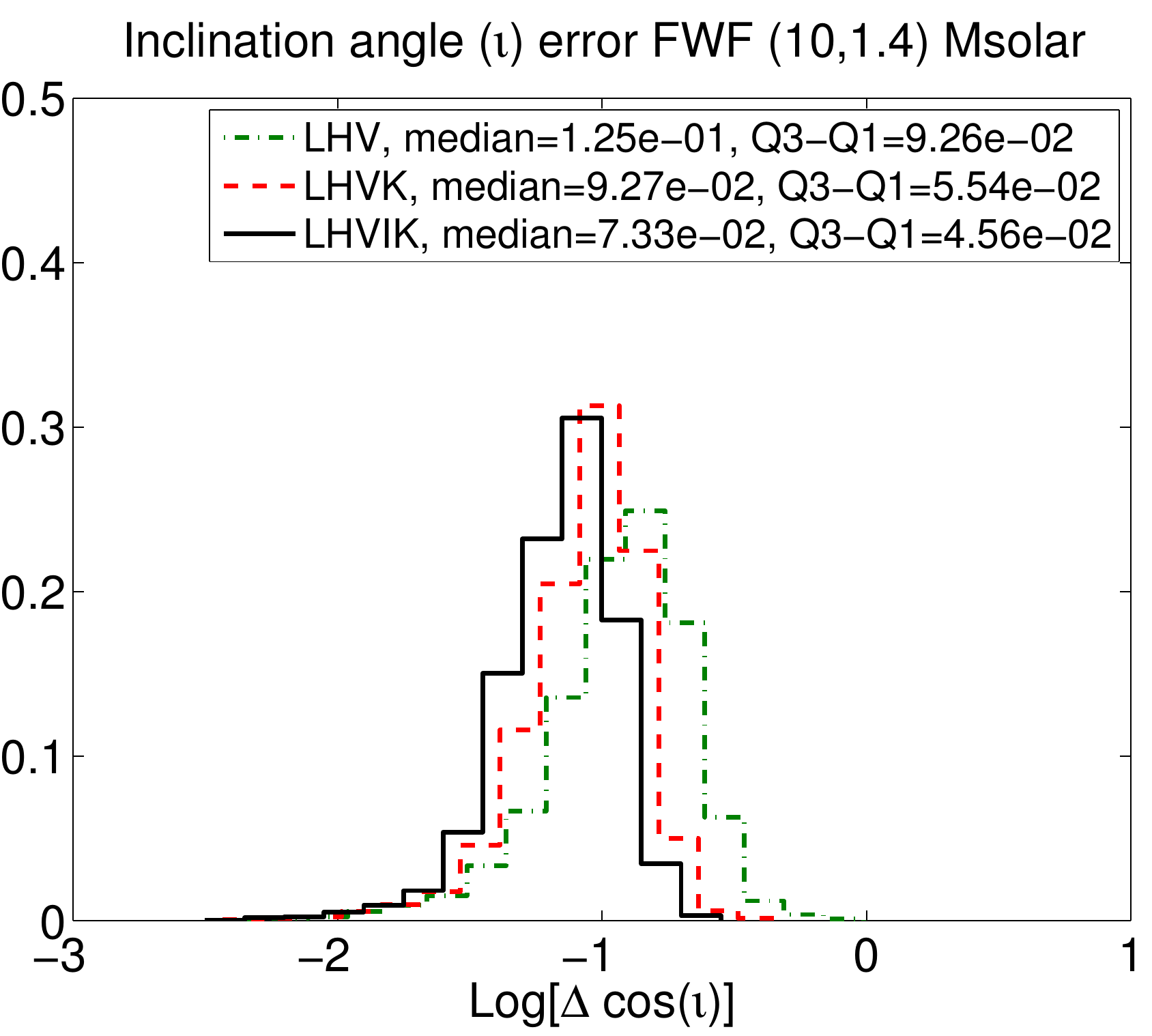} \vskip
0.2cm \includegraphics[width=0.40\textwidth,angle=0]{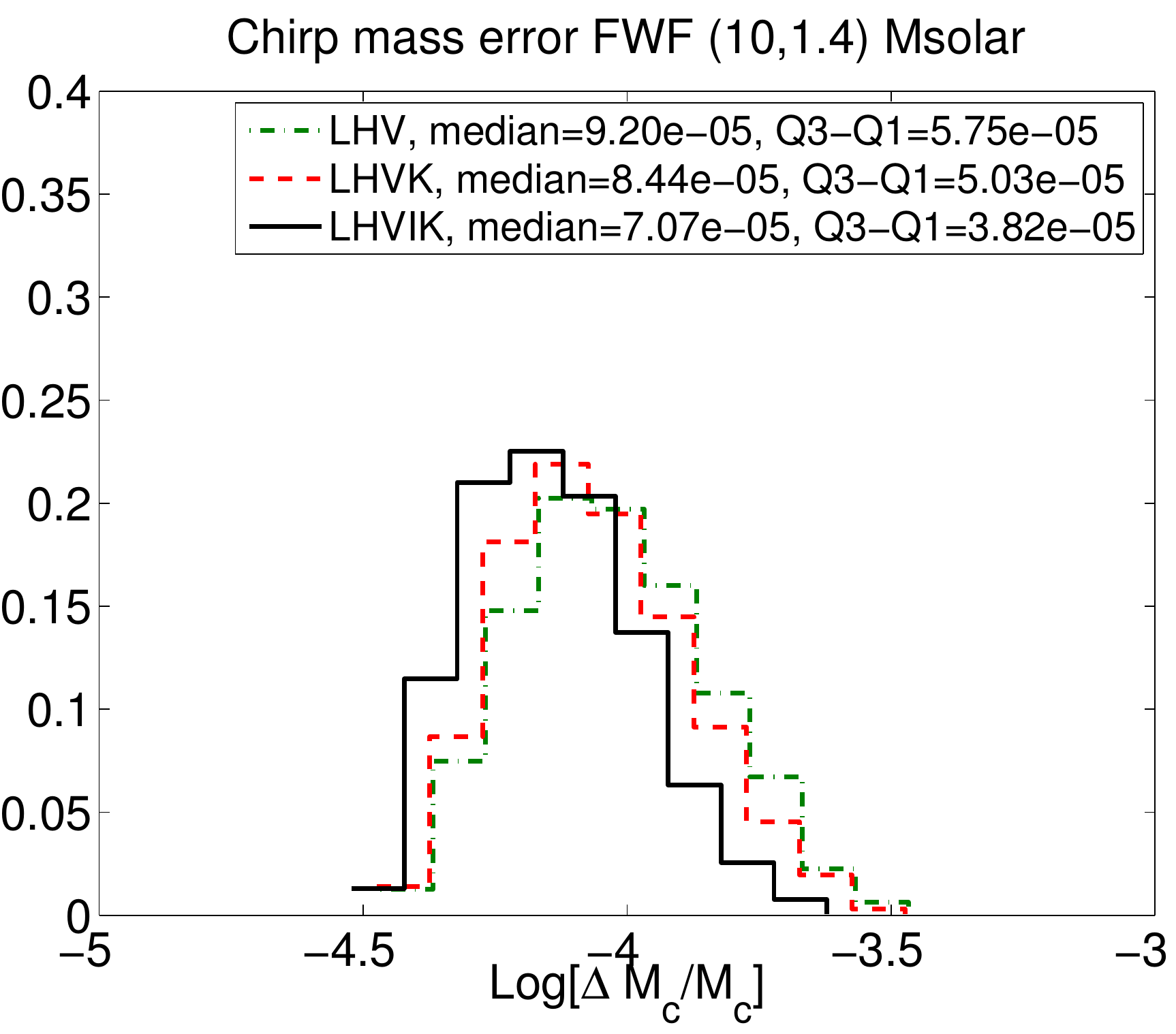}
\hskip 0.2cm
\includegraphics[width=0.40\textwidth,angle=0]{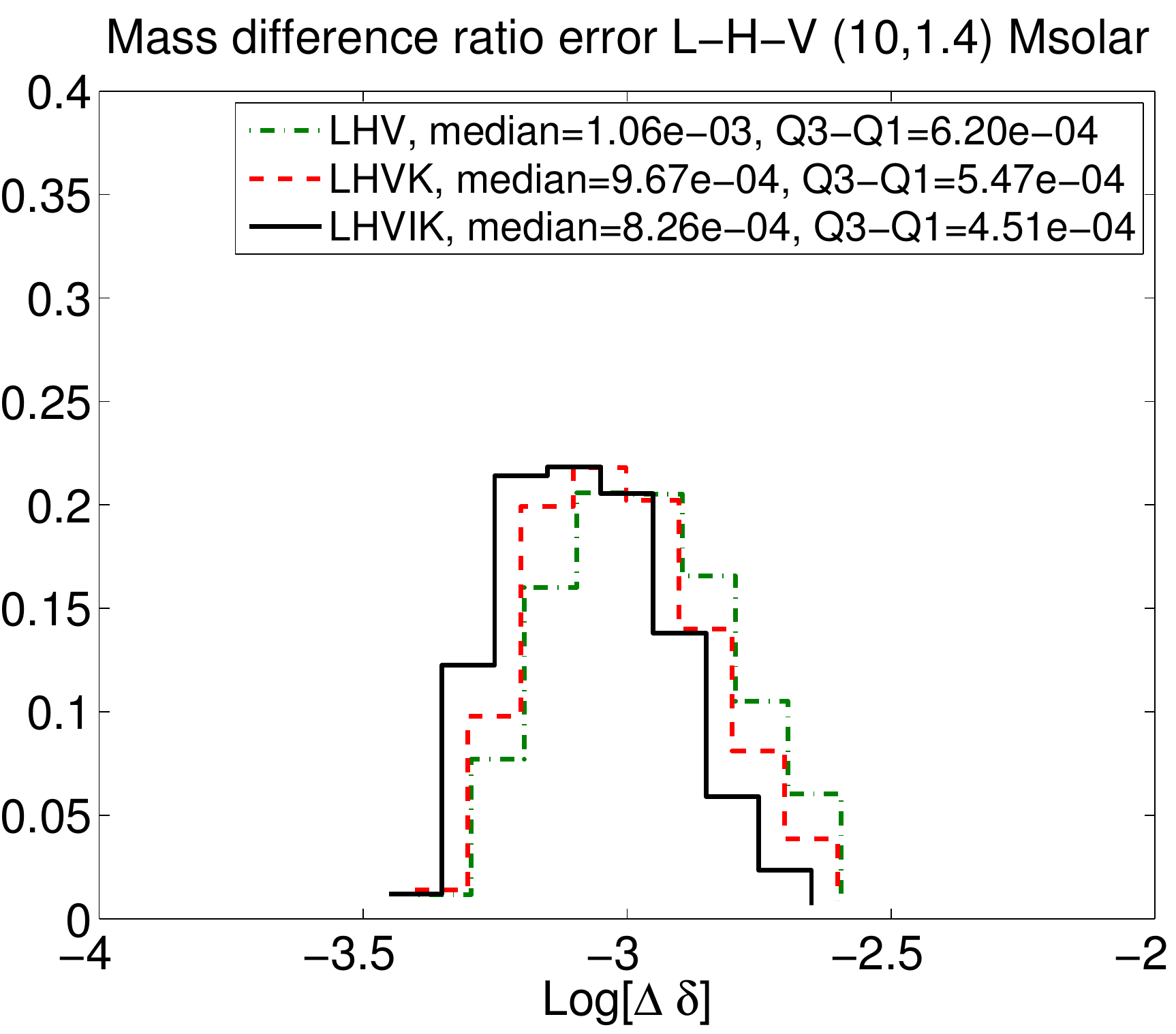}
\vskip 0.2cm
\includegraphics[width=0.40\textwidth,angle=0]{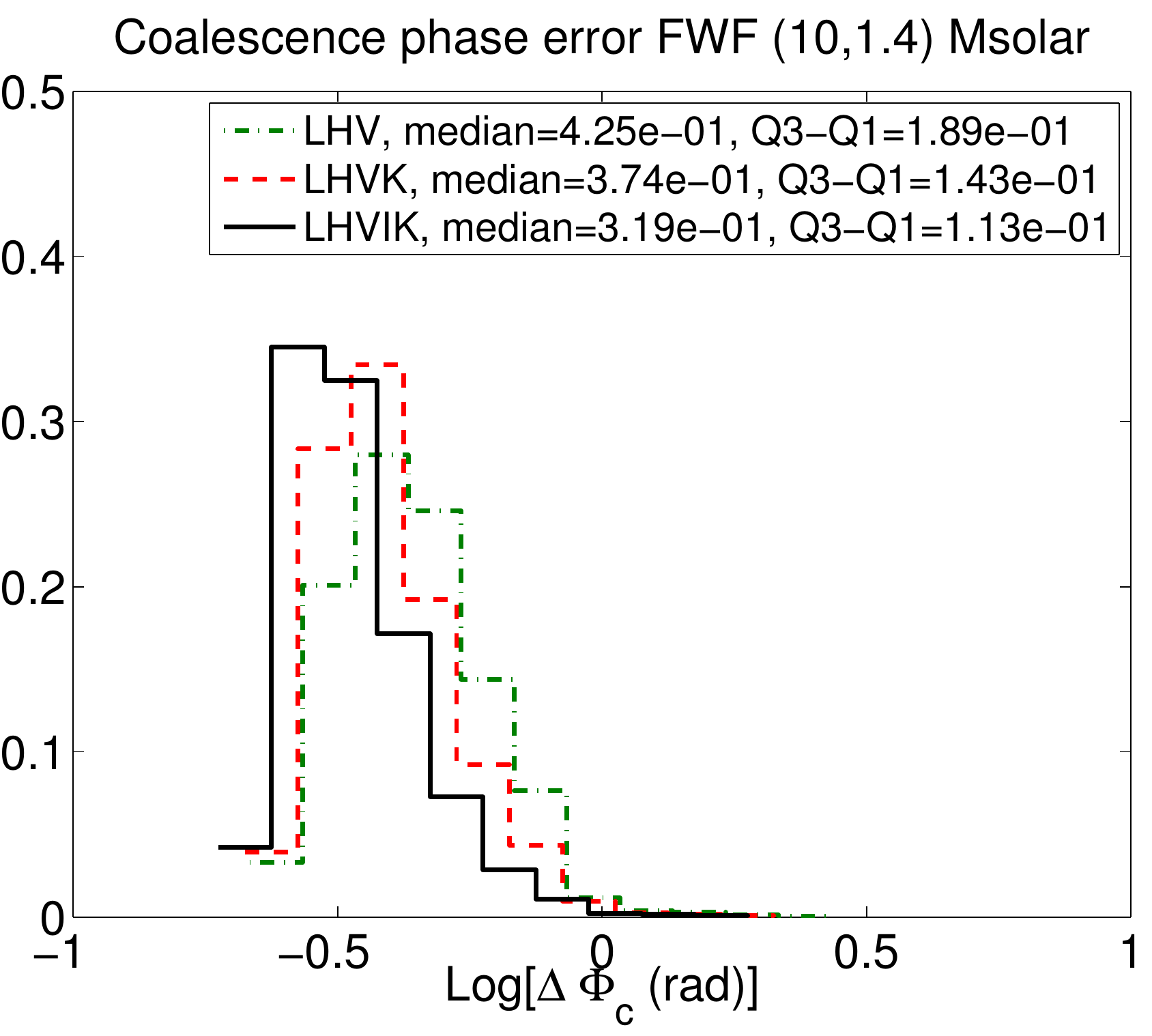} \hskip
0.2cm \includegraphics[width=0.40\textwidth,angle=0]{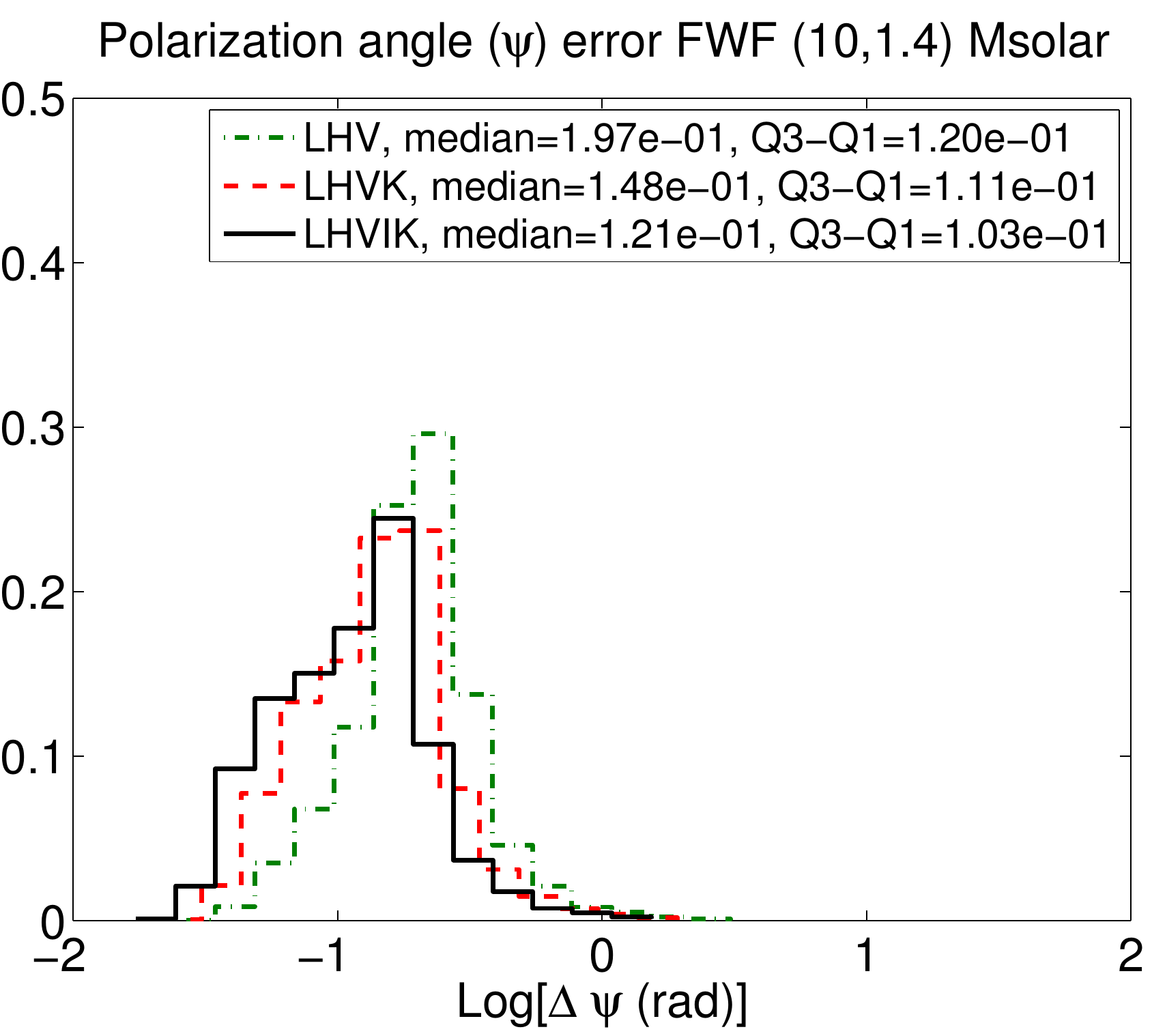}
\caption{Comparisons of error distributions of various parameters ($D_L$, $\cos
(\iota)$, $M_c$, $\delta$, $\Phi_c$ and $\psi$) with three different networks
(LHV, LHVK, and LHVKI) has been displayed. The study has been performed for a
population of (1.4-10)$M_\odot$ NS-BH binaries, all placed at 200 Mpc and
distributed uniformly over the sky surface. Error distributions displayed here
have been obtained using the amplitude corrected approximate inspiral waveform
of the GW signal expected from the source. Various panels also display median
and the inter-quartile range for each error distribution.}
\label{fig:compNetwork1} \end{figure*}

\begin{figure*}[t]
\includegraphics[width=0.40\textwidth,angle=0]{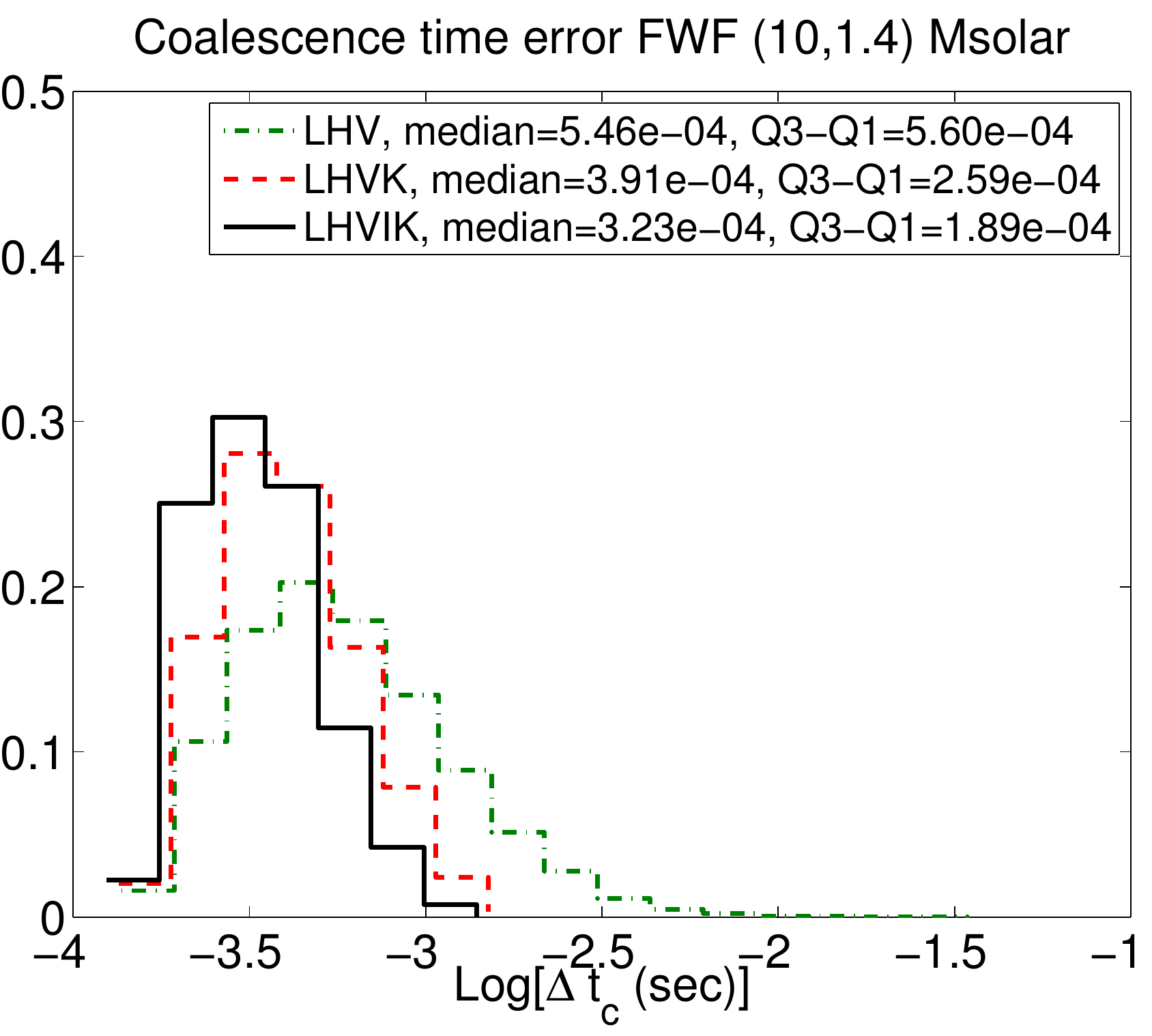} \hskip
0.2cm
\includegraphics[width=0.40\textwidth,angle=0]{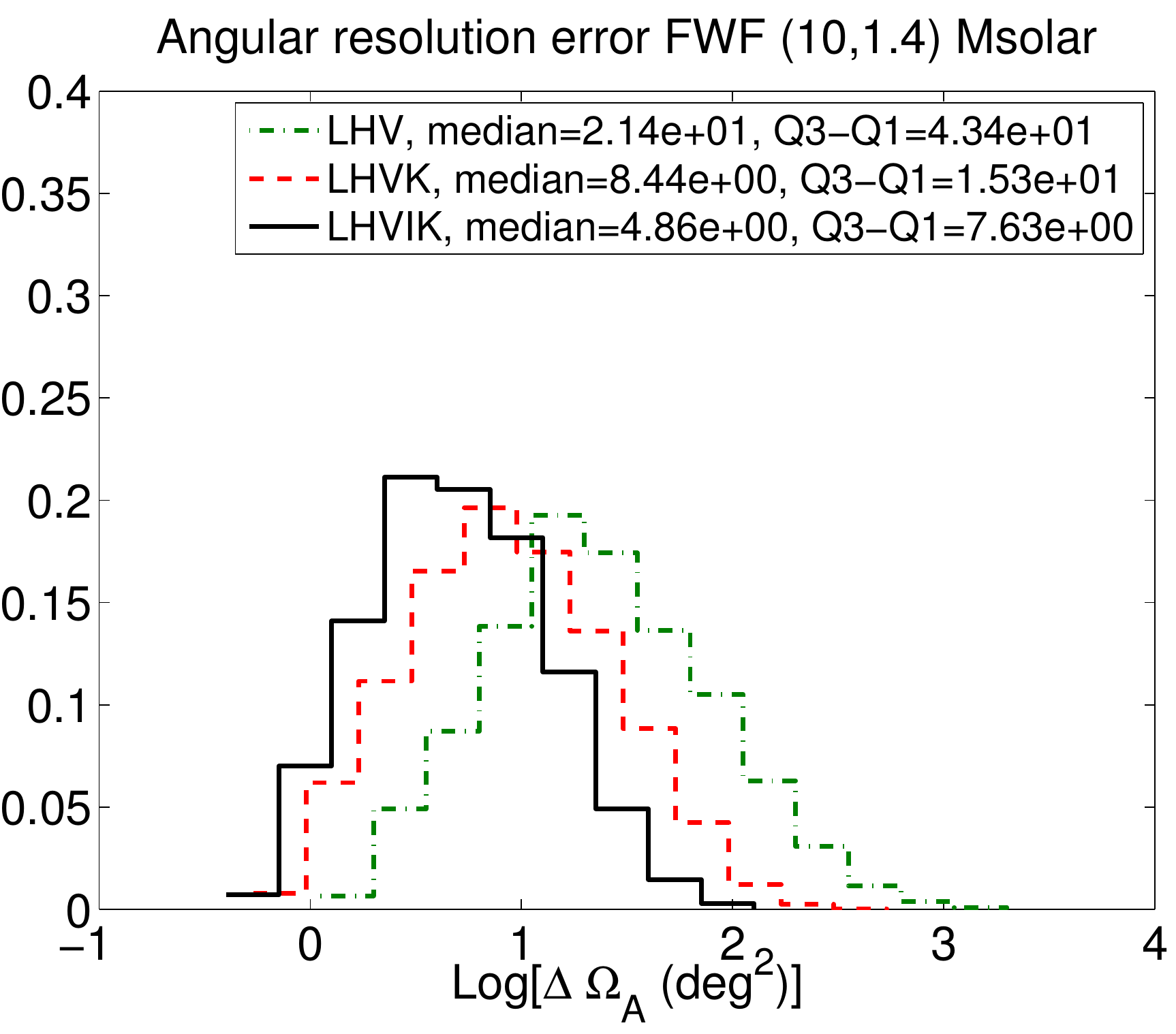}
\caption{Same as Fig.~\ref{fig:compNetwork1} but error distributions
correspond to the coalescence time $t_c$ and solid angle around source's
location $\Omega$.} \label{fig:compNetwork2} \end{figure*}

In previous subsections we discussed how the use of the FWF over the RWF
improves measurements of various parameters in context of three representative
network combinations which were chosen to be LHV, LHVK, and LHVKI. This choice
was mainly based on a time-line argument that when detectors would start
operating. However, we find that LHV, LHVK and LHVKI can also be assumed to be
representative configurations within the respective class of network
configurations as the error estimation within a class does not vary
significantly. Hence, in this section we aim to make rigorous comparisons of
our PE results in context of our three representative detector configurations
LHV, LHVK, and LHVKI. We shall refer to Table~\ref{tab:medianerror} for the
median errors in various parameters in context of all possible network
configurations. 

The parameter estimation accuracies for our three representative network
combinations are displayed in
Figs.~\ref{fig:compNetwork1}-\ref{fig:compNetwork2}.
Figure~\ref{fig:compNetwork1} displays error distributions for $D_L$,
$\cos(\iota)$, ${\cal M}_c$, $\delta$, $\Phi_c$ and $\psi$. On the other hand,
Fig.~\ref{fig:compNetwork2} displays the error distributions for $t_c$ and the
angular resolution $\Delta\Omega$. Note that here we chose to display the error
distribution for the angular resolution and not the ones related to the angular
parameters giving the location of the source ($\theta$, $\phi$). This is so
because the errors in $\theta$ and $\phi$ and the covariances between them can
be suitably combined to obtain the solid angle around the location of the
source (see Eq.~\eqref{eq:omegaAdef}) which precisely tells how well the source
can be localized by the given network (the angular resolution of the network).
Also note that while comparing different network we only use PE results
obtained using the FWF which is a better approximation to the actual signal.
Even a quick look at the shapes and respective positioning of error
distributions corresponding to various parameters appearing in
Figs.~\ref{fig:compNetwork1}-\ref{fig:compNetwork2} reveal that measurement
accuracies improve by the addition of the fourth and the fifth detector to the
three detector network.  This is true in general for all the detector
combinations (see Table~\ref{tab:medianerror}).  This is indeed what is
expected in general as the coherent SNR is larger for a network which consists
of more detectors which in turn improves the estimation of parameters. However,
it is not the end of the story. The unobvious is revealed when we look at the
fixed SNR case results listed in Table~\ref{tab:medianerror-fixedsnr}.
Comparing the FWF median errors corresponding to our three representative cases
we find that the improvement is not entirely due to the larger SNR for detector
networks with larger number of detectors but some other effects are also play
significant roles. Below we try to quantify these effects in the light of
results displayed in Table~\ref{tab:medianerror-fixedsnr}.

\bit
\item{{\bf Localization}:
Upon comparing median errors corresponding to the FWF cases in context of our 
representative network combinations listed in Table~\ref{tab:medianerror-fixedsnr}
we find that angular resolution improves by a factor of about 2.2 and 3.4 as one 
adds KAGRA and both KAGRA and LIGO-India to the LHV network, respectively. This 
can be understood in the following way.

Since both LHVK and LHVKI networks shall involve pairs of detectors with
baselines larger than the ones in the LHV network, an improvement in the
angular resolution is indeed expected as the angular resolution goes roughly as
the square of the distance between the two detectors. More precisely it is the
area of the triangle formed by three detectors in the network which decides
which 3-detector network shall give the best angular resolution
\cite{Pai:2000zt}.  For instance, we find that among the 3-detector networks
LVK has the largest area which is also the 3-detector network which can best
resolve sources with same SNR. However, by comparing LKI and LVK cases in
Table~\ref{tab:medianerror},  we can see that they both give comparable angular
resolution. This is because, LVK has larger geometrical area and smaller SNR
and LKI has larger SNR but smaller area.  It so happens that two different
effects give similar performance for these two cases.

In the case of detector networks with {\it four} or more detectors these areas
can be combined to get an ``effective'' area which shall decide which
combination gives the best estimate for the angular resolution. In
\cite{2010PhRvD..81h2001W}, similar results in the context of GW bursts are
obtained. Thus, as we include a different detector site, the effective area
increases and hence better angular resolution can be achieved using a network
with more detectors at different locations which is indeed true in the cases we
consider. Moreover, it was pointed out in Ref.~\cite{Sathya.LIGOIndia} that if
only time delays are used to triangulate the source, the source's location is
strictly bimodal for a three detector network.\footnote{Although, additional
information about the source position through difference in antenna pattern
functions, breaks this degeneracy even in 3-detector case over a large fraction
of sky, such degeneracy still persist in the significant region of the sky.}
However, with four or more detector sites, this degeneracy is completely
resolved which leads to better measurement to location angle parameters and
hence improves the angular resolution of the source.  }

\item{{\bf Luminosity Distance and the Orientation of the binary:} Inclusion of
detectors at the fourth and fifth site not only ensures better localization but
also improves the measurement of the inclination angle parameter as some of the
degeneracy among angular parameters are resolved which in turn lead to better
measurement of inclination angle of the binary. We find that the
$D_L$-$\cos(\iota)$ component of the median correlation coefficient matrix in
context of LHV, LHVK and LHVKI networks   are about 0.907, 0.898, and 0.889.
Since inclination angle is strongly correlated with the luminosity distance
($D_L$), an improvement in the measurement of inclination angle shall strongly
affect the distance measurements. However, it should be noted that correlations
do not vary much from case to case although there is a systematic decrease when
one goes from LHV to LHVK to LHVKI case.  This small decrease in correlations
is in fact responsible for small improvements we observe in measuring $D_L$,
and $\cos(\iota)$ as we do the analysis with detector networks with four or
five detectors. Note that the $\iota$-$D_L$ degeneracy, which we talked about
in Sec.~\ref{subsec:RWF&FWF}, is already resolved when one uses the FWF and
hence the inclusion of detector at fourth and fifth site further improves the
measurement of both inclination angle and the luminosity distance.}

\item{{\bf Mass parameters, Coalescence time and phase:} We find that
improvement in the measurement of mass parameters which is seen in
Fig.~\ref{fig:compNetwork1} is mostly due to the larger SNR for LHVK and LHVKI
case in comparison with the LHV case (this can be seen by comparing related
numbers provided in
Table~\ref{tab:medianerror}-\ref{tab:medianerror-fixedsnr}). However, in the
cases of errors corresponding to a fixed SNR=$20$,  we find an interesting
feature in many cases, that is, the detector network with more detectors gives
worse parameter estimation accuracy.  For example, for ${\cal M}_c$ and
$\delta$, LHVK and LHVKI cases are worse than LHV case.  Similar trend can be
seen between LHK and LHVK, between LHVI and LHVKI, and between LHKI and LHVKI.
We do not see these trends in other parameters. In order to investigate the
origin of this behaviour, we performed another simulation in which all 5
detectors have the same noise power spectrum of advanced LIGO.  The results are
summarized in Table \ref{tab:medianerror_allLIGO}.  In Table
\ref{tab:medianerror_allLIGO-fixedsnr}, errors corresponding to a fixed SNR=20
are given.  We find in Table \ref{tab:medianerror_allLIGO-fixedsnr} that we do
not see the trend found in Table \ref{tab:medianerror-fixedsnr}.  Indeed the
errors for $\ln{\cal M}_c$ and $\delta$ are nearly equal in all detector
combinations, and they are slightly better in 4 and 5 detector cases.  The
errors for $\ln{\cal M}_c$ (FWF) are ($8.22$-$8.24)\times 10^{-5}$ for 3
detector cases, and $8.22\times 10^{-5}$ for 4 and 5 detector cases, and the
error for $\delta$ (FWF) are about $1.01\times 10^{-3}$ in all cases.  These
facts suggest that the worse estimation errors of $\ln{\cal M}_c$ and $\delta$
for LHVK and LHVKI cases than LHV case are caused by the difference of shape of
the noise power spectrum density.  As we can see from Fig. \ref{fig:noisecurve}
that the noise curve used for advanced LIGO is wider bandwidth compared with
advanced Virgo and KAGRA.  This wider bandwidth, especially at low frequency
region, is effective to have a better estimation accuracy of mass parameters.
When we adopt the noise curve of advanced Virgo or KAGRA, we have a slightly
inferior estimation ability of mass parameter.  This effect becomes manifest
when we set the uniform network SNR.

It is interesting to note in Table \ref{tab:corrcoef} that, the median of the
correlation coefficients for the pairs, (ln${\cal M}_c$, $t_c$), (ln${\cal
M}_c$, $\Phi_c$), ($\delta$, $t_c$), and ($\delta$, $\Phi_c$),  systematically
increase as we go from LHV to LHVK or LHVKI case where as correlations between
mass parameters hardly change.  This would lead to small degradation in
measurement of mass parameters, $t_c$ and $\Phi_c$ when we go from LHV to LHVK
or LHVKI case.  Note however that, as we can see in Table
\ref{tab:corrcoef_allLIGO}, these feature remain even in the case when all of
the detector noise are given by that of advanced LIGO.  Thus, this is not the
main reason of larger errors of $\ln{\cal M}_c$ and $\delta$ in LHVK and LHVKI
cases than in LHV case.  Note also that the estimation errors of $t_c$ and
$\Phi_c$ systematically decrease from LHV to LHVK and LHVKI even for the fixed
SNR case, although the difference of $\Phi_c$ is very small.

On the other hand, we find in Table  \ref{tab:corrcoef} that the correlation
coefficients for pairs, (ln${\cal M}_c$, $\theta$), (ln${\cal M}_c$, $\phi$),
($\delta$, $\theta$), and ($\delta$, $\phi$) increases as we go from LHV to
LHVK, and from LHV to LHVKI.  For example, the median of correlation
coefficients of (ln${\cal M}_c$, $\theta$) are $5.38\times 10^{-3}$,
$1.62\times 10^{-2}$, and $1.47\times 10^{-2}$, for LHV, LHVK and LHVKI,
respectively.  Although these correlation coefficients are not very large, the
estimation errors of ln${\cal M}_c$, $\delta$, $\theta$ and $\phi$ might be
slightly affected as correlation coefficients change significantly from LHV to
LHVK or LHVKI case.  These feature are also explained with the difference of
the noise power spectrum used in the analysis.  In fact, this trend disappears
in the case when all of the detector noise are given by that of advanced LIGO.
As we can see in  Table \ref{tab:corrcoef_allLIGO}, the correlation
coefficients of (ln${\cal M}_c$, $\theta$) are $2.29\times 10^{-3}$,
$1.61\times 10^{-3}$, and $1.32\times 10^{-3}$, for LHV, LHVK and LHVKI,
respectively.} \eit

\begin{table*}
\centering
\caption{The table displays the median of various error distributions investigated in the paper 
which correspond to different detector combinations and to the use of different waveform models 
(RWF and FWF (2.5PN in amplitude)). Simulations performed for a population of BH-NS systems, all placed 
at a luminosity distance of 200 Mpc, and distributed uniformly on the sky surface.
The column, $\Delta \Omega_{95}$, show the median of the 95\% confidence region of the source localization error. 
The last column, SNR,  show the median of signal-to-noise ratio of the network. }
\vskip 0.5cm
\label{tab:medianerror}
\resizebox{16cm}{!}{
\begin{tabular*}{0.94\textwidth}{c|c|c|c|c|c|c|c|c|c|c|c|c}
\hline
\hline
\multicolumn{12}{l}{{$(m_1,m_2) = (1.4,\,10)M_\odot $};\,\,\,{$D_L=200$ Mpc};} \\ 
\hline
&Model&$\Delta D_L/D_L$ &$\Delta {\cal M}_c/{\cal M}_c$ &$\Delta \delta$ &$\Delta t_c$ 
&$\Delta \Phi_c$ &$\Delta \theta$ &$\Delta \phi$ &$\Delta \psi$ &$\Delta \cos(\iota)$ &$\Delta \Omega_{95}$ & SNR\\
&     &  &$(10^{-5})$ &$(10^{-3})$ &($10^{-4}$ sec) &(rad) &(arcmins) &(arcmins) &(rad) & &(deg$^2$)\\ 
\hline
LHV   
&FWF & 0.161 &  9.20 & 1.06 &  5.46 & 0.425 & 101 &  73.9 & 0.197 &0.125 & 21.5 &  19.8 \\ 
&RWF & 0.312 &  9.95 & 1.08 &  5.79 & 0.604 & 114 &  80.3 & 0.310 &0.236 & 26.1 &  20.6 \\ 
\hline
LHK   
&FWF & 0.166 &  8.75 & 1.05 &  6.01 & 0.427 & 136 & 107 & 0.196 &0.127 & 31.2 &  20.2 \\ 
&RWF & 0.315 &  9.50 & 1.07 &  6.25 & 0.604 & 150 & 119 & 0.319 &0.231 & 37.0 &  20.9 \\ 
\hline
LHI   
&FWF & 0.141 &  7.80 & 0.960 &  4.63 & 0.396 &  95.8 &  76.9 & 0.176 &0.105 & 16.9 &  21.1 \\ 
&RWF & 0.251 &  8.46 & 0.977 &  4.61 & 0.533 &   102 &  80.3 & 0.263 &0.188 & 19.1 &  21.9 \\ 
\hline
LVK   
&FWF & 0.148 & 10.6 & 1.18 &  4.75 & 0.447 &  72.5 &  50.3 & 0.189 &0.115 & 12.6 &  18.8 \\ 
&RWF & 0.246 & 11.5 & 1.19 &  4.63 & 0.577 &  75.7 &  51.2 & 0.250 &0.187 & 13.5 &  19.6 \\ 
\hline
LVI   
&FWF & 0.137 &  9.00 & 1.05 &  4.63 & 0.412 &  91.5 &  63.3 & 0.178 &0.104 & 15.0 &  19.8 \\ 
&RWF & 0.228 &  9.73 & 1.06 &  4.62 & 0.537 &  94.6 &  64.6 & 0.238 &0.173 & 16.0 &  20.7 \\ 
\hline
LKI   
&FWF & 0.135 &  8.76 & 1.05 &  4.47 & 0.423 &  76.9 &  51.5 & 0.176 &0.103 & 12.5 &  20.0 \\ 
&RWF & 0.225 &  9.42 & 1.06 &  4.39 & 0.534 &  79.0 &  52.3 & 0.231 &0.166 & 13.4 &  20.9 \\ 
\hline
HVK   
&FWF & 0.152 & 10.6 & 1.18 &  4.90 & 0.452 &  77.4 &  50.9 & 0.191 &0.117 & 14.3 &  18.8 \\ 
&RWF & 0.253 & 11.5 & 1.19 &  4.80 & 0.585 &  81.8 &  52.5 & 0.253 &0.194 & 15.8 &  19.7 \\ 
\hline
HVI   
&FWF & 0.143 &  8.87 & 1.04 &  4.55 & 0.402 &  81.3 &  56.4 & 0.177 &0.107 & 13.4 &  20.0 \\ 
&RWF & 0.229 &  9.59 & 1.05 &  4.49 & 0.527 &  85.8 &  57.7 & 0.237 &0.172 & 14.4 &  20.8 \\ 
\hline
HKI   
&FWF & 0.146 &  8.65 & 1.04 &  4.79 & 0.418 &  86.6 &  61.1 & 0.181 &0.110 & 15.2 &  20.2 \\ 
&RWF & 0.244 &  9.35 & 1.06 &  4.86 & 0.548 &  91.4 &  63.9 & 0.247 &0.179 & 17.3 &  21.0 \\ 
\hline
VKI   
&FWF & 0.172 & 10.6 & 1.17 &  5.39 & 0.456 &  86.3 &  67.9 & 0.226 &0.133 & 18.1 &  18.8 \\ 
&RWF & 0.390 & 11.5 & 1.19 &  5.68 & 0.697 &  101 &  81.2 & 0.421 &0.305 & 22.4 &  19.5 \\ 
\hline
LHVK  
&FWF & 0.123 &  8.44 & 0.967 &  3.91 & 0.374 &  56.9 &  40.0 & 0.148 &0.0927 &  8.44 &  22.3 \\ 
&RWF & 0.188 &  9.04 & 0.971 &  3.72 & 0.479 &  56.6 &  39.4 & 0.179 &0.138 &  8.15 &  23.4 \\ 
\hline
LHVI  
&FWF & 0.114 &  7.51 & 0.888 &  3.60 & 0.346 &  58.2 &  41.0 & 0.140 &0.0848 &  7.60 &  23.2 \\ 
&RWF & 0.172 &  8.07 & 0.894 &  3.41 & 0.437 &  56.9 &  39.2 & 0.167 &0.126 &  7.20 &  24.3 \\ 
\hline
LHKI  
&FWF & 0.116 &  7.39 & 0.891 &  3.75 & 0.359 &  60.0 &  43.0 & 0.140 &0.0859 &  8.28 &  23.4 \\ 
&RWF & 0.176 &  7.93 & 0.899 &  3.61 & 0.450 &  59.9 &  42.0 & 0.168 &0.128 &  8.15 &  24.5 \\ 
\hline
LVKI  
&FWF & 0.114 &  8.31 & 0.959 &  3.71 & 0.369 &  51.5 &  36.7 & 0.144 &0.0852 &  6.54 &  22.3 \\ 
&RWF & 0.173 &  8.89 & 0.963 &  3.51 & 0.455 &  50.5 &  35.0 & 0.169 &0.128 &  6.13 &  23.5 \\ 
\hline
HVKI  
&FWF & 0.120 &  8.24 & 0.952 &  3.81 & 0.364 &  54.3 &  38.2 & 0.146 &0.0879 &  7.10 &  22.4 \\ 
&RWF & 0.179 &  8.84 & 0.957 &  3.65 & 0.459 &  53.9 &  37.0 & 0.174 &0.131 &  6.89 &  23.5 \\ 
\hline
LHVKI 
&FWF & 0.101 &  7.07 & 0.826 &  3.23 & 0.319 &  44.0 &  31.3 & 0.121 &0.0733 &  4.86 &  25.5 \\ 
&RWF & 0.145 &  7.56 & 0.830 &  3.04 & 0.392 &  42.5 &  29.6 & 0.135 &0.104 &  4.39 &  26.8 \\ 
\hline
\hline
\end{tabular*}}
\end{table*}

\begin{table*}
\centering
\caption{Same as that in Table~\ref{tab:medianerror} but errors rescaled 
to values so as to correspond to a fix SNR or 20. Hence unlike median errors
displayed in Table~\ref{tab:medianerror} which correspond to a fixed distance 
of 200 Mpc, the errors displayed here correspond to a fixed SNR of 20.}
\vskip 0.5cm
\label{tab:medianerror-fixedsnr}
\resizebox{16cm}{!}{
\begin{tabular*}{0.94\textwidth}{c|c|c|c|c|c|c|c|c|c|c|c}
\hline
\hline
\multicolumn{12}{l}{{$(m_1,m_2) = (1.4,\,10)M_\odot $}; SNR=20} \\ 
\hline
&Model&$\Delta D_L/D_L$ &$\Delta {\cal M}_c/{\cal M}_c$ &$\Delta \delta$ &$\Delta t_c$ 
&$\Delta \Phi_c$ &$\Delta \theta$ &$\Delta \phi$ &$\Delta \psi$ &$\Delta \cos(\iota)$ &$\Delta \Omega_{95}$ \\
&     &  &$(10^{-5})$ &$(10^{-3})$ &($10^{-4}$ sec) &(rad) &(arcmins) &(arcmins) &(rad) & &(deg$^2$)\\ 
\hline
LHV   
&FWF &    0.163 &     9.01 &    1.05 &     5.14 &    0.395 &     98.4 &     76.7 &    0.206 &   0.130 &    20.5 \\ 
&RWF &    0.302 &     9.88 &    1.09 &     5.58 &    0.462 &    114 &     84.5 &    0.306 &   0.236 &    25.6 \\ 
\hline
LHK   
&FWF &    0.165 &     8.88 &    1.07 &     6.00 &    0.405 &    129 &    107.0 &    0.216 &   0.131 &    29.7 \\ 
&RWF &    0.306 &     9.77 &    1.11 &     6.25 &    0.472 &    140 &    119.0 &    0.319 &   0.237 &    36.3 \\ 
\hline
LHI   
&FWF &    0.153 &     8.24 &    1.01 &     4.45 &    0.388 &     94.9 &     87.2 &    0.199 &   0.120 &    18.1 \\ 
&RWF &    0.261 &     9.26 &    1.07 &     4.49 &    0.437 &    105 &     94.3 &    0.273 &   0.201 &    21.5 \\ 
\hline
LVK   
&FWF &    0.142 &    10.1 &    1.12 &     4.20 &    0.404 &     67.2 &     45.8 &    0.180 &   0.113 &    10.8 \\ 
&RWF &    0.233 &    11.2 &    1.17 &     4.15 &    0.440 &     73.0 &     49.4 &    0.237 &   0.182 &    13.0 \\ 
\hline
LVI   
&FWF &    0.142 &     8.97 &    1.05 &     4.34 &    0.384 &     87.6 &     67.7 &    0.182 &   0.110 &    14.6 \\ 
&RWF &    0.227 &     9.86 &    1.09 &     4.36 &    0.417 &     94.4 &     71.4 &    0.236 &   0.176 &    16.8 \\ 
\hline
LKI   
&FWF &    0.142 &     8.80 &    1.06 &     4.16 &    0.392 &     72.2 &     53.0 &    0.180 &   0.110 &    12.1 \\ 
&RWF &    0.226 &     9.71 &    1.10 &     4.11 &    0.423 &     79.0 &     55.8 &    0.230 &   0.172 &    14.3 \\ 
\hline
HVK   
&FWF &    0.145 &    10.1 &    1.12 &     4.33 &    0.404 &     71.3 &     45.8 &    0.183 &   0.114 &    12.6 \\ 
&RWF &    0.240 &    11.3 &    1.17 &     4.30 &    0.444 &     76.7 &     49.4 &    0.244 &   0.190 &    14.8 \\ 
\hline
HVI   
&FWF &    0.143 &     8.94 &    1.05 &     4.23 &    0.384 &     79.1 &     57.5 &    0.184 &   0.112 &    13.2 \\ 
&RWF &    0.232 &     9.83 &    1.09 &     4.24 &    0.421 &     85.1 &     60.8 &    0.241 &   0.178 &    15.3 \\ 
\hline
HKI   
&FWF &    0.148 &     8.79 &    1.06 &     4.56 &    0.395 &     83.9 &     61.0 &    0.189 &   0.116 &    15.5 \\ 
&RWF &    0.245 &     9.71 &    1.10 &     4.60 &    0.433 &     92.2 &     66.1 &    0.247 &   0.188 &    18.4 \\ 
\hline
VKI   
&FWF &    0.170 &    10.0 &    1.12 &     4.76 &    0.412 &     79.0 &     63.1 &    0.224 &   0.137 &    15.5 \\ 
&RWF &    0.371 &    11.1 &    1.16 &     5.05 &    0.531 &     90.7 &     77.2 &    0.394 &   0.292 &    19.5 \\ 
\hline
LHVK  
&FWF &    0.135 &     9.46 &    1.09 &     4.16 &    0.391 &     60.7 &     41.9 &    0.166 &   0.106 &    10.2 \\ 
&RWF &    0.206 &    10.5 &    1.13 &     4.10 &    0.418 &     63.8 &     43.0 &    0.206 &   0.160 &    11.0 \\ 
\hline
LHVI  
&FWF &    0.132 &     8.77 &    1.04 &     4.08 &    0.376 &     68.5 &     50.0 &    0.162 &   0.103 &    10.2 \\ 
&RWF &    0.195 &     9.65 &    1.08 &     4.03 &    0.400 &     70.1 &     50.7 &    0.196 &   0.151 &    10.7 \\ 
\hline
LHKI  
&FWF &    0.134 &     8.66 &    1.05 &     4.16 &    0.385 &     68.0 &     50.8 &    0.165 &   0.104 &    10.8 \\ 
&RWF &    0.199 &     9.59 &    1.09 &     4.09 &    0.408 &     70.2 &     52.4 &    0.199 &   0.154 &    11.8 \\ 
\hline
LVKI  
&FWF &    0.130 &     9.34 &    1.08 &     4.02 &    0.385 &     58.6 &     40.3 &    0.159 &   0.101 &     8.32 \\ 
&RWF &    0.191 &    10.3 &    1.12 &     3.93 &    0.408 &     60.5 &     40.8 &    0.190 &   0.148 &     8.88 \\ 
\hline
HVKI  
&FWF &    0.133 &     9.31 &    1.08 &     4.07 &    0.386 &     60.6 &     40.7 &    0.164 &   0.103 &     9.14 \\ 
&RWF &    0.201 &    10.3 &    1.12 &     3.98 &    0.412 &     61.6 &     42.0 &    0.200 &   0.155 &     9.62 \\ 
\hline
LHVKI 
&FWF &    0.126 &     9.08 &    1.06 &     4.00 &    0.380 &     55.6 &     39.0 &    0.152 &   0.0967 &    7.90 \\ 
&RWF &    0.180 &    10.1 &    1.11 &     3.90 &    0.399 &     56.1 &     39.0 &    0.178 &   0.138 &     8.09 \\ 
\hline
\hline
\end{tabular*}}
\end{table*}

\begin{table*}
\centering
\caption{The table displays the median of the absolute value of the correlation coefficient matrix
for FWF case. 
This is obtained from the same simulation of Table \ref{tab:medianerror}.}
\vskip 0.5cm
\label{tab:corrcoef}
\begin{tabular*}{0.62\textwidth}{c|c|c|c|c|c|c|c|c|c|}
\hline
\hline
\multicolumn{10}{l}{LHV} \\ 
\hline
                 & ln$D_L$ & ln${\cal M}_c$ & $\delta$ & $t_c$ & $\Phi_c$ & $\theta$ & $\phi$ & $\psi$ & $\cos(\iota)$\\
\hline
ln$D_L$          & 1.00    & 0.0133  & 0.0147  & 0.122  & 0.0352 & 0.211   & 0.198   & 0.164  & 0.912  \\
ln${\cal M}_c$   & 0.0134  & 1.00    & 0.897   & 0.530  & 0.721  & 0.00539 & 0.00606 & 0.0203 & 0.0193 \\
$\delta$         & 0.0147  & 0.897   & 1.00    & 0.658  & 0.871  & 0.0164  & 0.0164  & 0.0131 & 0.0198 \\
$t_c$            & 0.122   & 0.530   & 0.658   & 1.00   & 0.556  & 0.665   & 0.425   & 0.0896 & 0.0938 \\
$\Phi_c$         & 0.0352  & 0.721   & 0.871   & 0.556  & 1.00   & 0.0517  & 0.0528  & 0.454  & 0.0351 \\
$\theta$         & 0.211   & 0.00539 & 0.0164  & 0.665  & 0.0517 & 1.00    & 0.577   & 0.158  & 0.192  \\
$\phi$           & 0.198   & 0.00606 & 0.0164  & 0.425  & 0.0528 & 0.577   & 1.00    & 0.170  & 0.192  \\
$\psi$           & 0.164   & 0.0203  & 0.0131  & 0.0896 & 0.454  & 0.158   & 0.170   & 1.00   & 0.0780 \\
$\cos(\iota)$    & 0.912   & 0.0193  & 0.0198  & 0.0938 & 0.0351 & 0.192   & 0.192   & 0.0780 & 1.00   \\
\hline
\hline
\multicolumn{10}{l}{LHVK} \\ 
\hline
                 & ln$D_L$ & ln${\cal M}_c$ & $\delta$ & $t_c$ & $\Phi_c$ & $\theta$ & $\phi$ & $\psi$ & $\cos(\iota)$\\
\hline
ln$D_L$          & 1.00   & 0.0128 & 0.0146 & 0.0528 & 0.0277 & 0.160  & 0.126  & 0.145  & 0.899  \\
ln${\cal M}_c$   & 0.0128 & 1.00   & 0.895  & 0.680  & 0.753  & 0.0162 & 0.0179 & 0.0129 & 0.0172 \\
$\delta$         & 0.0146 & 0.895  & 1.00   & 0.852  & 0.905  & 0.0292 & 0.0307 & 0.0151 & 0.0190 \\
$t_c$            & 0.0528 & 0.680  & 0.852  & 1.00   & 0.744  & 0.366  & 0.132  & 0.0473 & 0.0536 \\
$\Phi_c$         & 0.0277 & 0.753  & 0.905  & 0.744  & 1.00   & 0.0522 & 0.0461 & 0.400  & 0.0311 \\
$\theta$         & 0.160  & 0.0162 & 0.0292 & 0.366  & 0.0522 & 1.00   & 0.254  & 0.142  & 0.187  \\
$\phi$           & 0.126  & 0.0179 & 0.0307 & 0.132  & 0.0461 & 0.254  & 1.00   & 0.127  & 0.158  \\
$\psi$           & 0.145  & 0.0129 & 0.0151 & 0.0473 & 0.400  & 0.142  & 0.127  & 1.00   & 0.0658 \\
$\cos(\iota)$    & 0.899  & 0.0172 & 0.0190 & 0.0536 & 0.0311 & 0.187  & 0.158  & 0.0658 & 1.00   \\
\hline
\hline
\multicolumn{10}{l}{LHVKI} \\ 
\hline
                 & ln$D_L$ & ln${\cal M}_c$ & $\delta$ & $t_c$ & $\Phi_c$ & $\theta$ & $\phi$ & $\psi$ & $\cos(\iota)$\\
\hline
ln$D_L$          & 1.00   & 0.0122 & 0.0139 & 0.0339 & 0.0227 & 0.139  & 0.111  & 0.122  & 0.890  \\
ln${\cal M}_c$   & 0.0122 & 1.00   & 0.897  & 0.700  & 0.764  & 0.0147 & 0.0166 & 0.0120 & 0.0139 \\
$\delta$         & 0.0139 & 0.897  & 1.00   & 0.875  & 0.913  & 0.0282 & 0.0307 & 0.0120 & 0.0182 \\
$t_c$            & 0.0339 & 0.700  & 0.875  & 1.00   & 0.779  & 0.312  & 0.125  & 0.0369 & 0.0382 \\
$\Phi_c$         & 0.0227 & 0.764  & 0.913  & 0.779  & 1.00   & 0.0476 & 0.0429 & 0.377  & 0.0289 \\
$\theta$         & 0.139  & 0.0147 & 0.0282 & 0.312  & 0.0476 & 1.00   & 0.313  & 0.128  & 0.163  \\
$\phi$           & 0.111  & 0.0166 & 0.0307 & 0.125  & 0.0429 & 0.313  & 1.00   & 0.118  & 0.141  \\
$\psi$           & 0.122  & 0.0120 & 0.0120 & 0.0369 & 0.377  & 0.128  & 0.118  & 1.00   & 0.0493 \\
$\cos(\iota)$    & 0.890  & 0.0139 & 0.0182 & 0.0382 & 0.0289 & 0.163  & 0.141  & 0.0493 & 1.00   \\
\hline
\hline
\end{tabular*}
\end{table*}

\begin{table*}
\centering
\caption{The table displays the median of the absolute value of the correlation coefficient matrix
for RWF case. 
This is obtained from the same simulation of Table \ref{tab:medianerror}.}
\vskip 0.5cm
\label{tab:corrcoefRWF}
\begin{tabular*}{0.62\textwidth}{c|c|c|c|c|c|c|c|c|c|}
\hline
\hline
\multicolumn{10}{l}{LHV} \\ 
\hline
                 & ln$D_L$ & ln${\cal M}_c$ & $\delta$ & $t_c$ & $\Phi_c$ & $\theta$ & $\phi$ & $\psi$ & $\cos(\iota)$\\
\hline
ln$D_L$          & 1.00    & 0.00688 & 0.0347  & 0.242  & 0.118 & 0.406   & 0.405   & 0.328   & 0.954   \\
ln${\cal M}_c$   & 0.00688 & 1.00    & 0.896   & 0.485  & 0.645 & 0.00786 & 0.00911 & 0.00720 & 0.00885 \\
$\delta$         & 0.0347  & 0.896   & 1.00    & 0.596  & 0.765 & 0.0336  & 0.0369  & 0.0360  & 0.0447  \\
$t_c$            & 0.242   & 0.485   & 0.596   & 1.00   & 0.462 & 0.722   & 0.474   & 0.254   & 0.267   \\
$\Phi_c$         & 0.118   & 0.645   & 0.765   & 0.462  & 1.00  & 0.195   & 0.197   & 0.619   & 0.159   \\
$\theta$         & 0.406   & 0.00786 & 0.0336  & 0.722  & 0.195 & 1.00    & 0.600   & 0.419   & 0.452   \\
$\phi$           & 0.405   & 0.00911 & 0.0369  & 0.474  & 0.197 & 0.600   & 1.00    & 0.438   & 0.458   \\
$\psi$           & 0.328   & 0.00720 & 0.0360  & 0.254  & 0.619 & 0.419   & 0.438   & 1.00    & 0.262   \\
$\cos(\iota)$    & 0.954   & 0.00885 & 0.0447  & 0.267  & 0.159 & 0.452   & 0.458   & 0.262   & 1.00    \\
\hline
\hline
\multicolumn{10}{l}{LHVK} \\ 
\hline
                 & ln$D_L$ & ln${\cal M}_c$ & $\delta$ & $t_c$ & $\Phi_c$ & $\theta$ & $\phi$ & $\psi$ & $\cos(\iota)$\\
\hline
ln$D_L$          & 1.00    & 0.00815 & 0.0257  & 0.0966  & 0.0729 & 0.318   & 0.266  & 0.257   & 0.939   \\
ln${\cal M}_c$   & 0.00815 & 1.00    & 0.895   & 0.682   & 0.732  & 0.0163  & 0.0196 & 0.00861 & 0.00999 \\
$\delta$         & 0.0257  & 0.895   & 1.00    & 0.851   & 0.868  & 0.0387  & 0.0319 & 0.0282  & 0.0344  \\
$t_c$            & 0.0966  & 0.682   & 0.851   & 1.00    & 0.640  & 0.387   & 0.133  & 0.103   & 0.112   \\
$\Phi_c$         & 0.0729  & 0.732   & 0.868   & 0.640   & 1.00   & 0.134   & 0.115  & 0.467   & 0.100   \\
$\theta$         & 0.318   & 0.0163  & 0.0387  & 0.387   & 0.134  & 1.00    & 0.256  & 0.335   & 0.376   \\
$\phi$           & 0.266   & 0.0196  & 0.0319  & 0.133   & 0.115  & 0.256   & 1.00   & 0.304   & 0.337   \\
$\psi$           & 0.257   & 0.00861 & 0.0282  & 0.103   & 0.467  & 0.335   & 0.304  & 1.00    & 0.186   \\
$\cos(\iota)$    & 0.939   & 0.00999 & 0.0344  & 0.112   & 0.100  & 0.376   & 0.337  & 0.186   & 1.00    \\
\hline
\hline
\multicolumn{10}{l}{LHVKI} \\ 
\hline
                 & ln$D_L$ & ln${\cal M}_c$ & $\delta$ & $t_c$ & $\Phi_c$ & $\theta$ & $\phi$ & $\psi$ & $\cos(\iota)$\\
\hline
ln$D_L$          & 1.00    & 0.00454 & 0.0202  & 0.0593  & 0.0451  & 0.253  & 0.207   & 0.187   & 0.932   \\
ln${\cal M}_c$   & 0.00454 & 1.00    & 0.898   & 0.709   & 0.756   & 0.0152 & 0.0182  & 0.00513 & 0.00540 \\
$\delta$         & 0.0202  & 0.898   & 1.00    & 0.879   & 0.893   & 0.0318 & 0.0369  & 0.0214  & 0.0265  \\
$t_c$            & 0.0593  & 0.709   & 0.879   & 1.00    & 0.723   & 0.320  & 0.126   & 0.0678  & 0.0703  \\
$\Phi_c$         & 0.0451  & 0.756   & 0.893   & 0.723   & 1.00    & 0.105  & 0.0933  & 0.419   & 0.0700  \\
$\theta$         & 0.253   & 0.0152  & 0.0318  & 0.320   & 0.105   & 1.00   & 0.284   & 0.269   & 0.298   \\
$\phi$           & 0.207   & 0.0182  & 0.0369  & 0.126   & 0.0933  & 0.284  & 1.00    & 0.247   & 0.263   \\
$\psi$           & 0.187   & 0.00513 & 0.0214  & 0.0678  & 0.419   & 0.269  & 0.247   & 1.00    & 0.114   \\
$\cos(\iota)$    & 0.932   & 0.00540 & 0.0265  & 0.0703  & 0.0700  & 0.298  & 0.263   & 0.114   & 1.00    \\
\hline
\hline
\end{tabular*}
\end{table*}


\begin{table*}
\centering
\caption{The table displays the median of various error distributions 
which correspond to different detector combinations and to the use of different waveform models 
(RWF and FWF (2.5PN in amplitude)). In this simulation,  
the noise power spectrum density of all detectors is 
that of advanced LIGO. The difference of the detectors comes from the location and orientation. 
Other parameters of the simulation are the same as Table \ref{tab:medianerror}. 
The column, $\Delta \Omega_{95}$, show the median of the 95\% confidence region of the source localization error. 
The last column, SNR,  is the median of signal-to-noise ratio of the network. }
\vskip 0.5cm
\label{tab:medianerror_allLIGO}
\resizebox{16cm}{!}{
\begin{tabular*}{0.94\textwidth}{c|c|c|c|c|c|c|c|c|c|c|c|c}
\hline
\hline
\multicolumn{12}{l}{{$(m_1,m_2) = (1.4,\,10)M_\odot $};\,\,\,{$D_L=200$ Mpc};} \\ 
\hline
&Model&$\Delta D_L/D_L$ &$\Delta {\cal M}_c/{\cal M}_c$ &$\Delta \delta$ &$\Delta t_c$ 
&$\Delta \Phi_c$ &$\Delta \theta$ &$\Delta \phi$ &$\Delta \psi$ &$\Delta \cos(\iota)$ &$\Delta \Omega_{95}$ & SNR \\
&     &  &$(10^{-5})$ &$(10^{-3})$ &($10^{-4}$ sec) &(rad) &(arcmins) &(arcmins) &(rad) & &(deg$^2$) & \\ 
\hline
LHV   
&FWF &    0.147 &     7.69 &    0.948 &     5.16 &    0.388 &     97.9 &     71.3 &    0.184 &   0.111 &    20.3 &     21.4 \\ 
&RWF &    0.279 &     8.36 &    0.965 &     5.52 &    0.540 &    110   &     77.5 &    0.272 &   0.205 &    24.5 &     22.2 \\ 
\hline
LHK   
&FWF &    0.154 &     7.68 &    0.946 &     5.50 &    0.385 &    127 &     99.1 &    0.184 &   0.115 &    26.5 &     21.4 \\ 
&RWF &    0.286 &     8.36 &    0.965 &     5.76 &    0.543 &    142 &      109 &    0.286 &   0.207 &    31.0 &     22.2 \\ 
\hline
LHI   
&FWF &    0.141 &     7.80 &    0.960 &     4.63 &    0.396 &     95.8 &     76.9 &    0.176 &   0.105 &    16.9 &     21.1 \\ 
&RWF &    0.251 &     8.46 &    0.977 &     4.61 &    0.533 &      102 &     80.3 &    0.263 &   0.188 &    19.1 &     21.9 \\ 
\hline
LVK   
&FWF &    0.127 &     7.58 &    0.933 &     4.03 &    0.369 &     65.2 &     45.6 &    0.163 &   0.0971 &    10.2 &     21.7 \\ 
&RWF &    0.205 &     8.22 &    0.950 &     3.96 &    0.470 &     67.6 &     45.8 &    0.204 &   0.152 &    10.7 &     22.5 \\ 
\hline
LVI   
&FWF &    0.128 &     7.72 &    0.949 &     4.39 &    0.382 &     89.1 &     61.6 &    0.166 &   0.0970 &    14.2 &     21.4 \\ 
&RWF &    0.209 &     8.34 &    0.963 &     4.37 &    0.496 &     91.8 &     63.2 &    0.214 &   0.156 &    15.2 &     22.2 \\ 
\hline
LKI   
&FWF &    0.128 &     7.74 &    0.952 &     4.10 &    0.389 &     70.3 &     47.4 &    0.164 &   0.0960 &    10.8 &     21.3 \\ 
&RWF &    0.209 &     8.35 &    0.964 &     4.05 &    0.490 &     71.8 &     47.8 &    0.211 &   0.153 &    11.3 &     22.2 \\ 
\hline
HVK   
&FWF &    0.131 &     7.54 &    0.928 &     4.15 &    0.371 &     69.9 &     46.2 &    0.164 &   0.0987 &    11.5 &     21.8 \\ 
&RWF &    0.212 &     8.19 &    0.946 &     4.11 &    0.476 &     73.6 &     46.8 &    0.206 &   0.158 &    12.6 &     22.6 \\ 
\hline
HVI   
&FWF &    0.132 &     7.64 &    0.941 &     4.33 &    0.374 &     79.2 &     55.1 &    0.165 &   0.0987 &    12.8 &     21.5 \\ 
&RWF &    0.208 &     8.27 &    0.956 &     4.30 &    0.486 &     84.1 &     56.7 &    0.213 &   0.154 &    13.7 &     22.4 \\ 
\hline
HKI   
&FWF &    0.137 &     7.66 &    0.943 &     4.44 &    0.387 &     78.8 &     55.8 &    0.169 &   0.102 &    13.0 &     21.5 \\ 
&RWF &    0.225 &     8.27 &    0.956 &     4.50 &    0.503 &     82.3 &     57.7 &    0.226 &   0.164 &    14.7 &     22.4 \\ 
\hline
VKI   
&FWF &    0.151 &     7.57 &    0.933 &     4.61 &    0.380 &     78.3 &     61.9 &    0.199 &   0.115 &    14.8 &     21.7 \\ 
&RWF &    0.333 &     8.28 &    0.956 &     4.91 &    0.577 &     90.8 &     72.1 &    0.346 &   0.252 &    18.2 &     22.4 \\ 
\hline
LHVK  
&FWF &    0.108 &     6.60 &    0.814 &     3.44 &    0.322 &     52.8 &     36.9 &    0.131 &   0.0803 &     7.11 &     24.8 \\ 
&RWF &    0.163 &     7.14 &    0.824 &     3.30 &    0.401 &     52.6 &     36.1 &    0.153 &   0.118 &     6.88 &     26.0 \\ 
\hline
LHVI  
&FWF &    0.108 &     6.70 &    0.825 &     3.44 &    0.325 &     56.9 &     40.2 &    0.131 &   0.0793 &     7.33 &     24.5 \\ 
&RWF &    0.159 &     7.21 &    0.833 &     3.29 &    0.409 &     56.2 &     38.8 &    0.153 &   0.116 &     6.97 &     25.7 \\ 
\hline
LHKI  
&FWF &    0.110 &     6.70 &    0.824 &     3.50 &    0.335 &     56.5 &     40.1 &    0.133 &   0.0811 &     7.39 &     24.6 \\ 
&RWF &    0.166 &     7.21 &    0.833 &     3.38 &    0.419 &     56.3 &     38.9 &    0.156 &   0.120 &     7.26 &     25.7 \\ 
\hline
LVKI  
&FWF &    0.104 &     6.61 &    0.815 &     3.31 &    0.323 &     46.8 &     33.6 &    0.130 &   0.0770 &     5.57 &     24.8 \\ 
&RWF &    0.155 &     7.14 &    0.824 &     3.15 &    0.397 &     46.2 &     32.2 &    0.150 &   0.114 &     5.25 &     25.9 \\ 
\hline
HVKI  
&FWF &    0.108 &     6.57 &    0.810 &     3.40 &    0.319 &     49.8 &     35.1 &    0.131 &   0.0791 &     6.12 &     25.0 \\ 
&RWF &    0.160 &     7.12 &    0.822 &     3.29 &    0.402 &     49.5 &     34.2 &    0.154 &   0.116 &     5.93 &     26.0 \\ 
\hline
LHVKI 
&FWF &    0.092 &     5.92 &    0.728 &     2.94 &    0.287 &     40.9 &     29.2 &    0.110 &   0.0669 &     4.28 &     27.7 \\ 
&RWF &    0.132 &     6.37 &    0.735 &     2.79 &    0.350 &     39.8 &     27.6 &    0.122 &   0.0946 &     3.90 &     29.1 \\ 
\hline
\hline
\end{tabular*}}
\end{table*}

\begin{table*}
\centering
\caption{The table displays the median of various error distributions 
in the case when the noise power spectrum density of all detectors is given by 
that of advanced LIGO. 
The errors displayed here correspond to a fixed SNR of 20.
Other parameters of the simulation are the same as Table \ref{tab:medianerror-fixedsnr}. }
\vskip 0.5cm
\label{tab:medianerror_allLIGO-fixedsnr}
\resizebox{16cm}{!}{
\begin{tabular*}{0.94\textwidth}{c|c|c|c|c|c|c|c|c|c|c|c}
\hline
\hline
\multicolumn{12}{l}{{$(m_1,m_2) = (1.4,\,10)M_\odot $};\,\,\,{SNR=20};} \\ 
\hline
&Model&$\Delta D_L/D_L$ &$\Delta {\cal M}_c/{\cal M}_c$ &$\Delta \delta$ &$\Delta t_c$ 
&$\Delta \Phi_c$ &$\Delta \theta$ &$\Delta \phi$ &$\Delta \psi$ &$\Delta \cos(\iota)$ &$\Delta \Omega_{95}$ \\
&     &  &$(10^{-5})$ &$(10^{-3})$ &($10^{-4}$ sec) &(rad) &(arcmins) &(arcmins) &(rad) & &(deg$^{2}$) \\ 
\hline
LHV   
&FWF &    0.163 &     8.24 &    1.01 &     5.41 &    0.390 &    104 &     78.9 &    0.205 &   0.127 &    23.0 \\ 
&RWF &    0.288 &     9.26 &    1.07 &     5.82 &    0.445 &    119 &     86.5 &    0.287 &   0.221 &    28.0 \\ 
\hline
LHK   
&FWF &    0.164 &     8.24 &    1.01 &     5.88 &    0.391 &    131 &    105 &    0.212 &   0.127 &    28.6 \\ 
&RWF &    0.295 &     9.26 &    1.07 &     6.14 &    0.450 &    142 &    117 &    0.308 &   0.225 &    34.7 \\ 
\hline
LHI   
&FWF &    0.153 &     8.24 &    1.01 &     4.45 &    0.388 &     94.9 &     87.2 &    0.199 &   0.120 &    18.1 \\ 
&RWF &    0.261 &     9.26 &    1.07 &     4.49 &    0.437 &    105 &     94.3 &    0.273 &   0.201 &    21.5 \\ 
\hline
LVK   
&FWF &    0.141 &     8.22 &    1.01 &     4.10 &    0.381 &     69.5 &     48.1 &    0.177 &   0.111 &    11.7 \\ 
&RWF &    0.222 &     9.26 &    1.07 &     4.05 &    0.411 &     74.2 &     51.2 &    0.222 &   0.170 &    13.1 \\ 
\hline
LVI   
&FWF &    0.142 &     8.23 &    1.01 &     4.40 &    0.381 &     90.2 &     69.6 &    0.181 &   0.110 &    15.7 \\ 
&RWF &    0.222 &     9.26 &    1.07 &     4.40 &    0.411 &     97.8 &     73.8 &    0.228 &   0.170 &    18.2 \\ 
\hline
LKI   
&FWF &    0.141 &     8.22 &    1.01 &     4.07 &    0.380 &     70.0 &     51.1 &    0.177 &   0.110 &    11.7 \\ 
&RWF &    0.222 &     9.26 &    1.07 &     4.02 &    0.410 &     75.8 &     54.4 &    0.223 &   0.167 &    13.3 \\ 
\hline
HVK   
&FWF &    0.144 &     8.22 &    1.01 &     4.27 &    0.381 &     74.8 &     48.9 &    0.179 &   0.112 &    13.6 \\ 
&RWF &    0.228 &     9.26 &    1.07 &     4.24 &    0.415 &     80.1 &     51.6 &    0.227 &   0.176 &    15.5 \\ 
\hline
HVI   
&FWF &    0.143 &     8.23 &    1.01 &     4.31 &    0.381 &     83.1 &     60.3 &    0.182 &   0.111 &    14.5 \\ 
&RWF &    0.226 &     9.26 &    1.07 &     4.30 &    0.412 &     88.7 &     64.5 &    0.232 &   0.171 &    16.7 \\ 
\hline
HKI   
&FWF &    0.147 &     8.22 &    1.01 &     4.49 &    0.382 &     81.2 &     58.4 &    0.186 &   0.115 &    14.8 \\ 
&RWF &    0.240 &     9.26 &    1.07 &     4.50 &    0.419 &     87.7 &     63.7 &    0.240 &   0.184 &    17.4 \\ 
\hline
VKI   
&FWF &    0.173 &     8.24 &    1.01 &     4.78 &    0.396 &     83.5 &     65.9 &    0.224 &   0.137 &    17.0 \\ 
&RWF &    0.355 &     9.26 &    1.07 &     5.08 &    0.498 &     95.1 &     78.3 &    0.369 &   0.273 &    20.8 \\ 
\hline
LHVK  
&FWF &    0.134 &     8.22 &    1.01 &     4.09 &    0.376 &     63.5 &     43.3 &    0.162 &   0.104 &    11.1 \\ 
&RWF &    0.197 &     9.26 &    1.07 &     4.03 &    0.400 &     66.6 &     44.6 &    0.194 &   0.151 &    11.7 \\ 
\hline
LHVI  
&FWF &    0.132 &     8.22 &    1.01 &     4.14 &    0.375 &     71.3 &     51.5 &    0.161 &   0.102 &    11.3 \\ 
&RWF &    0.192 &     9.26 &    1.07 &     4.09 &    0.397 &     73.2 &     52.4 &    0.190 &   0.147 &    11.8 \\ 
\hline
LHKI  
&FWF &    0.133 &     8.22 &    1.01 &     4.10 &    0.376 &     67.2 &     49.0 &    0.163 &   0.103 &    10.7 \\ 
&RWF &    0.195 &     9.26 &    1.07 &     4.03 &    0.399 &     69.3 &     50.3 &    0.194 &   0.151 &    11.5 \\ 
\hline
LVKI  
&FWF &    0.131 &     8.22 &    1.01 &     3.95 &    0.375 &     58.8 &     40.4 &    0.160 &   0.102 &     8.81 \\ 
&RWF &    0.190 &     9.26 &    1.07 &     3.87 &    0.396 &     60.3 &     41.0 &    0.187 &   0.145 &     9.24 \\ 
\hline
HVKI  
&FWF &    0.135 &     8.22 &    1.01 &     4.01 &    0.376 &     61.5 &     41.7 &    0.163 &   0.104 &     9.69 \\ 
&RWF &    0.198 &     9.26 &    1.07 &     3.93 &    0.400 &     62.4 &     43.0 &    0.196 &   0.151 &     10.0 \\ 
\hline
LHVKI 
&FWF &    0.127 &     8.22 &    1.01 &     3.95 &    0.372 &     56.3 &     38.9 &    0.151 &   0.0970 &     8.44 \\ 
&RWF &    0.177 &     9.26 &    1.07 &     3.86 &    0.391 &     56.7 &     39.1 &    0.175 &   0.136 &     8.54 \\ 
\hline
\hline
\end{tabular*}}
\end{table*}

\begin{table*}
\centering
\caption{The table displays the median of the absolute value of the correlation coefficient matrix
in the case when all of the detector noise is advanced LIGO. 
This is the case for FWF.
This is obtained from the same simulation of Table \ref{tab:medianerror_allLIGO}.}
\vskip 0.5cm
\label{tab:corrcoef_allLIGO}
\begin{tabular*}{0.68\textwidth}{c|c|c|c|c|c|c|c|c|c|}
\hline
\hline
\multicolumn{10}{l}{LHV} \\ 
\hline
                 & ln$D_L$ & ln${\cal M}_c$ & $\delta$ & $t_c$ & $\Phi_c$ & $\theta$ & $\phi$ & $\psi$ & $\cos(\iota)$\\
\hline
ln$D_L$          & 1.00    & 0.0120  & 0.00822 & 0.128   & 0.0312  & 0.214   & 0.201   & 0.171   & 0.912   \\
ln${\cal M}_c$   & 0.0120  & 1.00    & 0.900   & 0.513   & 0.725   & 0.00230 & 0.00231 & 0.0160  & 0.0150  \\
$\delta$         & 0.00822 & 0.900   & 1.00    & 0.640   & 0.871   & 0.00186 & 0.00198 & 0.00593 & 0.00912 \\
$t_c$            & 0.128   & 0.513   & 0.640   & 1.00    & 0.544   & 0.675   & 0.439   & 0.0993  & 0.104   \\
$\Phi_c$         & 0.0312  & 0.725   & 0.871   & 0.544   & 1.00    & 0.0518  & 0.0535  & 0.459   & 0.0336  \\
$\theta$         & 0.214   & 0.00230 & 0.00186 & 0.675   & 0.0518  & 1.00    & 0.570   & 0.167   & 0.192   \\
$\phi$           & 0.201   & 0.00231 & 0.00198 & 0.439   & 0.0535  & 0.570   & 1.00    & 0.180   & 0.202   \\
$\psi$           & 0.171   & 0.0160  & 0.00593 & 0.0993  & 0.459   & 0.167   & 0.180   & 1.00    & 0.0811  \\
$\cos(\iota)$    & 0.912   & 0.0150  & 0.00912 & 0.104   & 0.0336  & 0.192   & 0.202   & 0.0811  & 1.00    \\
\hline
\hline
\multicolumn{10}{l}{LHVK} \\ 
\hline
                 & ln$D_L$ & ln${\cal M}_c$ & $\delta$ & $t_c$ & $\Phi_c$ & $\theta$ & $\phi$ & $\psi$ & $\cos(\iota)$\\
\hline
ln$D_L$          & 1.00    & 0.0126  & 0.00879 & 0.0527  & 0.0223  & 0.157   & 0.126   & 0.140   & 0.898   \\
ln${\cal M}_c$   & 0.0126  & 1.00    & 0.900   & 0.674   & 0.754   & 0.00162 & 0.00131 & 0.0141  & 0.0109  \\
$\delta$         & 0.00879 & 0.900   & 1.00    & 0.841   & 0.901   & 0.00160 & 0.00141 & 0.00533 & 0.00675 \\
$t_c$            & 0.0527  & 0.674   & 0.841   & 1.00    & 0.732   & 0.370   & 0.118   & 0.0490  & 0.0568  \\
$\Phi_c$         & 0.0223  & 0.754   & 0.901   & 0.732   & 1.00    & 0.0425  & 0.0370  & 0.403   & 0.0296  \\
$\theta$         & 0.157   & 0.00162 & 0.00160 & 0.370   & 0.0425  & 1.00    & 0.252   & 0.147   & 0.190   \\
$\phi$           & 0.126   & 0.00131 & 0.00141 & 0.118   & 0.0370  & 0.252   & 1.00    & 0.130   & 0.161   \\
$\psi$           & 0.140   & 0.0141  & 0.00533 & 0.0490  & 0.403   & 0.147   & 0.130   & 1.00    & 0.0639  \\
$\cos(\iota)$    & 0.898   & 0.0109  & 0.00675 & 0.0568  & 0.0296  & 0.190   & 0.161   & 0.0639  & 1.00    \\
\hline
\hline
\multicolumn{10}{l}{LHVKI} \\ 
\hline
                 & ln$D_L$ & ln${\cal M}_c$ & $\delta$ & $t_c$ & $\Phi_c$ & $\theta$ & $\phi$ & $\psi$ & $\cos(\iota)$\\
\hline
ln$D_L$          & 1.00    & 0.0127  & 0.00889 & 0.0311  & 0.0199  & 0.140   & 0.110   & 0.125   & 0.890   \\
ln${\cal M}_c$   & 0.0127  & 1.00    & 0.900   & 0.699   & 0.763   & 0.00133 & 0.00107 & 0.0136  & 0.0100  \\
$\delta$         & 0.00889 & 0.900   & 1.00    & 0.872   & 0.910   & 0.00134 & 0.00117 & 0.00507 & 0.00592 \\
$t_c$            & 0.0311  & 0.699   & 0.872   & 1.00    & 0.771   & 0.301   & 0.09981 & 0.0350  & 0.0339  \\
$\Phi_c$         & 0.0199  & 0.763   & 0.910   & 0.771   & 1.00    & 0.0356  & 0.0297  & 0.384   & 0.0239  \\
$\theta$         & 0.140   & 0.00133 & 0.00134 & 0.301   & 0.0356  & 1.00    & 0.288   & 0.129   & 0.167   \\
$\phi$           & 0.110   & 0.00107 & 0.00117 & 0.0998  & 0.0297  & 0.288   & 1.00    & 0.119   & 0.141   \\
$\psi$           & 0.125   & 0.0136  & 0.00507 & 0.0350  & 0.384   & 0.129   & 0.119   & 1.00    & 0.0480  \\
$\cos(\iota)$    & 0.890   & 0.0100  & 0.00592 & 0.0339  & 0.0239  & 0.167   & 0.141   & 0.0480  & 1.00    \\
\hline
\hline
\end{tabular*}
\end{table*}


\subsection{Addition of LIGO-India and its benefits}
\label{subsec:LI}

In this section we aim to discuss in particular the benefits of including the
LIGO-India detector to the future LIGO-Virgo-KAGRA network. In previous
subsections we have already discussed benefits of having a full five detector
network that includes LIGO-India. In this section we argue how the presence of
the LIGO-India detector in the operational network would help us in achieving
better parameter estimation.  
In particular, we present the scenario 
when inclusion of LIGO-India helps achieving good parameter estimation accuracies 
even in a situation when one of the four detectors (LHVK) is not operational.
The discussion presented below is based on
comparisons of FWF errors in  various parameters in context of all possible
network combinations with three or more detectors displayed in
Table~\ref{tab:medianerror}. Although we discuss the benefits of including the
LIGO-India detector only in context of better localization and
distance-inclination angle measurements, the arguments presented below are in
general true to estimation of all parameters. 

\begin{itemize}

\item{{\bf Localization:} With 4-site LIGO-Virgo-KAGRA network, at times when
one of the detectors are not operational, the 4 possible 3-site networks, LHV,
LHK, LVK, and HVK, will be able to localize the source within 95\% confidence
region of about $12.6 - 21.4$ sqdeg.  However, if LIGO-India is included in the
network, all 6 possible 3-site networks including LIGO-India will be able to
localize the source within about $12.4-18.0$ sqdeg.  Among all possible 3-site
network, the best localization is achieved with LKI network.

 We mentioned in Sec.~\ref{subsec:RWF&FWF} that if only time delays are used to
triangulate the source, the source's location is strictly bimodal for a
3-detector network \cite{Sathya.LIGOIndia}. This degeneracy is completely
broken with the inclusion of the fourth detector. In contrast to the
LIGO-Virgo-KAGRA network, which has just one four detector combination (LHVK)
the future 5-site network (with the addition of LIGO-India) shall consist of 4
additional 4-site configurations (LHVI, LHKI, LVKI, and HVKI) which enhances
duty-cycle of 4-detector networks. Moreover, all the four detector combinations
involving LIGO-India have slightly better resolution as compared to the LHVK
combinations (see Table \ref{tab:medianerror} above).

 From the Table \ref{tab:medianerror} it should be clear that the 5-detector
combination (LHVKI) significantly improves the error estimation almost for all
the parameters of the source, in particular the angular resolution. As compared
to the best 3-detector (LKI) and the 4-detector (LVKI) which with 95\%
confidence can locate the source within about $\sim$12.4 sqdeg and about
$\sim$6.5 sqdeg, respectively, the LHVKI network should be able to resolve the
source within 4.8 sqdeg.  } \item{{\bf Distance and inclination angle
measurements :} With 4-site LIGO-Virgo-KAGRA network, at times when one of the
detector will not be operational, the 4 possible 3-site networks, LHV, LHK,
LVK, and HVK, will be able to determine the cosine of the inclination angle
with median errors of $(11.5$-$12.7)$\%.  However, if LIGO-India is included in
the network, all 6 possible 3-site networks including LIGO-India will be able
to constrain the cosine of the inclination angle with median errors of
$(10.3$-$13.3)$\%.  Among all possible 3-site network, the best determination
is achieved with LKI network, although the difference between the network is
very small.

We see exactly the same trend when comparing median errors in distance
measurements in context of various 3-detector networks (see
Table~\ref{tab:medianerror}). This is not surprising as distance and
inclination angle are strongly correlated with each other.  One can see  in
Table~\ref{tab:medianerror}, as compared with 3-site network without LIGO-India
when median errors in distance are $14.8-16.6$\%, the 3-site networks with
LIGO-India will measure the distance within median errors of about
$13.5-14.6$\% except for VKI case, for which the median error is 17.2\%.
Median error in the case of LHI network is about 14.1\%. 

 As discussed above, inclusion of LIGO-India to the LIGO-Virgo-KAGRA network
will allow 4 additional 4-detector networks which not only improve the duty
cycle for four or more detector networks but also will lead to better
localization than the one in case of LHVK network. As one can see in
Table~\ref{tab:medianerror}, this is also true in case of distance and
inclination angle measurements. Inclusion of LIGO-India will not only ensure
that more often we shall have an operational 4-detector network but also
measure these parameters with accuracies better than the one in case of LHVK
network. Finally, as can be seen in Table~\ref{tab:medianerror}, both distance
and inclination angle are best measured in the 5-detector network, with median
errors of about  10.1\% and 7.3\%, respectively.} \end{itemize}
\section{Discussion and future direction} \label{sec:discn} 

In this paper we presented our findings of the parameter estimation study which
was performed considering a population of NS-BH systems in context of the
network of future advanced detectors.  For the analysis we used 12800
realizations of the source (with fixed component masses of 1.4 and 10
M$_\odot$), obtained by randomizing all four angular parameters giving location
($\theta$, $\phi$) and orientation ($\iota$, $\psi$), all at a fixed luminosity
distance of 200 Mpc. Our prime focus in this paper has been to investigate the
quality of parameter estimation that can be achieved using amplitude-corrected
waveform  of inspiral signal from a nonspinning NS-BH system. For this purpose
we use a post-Newtonian waveform that is 2.5PN accurate in amplitude and 3.5PN
in phase given in \cite{ABIQ04}. Such a waveform is characterized in terms of
nine parameters given in Eq.~\eqref{eq:param}. We use the Fisher information
matrix approach to estimate all parameters of the source (see
Sec.~\ref{subsec:errest} for the discussion). Our findings have been presented
in Sec.~\ref{sec:results}. 

We discuss our results in three different subsections. In
Sec.~\ref{subsec:RWF&FWF} we compare the accuracies with which various
parameters of the source can be measured using both the RWF and FWF
approximation to the inspiral signal mainly in context of three representative
networks, namely, the LIGO-Virgo network (LHV), the LIGO-Virgo-KAGRA network
(LHVK) and the LIGO-Virgo-KAGRA network after including LIGO-India (LHVKI).
Although the median of the error distributions associated with each parameter for
all 16 possible combination of three, four, and five detectors has been displayed in
Table~\ref{tab:medianerror}, we find that for a given network the use of the
FWF in general improves the parameter estimation for various parameters.
However, the effect is more prominent in case of four parameters, namely, the
distance ($D_L$), the inclination angle of the binary ($\cos(\iota)$), the
polarization angle ($\psi$) and the phase at the coalescence epoch ($\Phi_c$).
The related error distributions have been presented in
Figs.~\ref{fig:compRvsFLHV}-\ref{fig:compRvFLHVKI}. Upon comparing the median
errors displayed in figures as well as in Table~\ref{tab:medianerror} we find
that, given the network under consideration, the errors in $D_L$ and
$\cos(\iota)$ improve roughly by a factor of $1.5-2$ whereas those related to
$\psi$ and $\Phi_c$ improve roughly by a factor of $1.2-1.6$. We also notice that
the factor of improvement is larger for detector networks with fewer
detectors. For instance, the factor of improvement in the LHV case reduces from the
value of about 2 to about 1.5 for LHVKI case.  This trend is in general true
for all parameters. This is not very surprising as the inclusion of additional
detector sites breaks the degeneracy in angular parameters such as $\psi$ and
$\Phi_c$, which in turn  improves the error estimation even for the RWF case,
diluting the importance of the use of FWF.  Measurement of other parameters
does not quite improve with the use of the FWF (see Table~\ref{tab:medianerror}
and the discussion presented in related subsection).

In Sec.~\ref{subsec:compNetwork} we compare our parameter estimation results
obtained using the FWF for three representative networks (LHV, LHVK, LHVKI). As
mentioned in the beginning of Sec.~\ref{subsec:compNetwork}, although the
choice of these networks for displaying our main results is mainly based on the
time line argument that when various detectors would start operating, we find
that they can indeed be chosen as representatives of the three, four, and five-detector
networks. As should be clear from 
Figs.~\ref{fig:compNetwork1}-\ref{fig:compNetwork2} and the median errors
displayed there, although in general the parameter estimation improves for all
parameters when we add KAGRA and LIGO-India to the LIGO-Virgo network, the
improvement is most significant in the case of angular resolution. The angular
resolution improves almost by a factor of 2.5 with the addition of KAGRA to the
LHV network where as the same improves almost by a factor of 4.5 when
LIGO-India is added to the LHVK network. Again we refer to
Table~\ref{tab:medianerror} for comparing the parameter estimation accuracies
for all 16 possible combinations of three, four, and five-detector networks.

Finally, in Sec.~\ref{subsec:LI}, we discuss in particular the benefits of
adding the LIGO-India detector to the LHVK network. In addition to our
conclusions based on comparisons of different networks presented in
Sec.~\ref{subsec:compNetwork}, in this section we basically argue how the
addition of LIGO-India detector would help achieving scientific objectives.

Table~\ref{tab:medianerror-fixedsnr} corresponds to a case when the errors
listed in Table~\ref{tab:medianerror} has been rescaled so that all errors
would correspond to a SNR of 20. The reasons  behind displaying such a table are
manyfold. First and foremost it helps us quantifying various effects which
play an important role in the measurement of various parameters apart from the
SNR. For instance, after comparing the FWF and RWF numbers for $D_L$ and
$\cos(\iota)$ errors in the two tables we find that the improvement is actually
coming from the fact that the use of FWF helps breaking the $D_L$-$\iota$
degeneracy which persists in the case of RWF and SNR indeed plays no role here.
Similarly, it also helps in quantify effects of having a detector network with
larger areas while comparing different networks. The other reason for including
the table is related to the fact that although different networks have
different distance reach in our main analysis we choose to keep the sources at
200 Mpc for all network configurations. Ideally one should keep sources at
different distances for different detector networks as the horizon distance for
each network is different.  By fixing the SNR, this issue is automatically
resolved since for networks with larger horizon distance, the errors would be
rescaled to values that actually correspond to the source at larger distance
and vice versa. Finally, in practice sources will probably be observed with an
SNR of about 20 or so. The errors displayed in
Table~\ref{tab:medianerror-fixedsnr} present a more realistic scenario, which we
might witness in the coming years of GW astronomy.

In Table~\ref{tab:medianerror_allLIGO} we show the median errors with
hypothetical detector networks in the case when all of the detector noise power
spectrum is given by that of advanced LIGO and all sources are located at
200 Mpc.  In Table~\ref{tab:medianerror_allLIGO-fixedsnr}, we also show the
median errors with hypothetical detector networks in the case when all of
the detector noise power spectrum is given by that of advanced LIGO, and the
SNR is rescaled to 20.  Differences between Table~\ref{tab:medianerror} and
\ref{tab:medianerror_allLIGO}, and between
Table~\ref{tab:medianerror-fixedsnr} and \ref{tab:medianerror_allLIGO-fixedsnr}
are caused by the differences in the noise power spectrum of Virgo and KAGRA.
In Sec.~\ref{subsec:compNetwork}, by comparing
Table~\ref{tab:medianerror-fixedsnr} and
\ref{tab:medianerror_allLIGO-fixedsnr}, we found that some unusual trends of
the median errors in the mass parameters, ${\cal M}_c$ and $\delta$, in
Table~\ref{tab:medianerror-fixedsnr} were caused by the differences in the noise
power spectrum of LIGO, Virgo, and KAGRA.

Although the Fisher analysis can be used to get a fair idea about the quality
of the parameter estimation that can be achieved in future, it assumes ideal
situations (such as the use of Gaussian noise) and merely provides the lower
bound on errors with which various parameters can be measured. Moreover, the
method is limited to the signals of high strength. In order to have a more
realistic estimate of parameters of the GW source, one has to perform more
realistic simulations, such as those based on Bayesian inference with real data,
which are applicable to signals with arbitrary strength. However, such methods
are quite expensive especially since one has to repeat the exercise for
different noise realizations.  Proposed variants of the Fisher matrix, such as 
effective Fisher matrix~\cite{Cho:2012ed}, can also be used to carry out
similar studies.  In addition, Ref.~\cite{Vallisneri:2011ts} provides a
semianalytical technique to perform parameter estimation for signals of
arbitrary strength. One can expect that this approach might be computationally
bit cheaper, but an actual analysis based on this proposal is yet to be made. 

Besides these limitations of the Fisher analysis, the importance of the effect of 
 abrupt termination of the waveform at $f_{\rm LSO}$ was pointed out recently in
\cite{MBOFF2014}. 
The LSO frequency of gravitational waves is given as $1/(6^{3/2}\pi M)$ for RWF
and  $k/(6^{3/2}\pi M)$ for the $k$th harmonic mode of FWF, respectively. 
Since these values depend on the total mass, 
when we approximate the likelihood function by using the Taylor expansion 
around the true value of parameters, 
we have to take into account the dependence of $f_{\rm LSO}$ on the total mass.
In this paper, we have not taken into account such an effect. 
However, as discussed in \cite{MBOFF2014}, 
if the detector's noise dominates the signal at the frequency of $f_{\rm LSO}$,
this effect can be neglected. 
In Fig. 3 of \cite{MBOFF2014}, they compare the statistical uncertainty of the chirp mass 
$\sigma_{M_c}/M_c$ and the systematic bias, $\Delta{\hat{M}_c}/M_c$, produced 
by the mass-dependent LSO frequency.
They consider the case of RWF, $m_1=1.35M_\odot$ and $m_2= 5-20M_\odot$, 
and the advanced LIGO noise spectrum which is the same as this paper.
They found that, in the case of signal-to-noise ratio of 10, 
the systematic bias due to $f_{\rm LSO}$ dominates the statistical uncertainty 
if $m_2\gtrsim 11M_\odot$. 
In the case of the source distance of 200Mpc in our simulation, 
the average signal-to-noise ratio at three LIGO detectors is around $12\sim 13$ 
which is similar to the value. 
Thus, the effect of the cutoff at $f_{\rm LSO}$ might marginally affect the value of the parameter estimation 
errors in this paper. 
We expect that, as far as we are comparing the cases for FWF and RWF, and 
are comparing  combinations of various detectors, 
the effect of the cutoff at $f_{\rm LSO}$ will not change the trend we observed in this paper,
since the cutoff at $f_{\rm LSO}$ may affect the results of all cases in a similar way. 
Nevertheless, in order to obtain the definite answer to this, 
we need to investigate the effect of $f_{\rm LSO}$ for FWF and for the case of the network of detectors. 

Finally, we want to point out two important effects in the waveform modeling
that we have not accounted for, which can significantly affect our estimates.
First is the neglect of the spin effects in modeling the binary system. Though
it may be safe to neglect the spin of the NS, the BH in the binary system may
be spinning in which case our nonspinning waveforms are not adequate to
describe such a system. If the BH spins are not aligned with respect to the
orbital angular momentum axis of the binary, there can be precessional effects
as well. One may want to revisit the problem accounting for the spin effects,
say using the waveforms of ~\cite{ABFO08},  in future. The second effect we
have completely ignored is the finite size effects related to the NS in the
binary. Though formally the finite size effects are a 5PN in the phasing (1.5PN
higher than our current 3.5PN accuracy), these effects may become significant
towards the late stages of the inspiral~\cite{FH08,1PNTidal2011}. 
Inspiral-merger-ringdown waveform models which take into account the tidal effect
in NS-BH binaries have been developed~\cite{Lackey:2013axa},
and the prospect of extracting equation of state parameters from the waveform is discussed~\cite{Lackey:2013axa}.
The waveforms are calibrated to the results of the numerical relativity simulations~\cite{lrr-2011-6}.
The tidal effects and the merger-ringdown phases are completely ignored in our analysis, 
and it may be worth revisiting the parameter estimation problem 
with the network of detectors 
by using the above-mentioned waveform models.
\begin{acknowledgments} This work is supported by the Department of Science and
Technology (DST) and the Japan Society for the Promotion of Science (JSPS),
Indo-Japan international cooperative program, Grant No. DST/INT/JSPS/P-127/11.
C. K. M. was supported in part by the MPG-DST Max Planck Partner Group on Gravitational
Waves.  H. T. was also supported in part by Grant-in-Aid for Scientific Research
(C) No. 23540309, Grant-in-Aid for Scientific Research (A) No. 24244028, and
Grant-in-Aid for Scientific Research on Innovative Areas No. 24103005.  
This work was also supported by the JSPS Core-to-Core Program, A. Advanced
Research Networks. 
The project used the octave-based codes developed by Roby Chacko, a project
assistant under the AP's SERC Fast-Track Scheme.  We gratefully acknowledge
useful discussions with S. Dhurandhar, B. S. Sathyaprakash, I. Mandel,
H.Takahashi, N.Kanda and Vivien Raymond. We thank C. Capano for reviewing our
paper and for his suggestions.  A. P. and K. G. A. would like to thank Osaka University
for hospitality during the spring of 2013. K. G. A. thanks IISER-TVM for hospitality during
different phases of the project. H. T.  would like to thank IISER-TVM for
hospitality.  
This is a LIGO document, LIGO-P1400030. 
\end{acknowledgments}
\bibliography{ref-list} \end{document}